\documentclass[twocolumn]{aastex631} 
\usepackage[utf8]{inputenc}
\usepackage{savesym}
\usepackage{amsmath,amssymb}
\savesymbol{longtable}
\usepackage{xfrac}
\usepackage{numprint}
\usepackage{rotating}
\usepackage{graphicx}
\usepackage{float}
\usepackage{multirow}
\usepackage[T1]{fontenc}
\usepackage[caption=false]{subfig}
\usepackage{booktabs}
\usepackage{hyperref}
\graphicspath{{./}}
\def\trunc\ignorespaces#1\\{%
        \FPset\a{#1}
        \FPround\a\a4
        \a
        \\
}
\begin{document}

\title{Optical Variability Properties of Southern TESS Blazars}

\author[0009-0007-0454-6245]{Ryne Dingler}
\affiliation{Texas A\&M University,
576 University Dr,
College Station, TX 77840, USA}
\affiliation{Southern Methodist University,
3215 Daniel Ave,
Dallas, TX 75205, USA}

\author{Krista Lynne Smith}
\affiliation{Texas A\&M University,
576 University Dr,
College Station, TX 77840, USA}
\affiliation{Southern Methodist University,
3215 Daniel Ave, 
Dallas, TX 75205, USA}

\begin{abstract}
We present a study of high-cadence, high-precision optical light curves from the TESS satellite of 67 blazars in the southern sky. We provide descriptive flux statistics, power spectral density model parameters, and characteristic variability timescales. We find that only 15 BL Lacertae objects (BLLs) and 18 Flat Spectrum Radio Quasars (FSRQs) from the initial 26 and 41, respectively exhibit statistically-significant variability. We employ an adapted Power Spectral Response method to test the goodness of fit for the power spectral density function (PSD) to 3 power-law variant models. From our best-fitting description of the PSD, we extract the high-frequency power-spectral slopes and, if present, determine the significant bend or break in the model to identify characteristic timescales. We find no significant difference in the excess variance or rms-scatter between blazar subpopulations. We identify a linear rms-flux relation in $\sim69\%$ of our sample, in which $\sim$20\% show a strong correlation. We find that both subpopulations of blazars show power spectral slopes of $\alpha\sim$2 in which a broken power-law best fits 5 BLL \& 6 FSRQ and a bending power-law best fits 1 BLL \& 5 FSRQ. The shortest timescales of variability in each light curve range widely from minutes to weeks. Additionally, these objects' characteristic timescales range from $\sim$0.8-8 days, consistent with optical variability originating in the jet.
\end{abstract}


\section{Introduction}
    \label{sec:section1}
    
    Radio-loud Active Galactic Nuclei (AGN) with near line-of-sight oriented jets form a subclass  known as blazars. These objects demonstrate Doppler-boosted, rapid, high-amplitude variability in flux over timescales from minutes to years at all wavelengths (radio-TeV) and in polarization in the bands where it can be measured. The small-viewing angle ($\theta\lesssim10^{\circ}$) of blazars results in relativistic beaming of jet emission and blue-shifted, Doppler amplified luminosity and shortened apparent timescales. These offer a direct probe into the inner regions of AGN and the blazar central engine \citep[e.g., ][]{Urry1995, Blandford2019, Hovatta2019, Raiteri2020}.

Blazars are further categorized into two sub-populations: in the rest frame, Flat Spectrum Radio Quasars (FSRQs), whose spectra show only broad emission lines with equivalent width $> 5\text{\AA}$ and BL Lacertae objects (BLLs) whose spectra show  no emission lines or weak ones of equivalent width $< 5\text{\AA}$, \citep{Stickel1991,Stocke1991,Urry1995, Ghisellini2017, Blandford2019, Hovatta2019}. FSRQs form the majority of \textit{Fermi}-detected blazars and generally display a greater amplitude of $\gamma$-ray variability; however, they also tend to have soft $\gamma$-ray spectra, and few emit in the TeV range. BLLs have harder $\gamma$-ray spectra and show  greater amplitudes of variation in the very high-energy (VHE) regime. The key differences between these classifications is expected due to differences in the observed emission mechanism: the emission is likely purely jet-generated in BLL objects, whereas FSRQs may have additional contributions from the accretion disk or torus \citep{Fossati1998, Ghisellini2017, Blandford2019}.

Of particular interest in the study of optical variability properties is the Power Spectral Density (PSD). The stochastic nature of blazar light curves elicits the need to describe variability amplitude as a function of the Fourier frequencies (inverse of variability timescales). In most cases, a power-law $\mathcal{P}(\nu) \propto \nu^{-\alpha}$ models this function quite well. The value of $\alpha$, the power spectral index, describes the color-noise type of stochastic processes in which $\alpha \rightarrow 0$ represents a white-noise component that dominates at those timescales corresponding to random, uncorrelated processes. Random variations within these processes give rise to statistical fluctuations, white noise, and non-stationarity within time series. Such fluctuations are intrinsic to red-noise signals, which, along with Poisson noise, increase the variance about the underlying PSD. However, the PSD can still be described predictably by a probability distribution, allowing for a statistical description and study of average variability through power spectral replication and simulation of artificial light curves \citep[e.g., ][]{Vaughan2003, Emmanoulopoulos2013, Smith2018,Sartori2019}. 

A power spectral index $\alpha \approx 2$ describes the red-noise behavior consistently observed in X-ray and optical/infrared power spectra; however a simple power law cannot always fully describe the PSD. Some PSDs exhibit a flattening of $\alpha$ in the low-frequency regime, causing a ``break'' or ``bend'' in the power-law model, which may be indicative of characteristic variability timescales related to physical parameters of the system \citep[e.g., ][]{Uttley2002, Vaughan2003, Edelson2013, Kasliwal2015, Goyal2021}. Analyses of PSDs over multiple wave bands have reported $\alpha \approx$ [1-3] for short- and long-term variability showing general consistency with the ``damped random walk" model \citep{Chatterjee2012, Ackermann2016,Nilsson2018, Ryan2019, Raiteri2020,Bhattacharyya2020,Pininti2022,Wehrle2023}.

For years, $\gamma$-ray and X-ray observations have received a great deal of attention in analyzing blazar light curves, particularly over long timescales. However, AGN are variable over all wavelengths and observations in optical bands have been frequent even if they can rarely be continued for long periods \citep{Carini2004, Kurtandize2005, Abdo2010, Patino2013, Sandrinelli2014, Ackermann2016, Raiteri2020}. The need for a complete understanding of the physical mechanisms of variability in the accretion disk and jet has prompted increasing interest and importance on short-term, intra-day, timescales in the optical and infrared bands \citep{Kurtandize2005,Bonning2012,Giommi2013,Abrahamyan2019,Carini2020,Pininti2022,Wehrle2023}. Recently, light curves from the Transiting Exoplanet Survey Satellite \citep[TESS;][]{Ricker2015} have become of interest due to the fast sampling rate and months to years long, observations that access both short- and sometimes also long-timescales. The potential for such analyses increases as TESS continues its full-sky observation at nominal cadence intervals \citep{Raiteri2020,Kishore2023}. 

A recent study \citep{Pininti2022} has investigated the short-term optical variability in 29 TESS blazars. They have, as previously, observed significant short-timescale variability, but found no significant breaks or bends beyond a simple power-law model. However, \citet{Pininti2022} utilizes uncorrected, Simple-Aperture Photometry (SAP) fluxes in an attempt to retain intrinsic variability. While a single ground-based observation validated some reliability of these light curves, the effects of long-term systematics by backgrounds and TESS instrumentation\footnote{\url{https://heasarc.gsfc.nasa.gov/docs/tess/documentation.html}} are non-negligible. These effects prompt the necessity of a pipeline that can correct SAP fluxes through Principal Component Analysis (PCA) and/or use of Co-trending Basis Vectors (CBV) without removing intrinsic source variability\citep{Brasseur2019,Feinstein2019,Burke2020,Ridden-Harper2021,Powell2022,Smith2023}.
    
If FSRQs and BLLs are driven by physically distinct mechansims, or have a different balance between disk- and jet-driven emission, we would expect this to manifest in statistically distinct variability properties. This has been explored in gamma rays, where FSRQs are found to be more variable, and to vary with higher amplitudes, than BLLs \citep{Abdo2010,Rajput2020,Tarnopolski2020}. The rapid, continuous cadence and high photometric precision of the TESS mission allows us to explore this in detail in the optical. This study presents a detailed analysis of the optical variability properties of the TESS light curves of a sample of BLLs and FSRQs. Unlike previous studies, we meticulously correct our light curves using sophisticated spacecraft systematics modeling. We present variability statistics and, for all objects within our sample that demonstrate a sufficient level of variability above the photometric errors, power spectral modeling for both subsamples, and discuss the physical implications of our results.

This paper shall first describe our sample and data collection processes in Section \ref{sec:section2}. Section \ref{sec:section3} provides a comprehensive description of our analyses, each beginning with our methodology before presenting the results. A discussion is in Section \ref{sec:discussion} and a summary of our conclusions are in Section \ref{sec:summary}. Lastly, within the appendices, we provide:  (\ref{sec:lightcurves}) the flattened light curves prior to mean subtraction as used in the PSD analysis for our sufficiently variable sub-sample; (\ref{sec:errors}) a discussion of errors;  (\ref{sec:timescales}) examples of our investigation of the shortest rise/decay times for significant change in flux; (\ref{sec:CARMA}) a comparison of our method with CARMA models.

\section{Sample \& Data Collection}
    \label{sec:section2}
    
    \subsection{Observations by TESS}
    \label{sec:TESS}
    
    The TESS mission \citep{Ricker2015} conducts long-term monitoring with high photometric precision of stars across nearly the entire sky with a rapid cadence: every 30 minutes in the early cycles (2018 -- 2019) and every 10 minutes or 200 seconds in later and current cycles (2020 -- present). The coverage is continuous except for a $\sim$1-day gap every orbit (approximately every two weeks). The 4 TESS cameras provide a combined field of view (FOV) of $24^{\circ}\times96^{\circ}$ in which one pixel covers $\sim$21 arcseconds across the sky.
    
    The monitoring baseline for a TESS source depends upon its ecliptic latitude. Most sources at low-to-middle ecliptic latitudes are monitored for 27 days before the spacecraft shifts to the next wedge of the sky. Each 27-day monitoring segment is referred to as a ``Sector''. As the latitudes approach the ecliptic poles, these Sectors overlap, resulting in a more extended monitoring baseline. At the poles,  $\sim$1 year of continuous monitoring is conducted. After this year of exposure, the spacecraft turns over and surveys the other ecliptic hemisphere for one year with the same pattern. Each year is referred to as a ``Cycle''.
    
    \subsection{Blazar Sample and Data Reduction}
    \label{sec:sample}
    Our initial target list consists of 67 likely-beamed AGN selected from Roma-BZCAT\footnote{\url{http://www.asdc.asi.it/bzcat}} \citep{Massaro2015} and spectroscopic classification in the \textit{Fermi} catalogs \citep{Ballet2023} if available. If spectroscopic classification is not provided in the \textit{Fermi} catalogs, we refer to the classification in BZCAT or the CRATES survey \citep{Healey2007}. Redshifts and R-magnitudes are obtained from BZCAT if available. If not we defer to the available value on NED as taken from either \citet{Wisotzki2000,Jones2009} or \citet{Makarov2019}. Black hole mass ($M_{BH}$) estimates are provided by \citet{Paliya2021} if available. If no estimate is provided, we assume the average mass of blazar black holes, $M_{BH}\sim 10^{8.5}$, as calculated by \citet{Xiong2015} and \citet{Paliya2019}.
    
\begin{figure}
        \centering
        \includegraphics[width=0.5\textwidth]{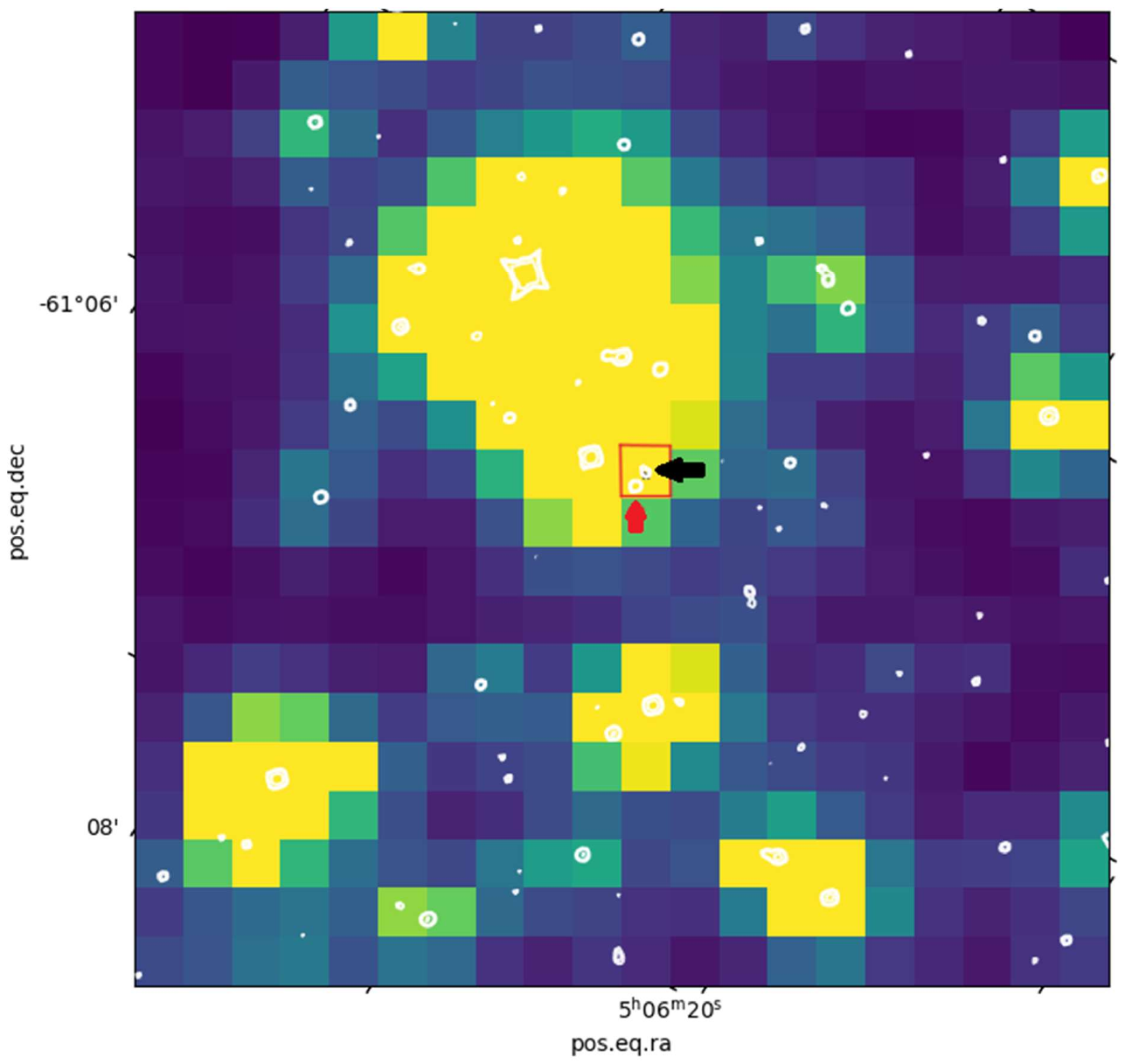}
        \caption{$20\times20$p TPF for sector 5 observation of object PKS$~0506-61$ with provided DSS contours. The provided contours demonstrate a brightness of $0.4\times$ the brightest pixel. The red box indicates an optimal extraction region while the arrows indicate the target (black) a contaminating object (red) which \textit{cannot} be consistently avoided when selecting an aperture. We do not consider objects like this, which do not fit our criterion for analysis.}
        \label{fig:PKS0506TPF}
    \end{figure}

    This study presents the largest sample of blazars for time-series and power spectral analysis, with optical light curves at a sampling rate and baseline comparable to X-ray and $\gamma$-ray observations. These targets must be: 1) bright enough for TESS observation (R < 19.8) \citep{Fausnaugh2021}; 2) located in the southern ecliptic hemisphere TESS Cycle 1 viewing zone in 2018-2019; 3) not seriously affected by TESS systematics; and 4) be in a detector region that is not significantly overcrowded by contaminating sources; an example of a source rejected due to source contamination in TESS's large pixels is shown in Figure~\ref{fig:PKS0506TPF}. By these criteria, we produce an original sample of 22 BL Lacertae objects (BLLs) and 45 Flat Spectrum Radio Quasars (FSRQs), of which 13 are not \textit{Fermi}-detected. These were eventually reduced to 33 clearly variable objects (20 FSRQs and 13 BLLs) for our final analysis. The mechanism which differentiates these subclasses is currently unclear, but is likely to arise due to a difference in the prominence of jet emission versus disk emission; indeed, objects have even been known to transition between subclasses across different epochs \citep[e.g., ][]{Ghisellini2011}. Results of ground- and space-based observations from gamma-rays to optical, have shown disagreement on whether the variability properties between these subclasses are statistically different, as would be expected if the accretion disk or jet, which are best-modelled by different underlying phenonema, were dominant.\citep{Ghisellini2009,Bauer2009,Giommi2013,Ryan2019,Wehrle2023}. We divide our sample into the sub-types based on the best available archival classification, but note that these classifications may not be fundamental. We provide a complete list of this sample in Table \ref{tab:sample}.

    \begin{widetext}
        \begin{longdeluxetable}{lrrccrcccc}
    \centering
    \tabletypesize{\scriptsize}
    \tablewidth{0pt} 
    \tablenum{1}
    \tablecaption{\label{tab:sample} Table of all beamed AGN analyzed in this study. The sample is separated by a horizontal line to indicate which objects were found to be significantly variable (above) and those with $\sigma^{2}_{XS}$ consistent with zero (below). Targets marked as $^{*}$ indicate that the ``simple" light curve may have been affected by background contamination or systematic effects and was thereby re-extracted using the ``full" hybrid method.}
    \tablehead{
    \colhead{Target} &
    \colhead{R.A.} & \colhead{Dec.}  & \colhead{z} & \colhead{$\delta$} & \colhead{Type} & \colhead{R-Magnitude} & 
    \colhead{$F_\gamma$ (1-100 GeV)}\\
    \colhead{} & \multicolumn{2}{c}{[deg]} & \colhead{[NED]} & \colhead{} & \colhead{} & \colhead{} & 
    \colhead{[$\times10^{-10}$ photons/s/cm$^{2}$]}
    }
    \startdata 
    1ES2322-409 & 351.186 & -40.680 & 0.174 & - & BLL & 15.7 & 9.98\\
    1RXSJ054357.3-553206 & 85.988 & -55.535 & 0.273 &  - & BLL & 16.5 & 23.91\\
    3C120 & 68.296 & 5.354 & 0.033 & 4.16 & FSRQ & 13.8 & 7.59\\
    NGC1218$^{*}$ & 47.109 & 4.111 & 0.028 &  - & BLL & 10.2 & 8.32\\
    PKS0035-252 & 9.561 & -24.984 & 0.498 &  - & FSRQ & 19.1 & 36.31\\
    PKS0130-17 & 23.181 & -16.913 & 1.020 & 12.53 & FSRQ & 18.6 & 21.07\\
    PKS0208-512 & 32.693 & -51.017 & 0.999 &  - & FSRQ & 14.8 & 90.16\\
    PKS0226-559 & 37.090 & -55.768 & 2.464 &  - & FSRQ & 19.8 & 39.71\\
    PKS0235-618 & 39.222 & -61.604 & 0.467 &  - & FSRQ & 17.6 & 8.18\\
    PKS0301-243 & 45.860 & -24.120 & 0.266 &  - & BLL & 15.4 & 46.34\\
    PKS0336-177 & 54.807 & -17.600 & 0.066 &  - & BLL & 11.0 & 7.55\\
    PKS0346-27 & 57.159 & -27.820 & 0.991 &  - & FSRQ & 17.7 & 186.37\\
    PKS0420-01 & 65.816 & -1.343 & 0.916 & 5.31 & FSRQ & 16.7 & 32.35\\
    PKS0422+00 & 66.195 & 0.602 & 0.310 & 5.73 & BLL & 14.3 & 15.78\\
    PKS0426-380 & 67.168 & -37.939 & 1.105 &  - & BLL & 16.3 & 194.05\\
    PKS0454-234 & 74.263 & -23.414 & 1.003 &  - & FSRQ & 17.9 & 189.61\\
    PKS0521-36 & 80.742 & 36.459 & 0.057 &  - & BLL & 14.6 & 40.15\\
    PKS0637-75$^{*}$ & 98.944 & -75.271 & 0.653 &  - & FSRQ & 15.7 & 6.43\\
    PKS0736+01 & 114.825 & 1.618 & 0.189 & 11.44 & FSRQ & 16.0 & 58.12\\
    PKS0829+046 & 127.954 & 4.494 & 0.174 & 5.94 & BLL & 15.4 & 29.86\\
    PKS1244-255 & 216.985 & -42.105 & 0.633 &  - & FSRQ & 15.9 & 71.69\\
    PKS2155-304 & 329.717 & -30.226 & 0.117 &  - & BLL & 12.8 & 190.94\\
    PKS2155-83 & 330.580 & -83.637 & 1.865 &  - & FSRQ & 19.3 & 17.14\\
    PKS2255-282 & 344.525 & -27.973 & 0.926 &  - & FSRQ & 17.2 & 36.90\\
    PKS2326-502 & 352.337 & -49.928 & 0.518 &  - & FSRQ & 19.7 & 86.58\\
    PKS2345-16 & 357.011 & -16.520 & 0.576 & 16.58 & FSRQ & 18.1 & 48.96\\
    PMNJ0948+0022 & 147.239 & 0.374 & 0.584 & 29.77 & FSRQ & 18.4 & 19.00\\
    PMNJ2141-6411 & 325.444 & -64.187 & 0.959 &  - & FSRQ & 18.9 & 23.10\\
    PMNJ2345-1555 & 356.302 & -15.919 & 0.621 & 30.23 & FSRQ & 18.5 & 76.38\\
    WISEJ031612.72+090443.3 & 49.053 & 9.079 & 0.372 &  - & BLL & 17.2 & 18.08\\
    WISEJ085009.64-121335.3 & 132.540 & -12.226 & 0.566 & 17.87 & FSRQ & 18.4 & 18.00\\
    WISEJ095302.70-084018.2 & 148.261 & -8.672 & 0.400 &  - & BLL & 17.4 & 23.58\\
    WISEJ105912.43-113422.6 & 164.802 & -11.573 & - & 27.29 & BLL & 18.0 & 32.73\\
    \hline
    1RXSJ120417.0-070959$^{*}$ & 181.069 & -7.169 & 0.184 &  - & BLL & 14.6 & 7.08\\
    2MASSJ02271658+0202005$^{*}$ & 36.819 & 2.033 & 0.456 &  - & BLL & 18.9 & 6.18\\
    2MASXJ04010663-1606391 & 60.278 & -16.111 & 0.032 &  - & FSRQ & 10.2 & -\\
    2MASXJ04390223+0520443$^{*}$ & 69.759 & 5.345 & 0.208 &  - & FSRQ & 15.3 & -\\
    2MASXJ08471294+1133497 & 131.804 & 11.564 & 0.198 &  - & FSRQ & 16.6 & 4.04\\
    2MASXJ11255199-0742215$^{*}$ & 171.467 & -7.706 & 0.279 &  - & BLL & 15.2 & 4.40\\
    2MASXJ12571157-1724344$^{*}$ & 194.298 & -17.410 & 0.047 &  - & FSRQ & 12.6 & -\\
    4C+14.23$^{*}$ & 111.320 & 14.420 & 1.039 & 55.86 & BLL & 17.5 & 33.55\\
    NRAO190 & 70.661 & -0.295 & 0.844 & 12.61 & BLL & 17.9 & 28.31\\
    NVSSJ114117-140447$^{*}$ & 175.324 & -14.080 & 0.886 &  - & FSRQ & 15.3 & 2.45\\
    PKS0002-170 & 1.325 & -16.801 & 0.775 &  - & FSRQ & 18.4 & -\\
    PKS0202-17 & 31.240 & -17.022 & 1.739 & 3.19 & FSRQ & 17.2 & 19.12\\
    PKS0217-133$^{*}$ & 35.118 & -13.089 & 1.445 &  - & FSRQ & 16.1 & -\\
    PKS0217-189 & 34.838 & -18.711 & 0.600 &  - & BLL & 18.3 & -\\
    PKS0244-470 & 41.500 & -46.855 & 1.385 &  - & FSRQ & 17.9 & 19.22\\
    PKS0310+013 & 48.182 & 1.555 & 0.664 & 11.77 & FSRQ & 18.0 & 4.34\\
    PKS0403-13 & 61.392 & -13.137 & 0.571 & 6.53 & BLL & 16.6 & 6.68\\
    PKS0406-127 & 62.274 & -12.647 & 1.563 & 39.61 & FSRQ & 18.4 & -\\
    PKS0454+039$^{*}$ & 74.197 & 4.015 & 1.345 &  - & FSRQ & 16.3 & -\\
    PKS0855-19 & 134.522 & -19.844 & 0.659 &  - & FSRQ & 16.3 & 5.40\\
    PKS0859-14$^{*}$ & 135.570 & -14.259 & 1.332 & 7.72 & FSRQ &-\\
    PKS1004-217$^{*}$ & 151.693 & -21.989 & 0.331 &  - & FSRQ & 16.3 & 27.16\\
    PKS1005+007 & 152.048 & 0.500 & 0.098 &  - & BLL & 13.0 & 4.20\\
    PKS1045-18$^{*}$ & 162.028 & -19.160 & 0.595 & 12.47 & FSRQ & 17.1 & 1.35\\
    PKS1424-41$^{*}$ & 216.985 & -42.105 & 1.522 &  - & FSRQ & 16.3 & 360.38\\
    PKS1451-375$^{*}$ & 223.614 & -37.793 & 0.314 &  - & FSRQ & 16.1 & 3.79\\
    PKS2142-75$^{*}$ & 326.803 & -75.604 & 1.139 &  - & FSRQ & 15.9 & 32.72\\
    PKS2354-11 & 359.380 & -11.428 & 0.960 & 5.05 & FSRQ & 17.3 & -\\
    WISEJ023951.27+041621.2$^{*}$ & 39.964 & 4.273 & 0.372 & 18.71 & FSRQ & 17.3 & 3.46\\
    WISEJ025741.00-121201.3 & 44.421 & -12.200 & 1.391 &  - & FSRQ & 17.7 & 2.22\\
    WISEJ030923.28-055920.4 & 47.347 & -5.989 & 0.745 &  - & FSRQ & 18.6 & -\\
    WISEJ032613.94+022514.6 & 49.053 & 9.079 & 0.147 &  - & BLL & 15.3 & 6.76\\
    WISEJ102243.73-011302.2$^{*}$ & 155.682 & -1.217 & 1.285 &  - & FSRQ & 16.9 & 6.91\\
    WISEJ115217.21-084103.1$^{*}$ & 178.072 & -8.684 & 2.370 &  - & FSRQ & 18.0 & 13.28
    \enddata
\end{longdeluxetable}

    \end{widetext}
    
    \subsection{Light Curve Extraction and Correction}
    \label{sec:quaver}
    
    Public archival data from the TESS mission is available in full-frame images (FFIs), which are not photometrically calibrated. To extract the light curves for our study, we use the publicly available pipeline \texttt{quaver}, developed by our group specifically for TESS light curves of AGN (or other stochastically-varying sources) \citep{Brasseur2019,Ginsburg2019b, Smith2023}. 
    
    \texttt{Quaver} \citep{Smith2023} allows for interactive aperture selection, mitigating the effects of extended sources like host galaxies, and interactive cadence masking, in which the user can manually exclude times of poor spacecraft pointing or cosmic ray hits. Once extracted from the custom aperture, a combination of simple background subtraction (to remove the dominant additive effects of scattered light from the sun and moon) and linear matrix regression (to remove multiplicative effects that typically arise from electronic noise and other spacecraft systematics) correct the light curve.

    One user-adjustable parameter of \texttt{quaver} is the choice of either ``simple'' hybrid, ``full'' hybrid, or a straightforward principal component analysis method. The simple method matches ground-based data better but may only be partially capable of removing signals from background or systematics \citep{Poore2023, Smith2023}. 
    
    We initially extracted our entire sample under the ``simple'' method; however, when we believed we had identified residual contamination that provided potentially unreal variability, those light curves were re-extracted under the ``full'' method. Complete information about the different \texttt{quaver} methods can be found online\footnote{DOI:\dataset[10.5281/zenodo.8400525]{https://zenodo.org/records/8400525}} and in \citet{Smith2023}. A complete collection of final light curves for all sufficiently variable targets, can be found in Appendix \ref{sec:lightcurves}.

\section{Analysis \& Results}
    \label{sec:section3}

    \subsection{Time-domain Analysis}
    \label{sec:variability}
    To determine the amount of variability in our sample, we analyze the statistical properties, assuming a normal flux distribution, of two standard variability metrics,  excess variance, and characteristic timescales for each light curve. These are found in Appendix \ref{sec:lightcurves} (although before the removal of the longest-term linear trend). To provide a level of significance to the variability, we calculate the $\chi^{2}$ per degree of freedom,
    \begin{equation}
        \label{eqn:chi-sqr_dof}
        \chi^{2}_{d.o.f.} = \frac{1}{N-1} \sum_{i=1}^{N} \frac{(F_{i} - \overline{F})^{2}}{\xi_{i}^{2}} ,
    \end{equation}
    where $\overline{F}$ is the mean flux, \textit{N} is the number of data points, and $\xi_{i}$ are the photometric flux errors \citep{Sesar2007}.
    
    A considerable value of $\chi^{2}_{d.o.f.}$ demonstrates that the light curve significantly varies from its mean flux. We provide these measurements over three timescales: 30-minute cadences, 6-hour bins, and 12-hour bins. We offer consistent descriptive statistics for light curve flux distributions that are generally not normal when unbinned. This way, variability at short timescales can be examined in context to the flux level over long timescales via the rms-flux relation. Thereby, those parameters determined for the light curve as a whole are, at the same time, also calculated \textit{within} each 6- and 12-hour bin. A complete collection of figures for all light curves, excess variance calculations with confidence intervals, rms-flux plots, and flux distributions have been made publicly available as a compressed file\footnote{DOI:\dataset[10.5281/zenodo.11222806]{https://doi.org/10.5281/zenodo.11222806}}. Table \ref{tab:variability1} contains the results for the following analyses, including the means and 1$\sigma$ widths of the flux and subsequent flux distributions assuming normality. This table also provides the values of $\sigma^{2}_{XS}$ and error, the shortest timescales of variability for increasing and decreasing flux (along with their associated time separation) provided in days, and the $\chi^{2}$/d.o.f. according to Equation \ref{eqn:chi-sqr_dof} as an insight to the significance of the variability. These same results, excluding the shortest timescales of variability, are also provided for the light curves binned into 6- and 12-hour segments and given in Table \ref{tab:variability2}.
    
    \subsubsection{Excess Variance and RMS-scatter}
        \label{sec:excess_var}
        
        We utilize the excess variance, $\sigma^{2}_{XS}$, and rms-scatter, $\sigma_{rms}$, where   
        \begin{equation}
            \label{eqn:sigmaXS}
            \sigma^{2}_{XS} = \sigma_{F}^{2}-\overline{\xi^{2}};~~~\sigma_{rms} = \sqrt{\sigma^{2}_{XS}},
        \end{equation}
           to determine the level of variability above the flux uncertainty, in which $\sigma_{F}^{2}$ is the sample variance, and $\overline{\xi^{2}}$,    
        \begin{equation}
            \label{eqn:MSE}
            \overline{\xi^{2}} = \frac{1}{N}\sum_{i=1}^{N}\xi^{2}_{i},
        \end{equation}
        is the mean square error. This measure provides for an intrinsic variance of the source such that a value of $\sigma^{2}_{XS} \leq 0$ implies that no significant variability is observed \citep{Vaughan2003,Sesar2007}. Thereby, $\sigma^{2}_{XS}$ additionally serves as our limit parameter in that all objects considered variable enough for further analysis must have an excess variance inconsistent with 0 at a $5\sigma$ level, i.e., $\sigma^{2}_{XS} - 5 \times err(\sigma^{2}_{XS}) > 0$. Discussion of the errors on $\sigma^{2}_{XS}$ and $\sigma_{rms}$ is provided in Appendix \ref{sec:errors}. Note that all objects passing this criterion, i.e., all objects that have any statistically significant variability, are included in the upcoming analysis.

        \begin{figure}
            \centering
            \includegraphics[width=0.45\textwidth]{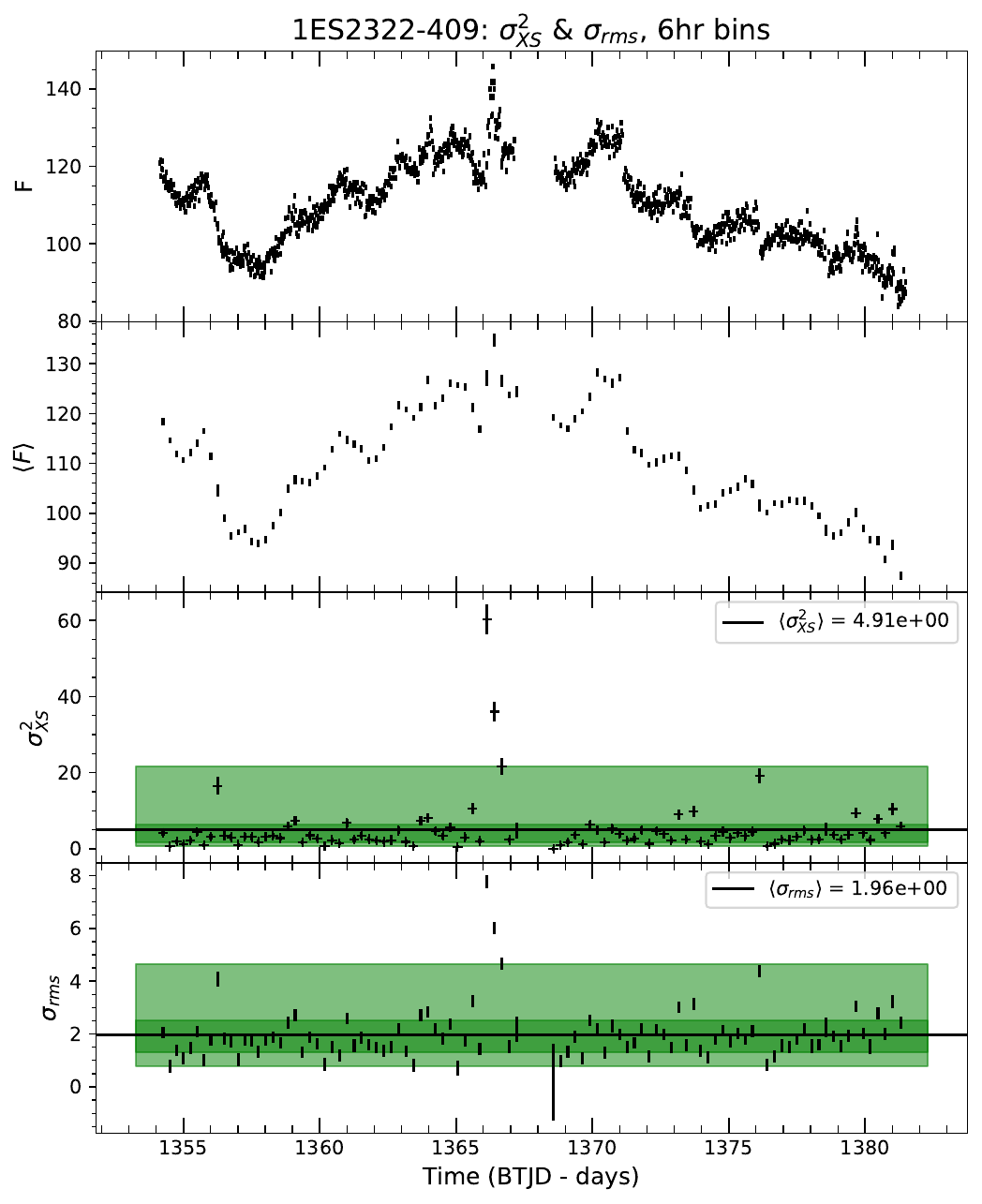}             
            \includegraphics[width=0.45\textwidth]{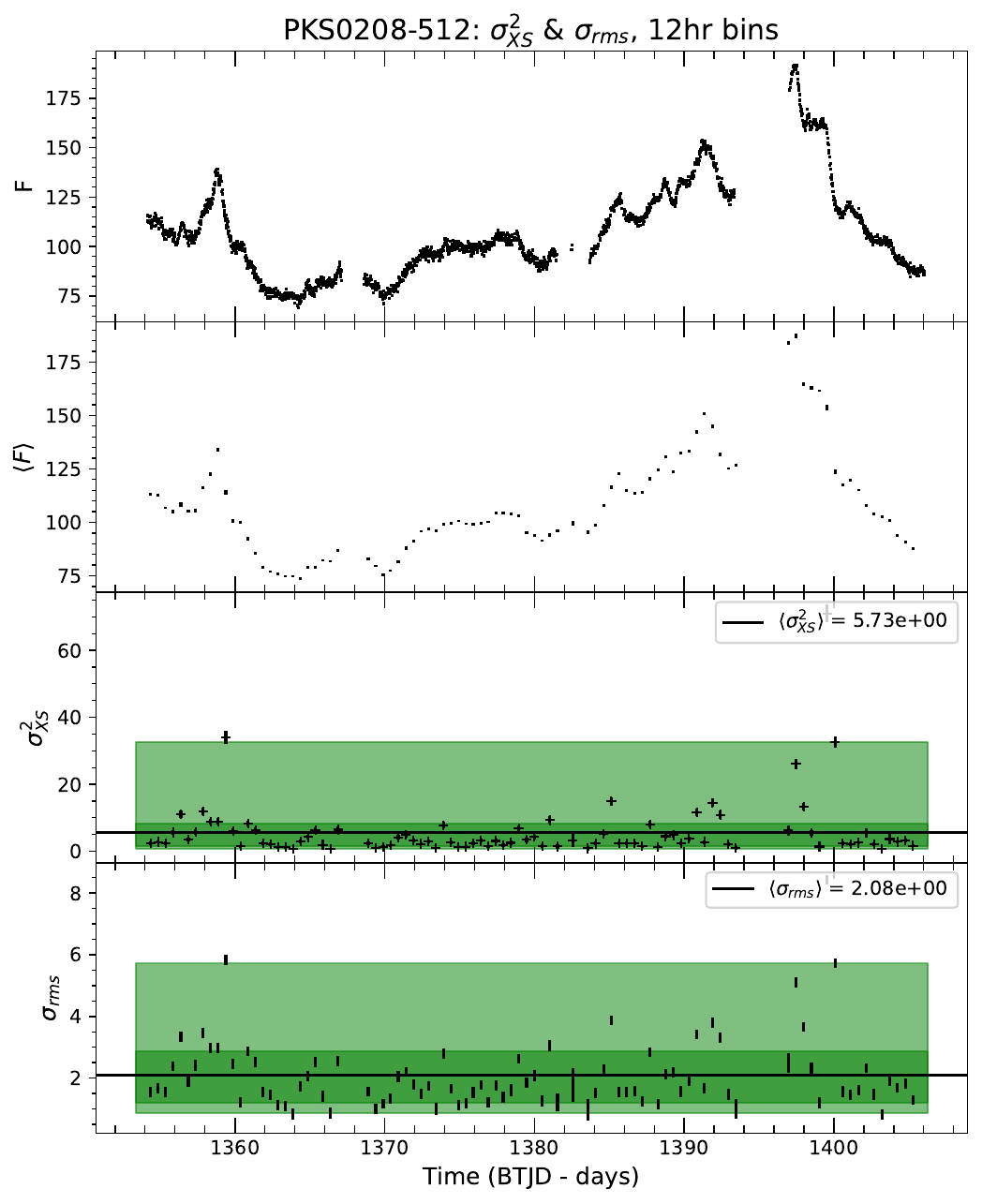}
            \caption{1ES2322-409 (top) under 6-hr binning scheme and PKS~208-512 (bottom) under 12-hr binning scheme. Light curve binning and intra-bin $\sigma^{2}_{XS}$ and $\sigma_{rms}$ (segments 1 \& 2); 68\% and 95\% confidence intervals (green) for intra-bin $\sigma^{2}_{XS}$ and $\sigma_{rms}$ (segments 3 \& 4).}
            \label{fig:excess_var_intrabin}
        \end{figure}

        Figure \ref{fig:excess_var_intrabin} illustrates our process of binning light curves and determining the level of intra-bin $\sigma^{2}_{XS}$ and $\sigma_{rms}$ when compared to the associated mean flux value of each bin.
        
        As determined by the excess variance of each light curve, produced by ``simple'' extraction, we found that 52 blazars from our original sample of 67 demonstrated a statistically significant level of variability above error within our stringent 5$\sigma$ limit. However, 21 of these light curves contained trends that suspiciously resembled systematic effects, such as thermal rising, which significantly skewed the variance. Additionally, they produced relatively flat, noisy power spectra. As a result, we then chose to re-extract these light curves under the ``full'' hybrid method (see Section~\ref{sec:quaver}). These 21 targets are identified in Table \ref{tab:sample} by an asterisk, of which we recovered light curves from only PKS~00637$-$75 and NGC~1218, reducing our selection to a final count of 33 objects (20 FSRQs and 13 BLLs), or 35 light curves
                
\begin{figure}
            \centering
            \includegraphics[width=0.5\textwidth]{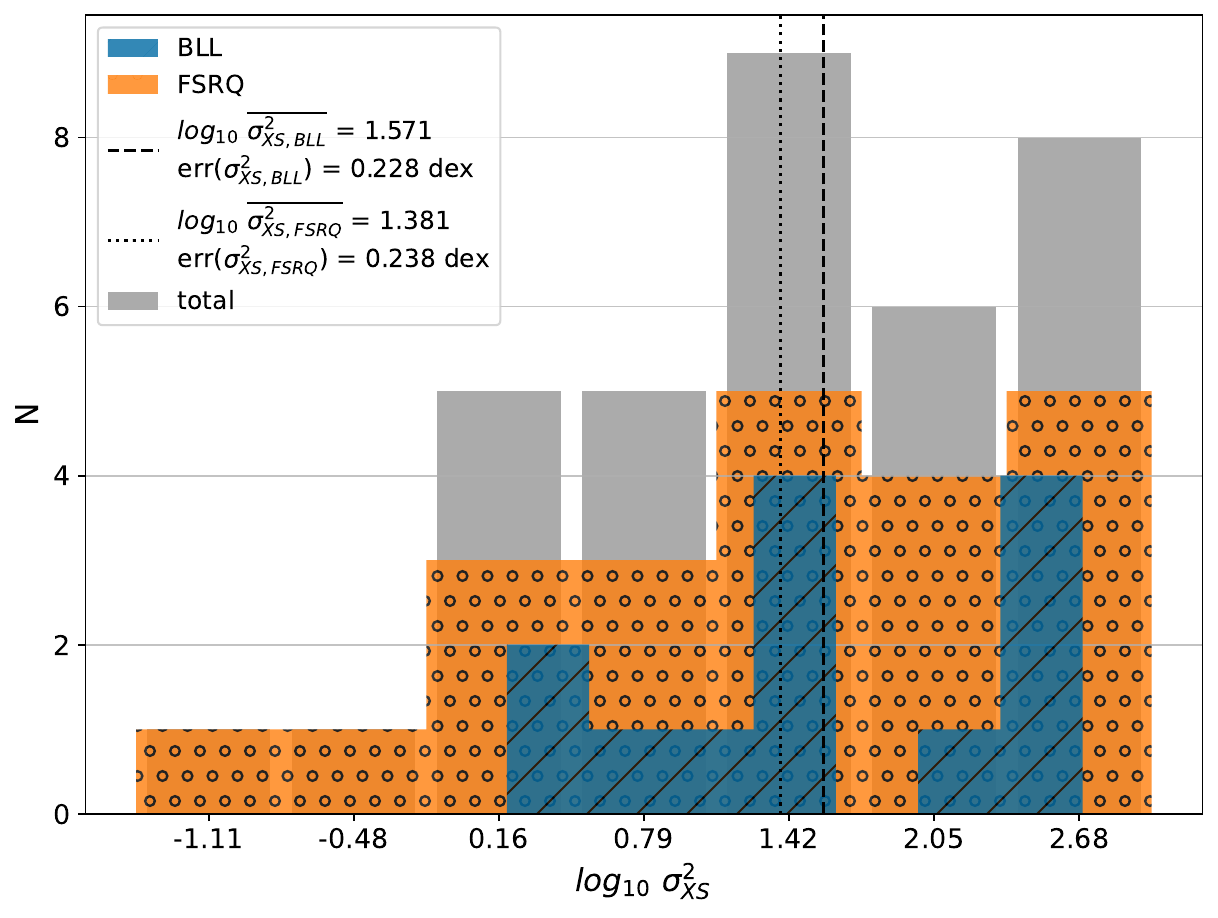}
            \includegraphics[width=0.5\textwidth]{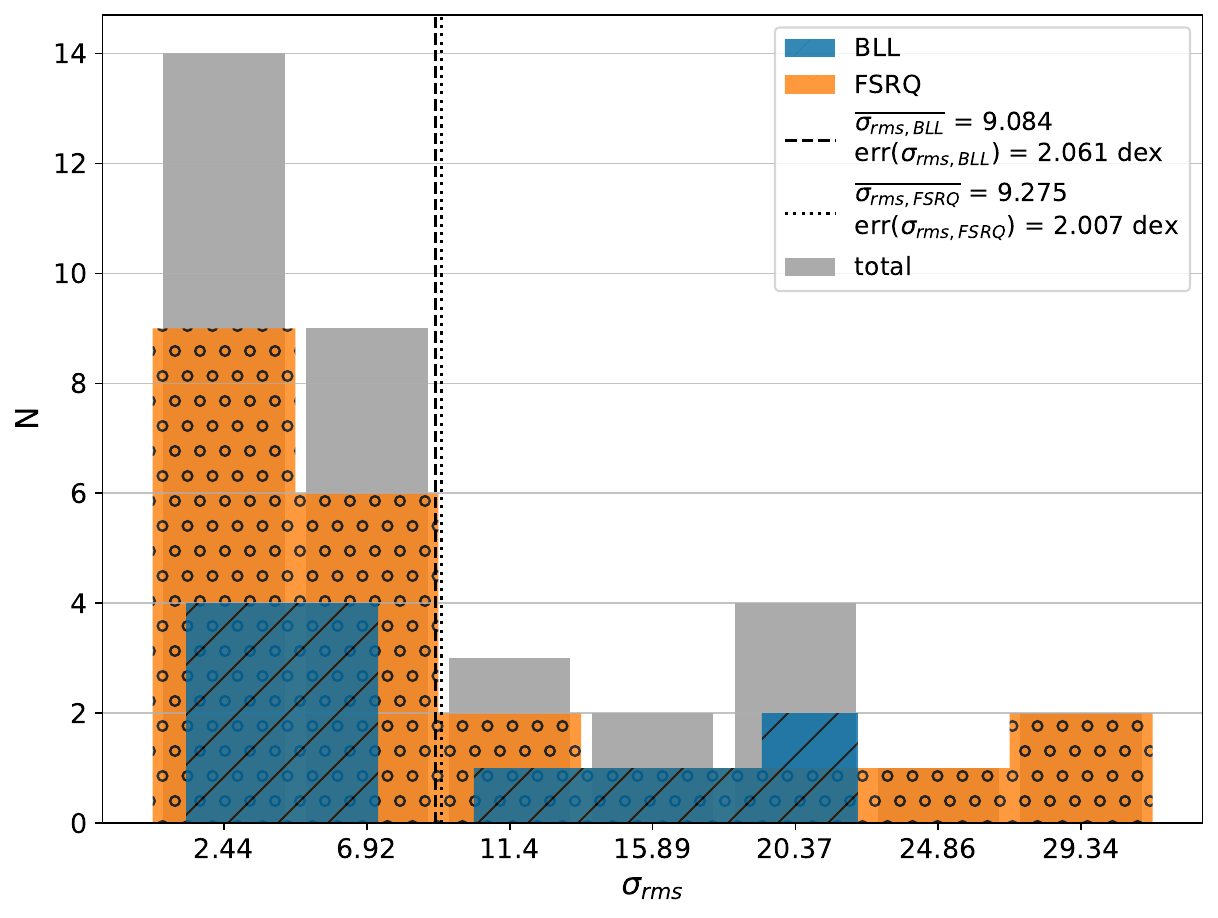}
            \caption{Histogram of $\log{_{10}(\sigma^{2}_{XS})}$ (top) and $\sigma_{rms}$ (bottom) for all variable light curves by activity type. Provided are the mean values and standard errors.}
            \label{fig:excess_var}
        \end{figure}

        Neither the excess variance nor the RMS-scatter for activity type subpopulations are well fit by normal, lognormal, or bimodal-normal distributions with greater than 20\% confidence. However, we can compare these subpopulations by each distribution's simple mean and standard deviations. Figure \ref{fig:excess_var} shows the logarithms of the $\sigma^{2}_{XS}$ (top) and $\sigma_{rms}$ (bottom) from our sample, including values for all light curves, binned and unbinned. Comparing $\sigma_{rms}$ to the average photometric error, $\overline{\xi^{2}}$, we find intrinsic variability levels at $\sigma_{rms,FSRQ} \simeq 11.47\cdot\overline{\xi^{2}_{FSRQ}}$, $\sigma_{rms,BLL} \simeq 11.37\cdot\overline{\xi^{2}_{BLL}}$ for BLL, and $\sigma_{rms} \simeq 11.42\cdot\overline{\xi^{2}}$ for our sample overall; in other words, the variability of all objects in our final variable sample typically exceeds the photometric error by an order of magnitude.

    \subsubsection{Flux distributions}
        \label{sec:flux_dist}       
        \begin{figure}
            \centering
            \includegraphics[width=0.5\textwidth]{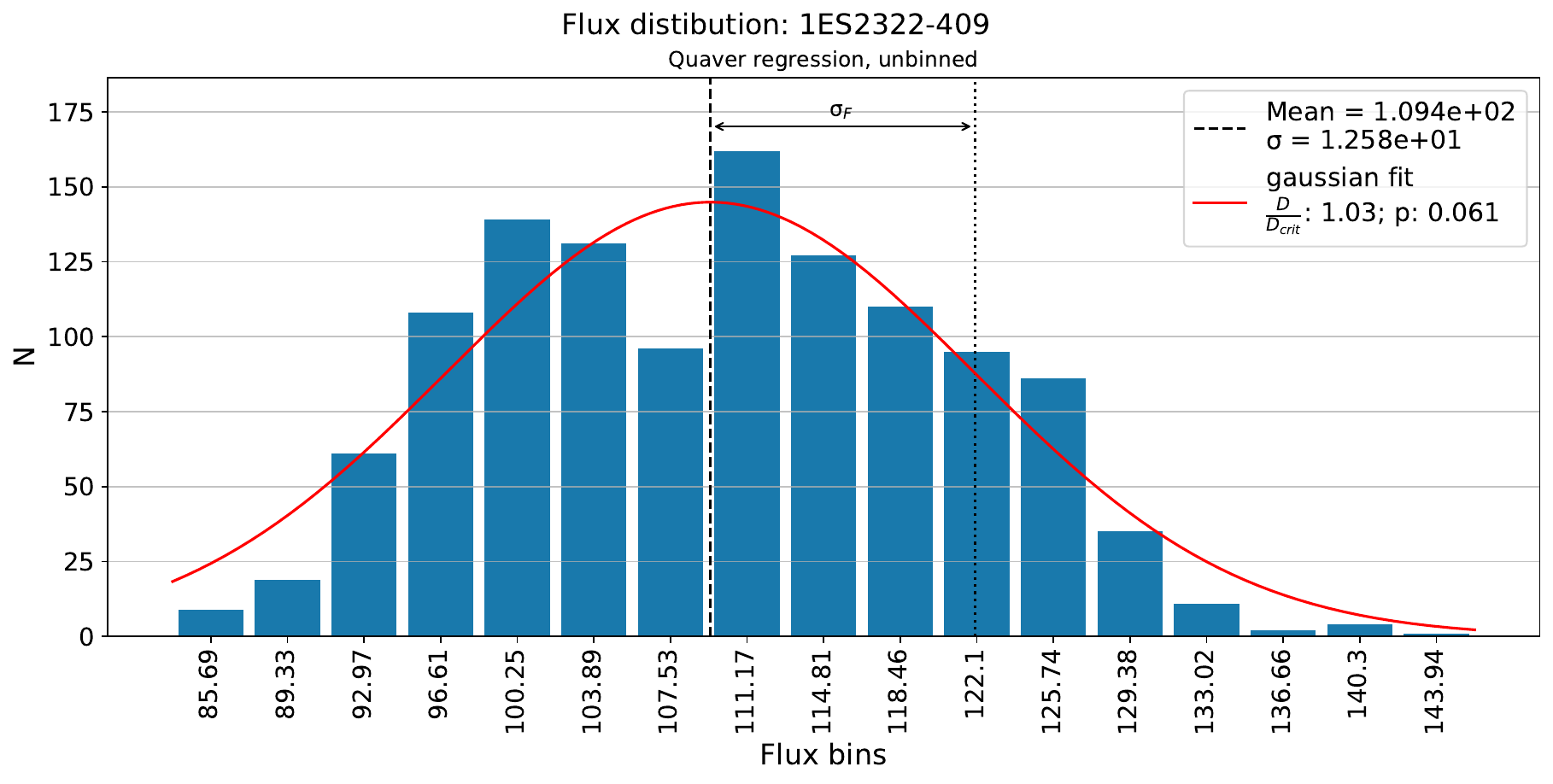}
            \includegraphics[width=0.5\textwidth]{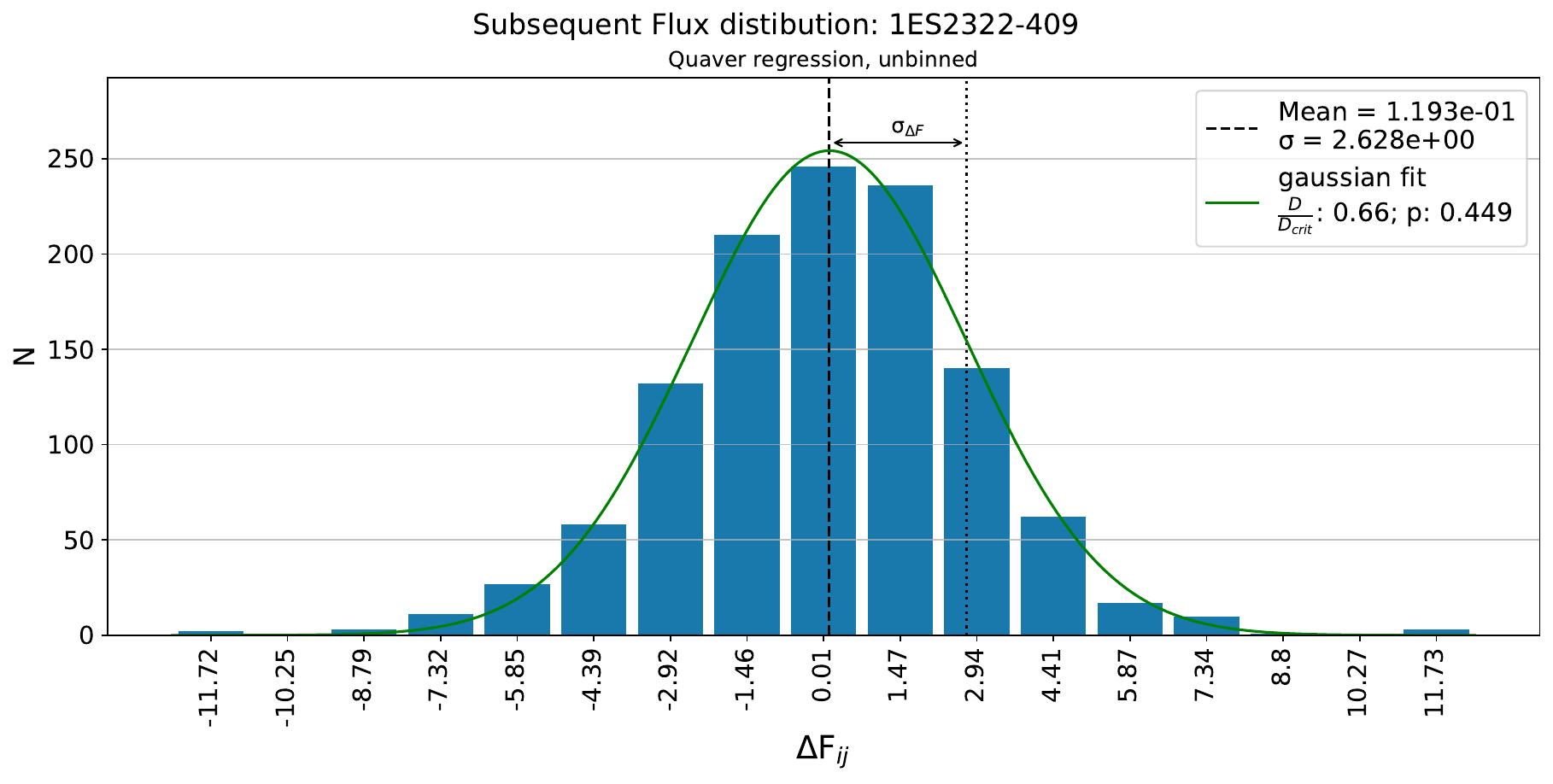}
            \caption{Flux (top) and subsequent flux (bottom) distributions and associated Gaussian fits for BLL 1ES2322+409.}
            \label{fig:1ES2322_fluxdist}
        \end{figure}
\begin{figure}
            \centering
            \includegraphics[width=0.5\textwidth]{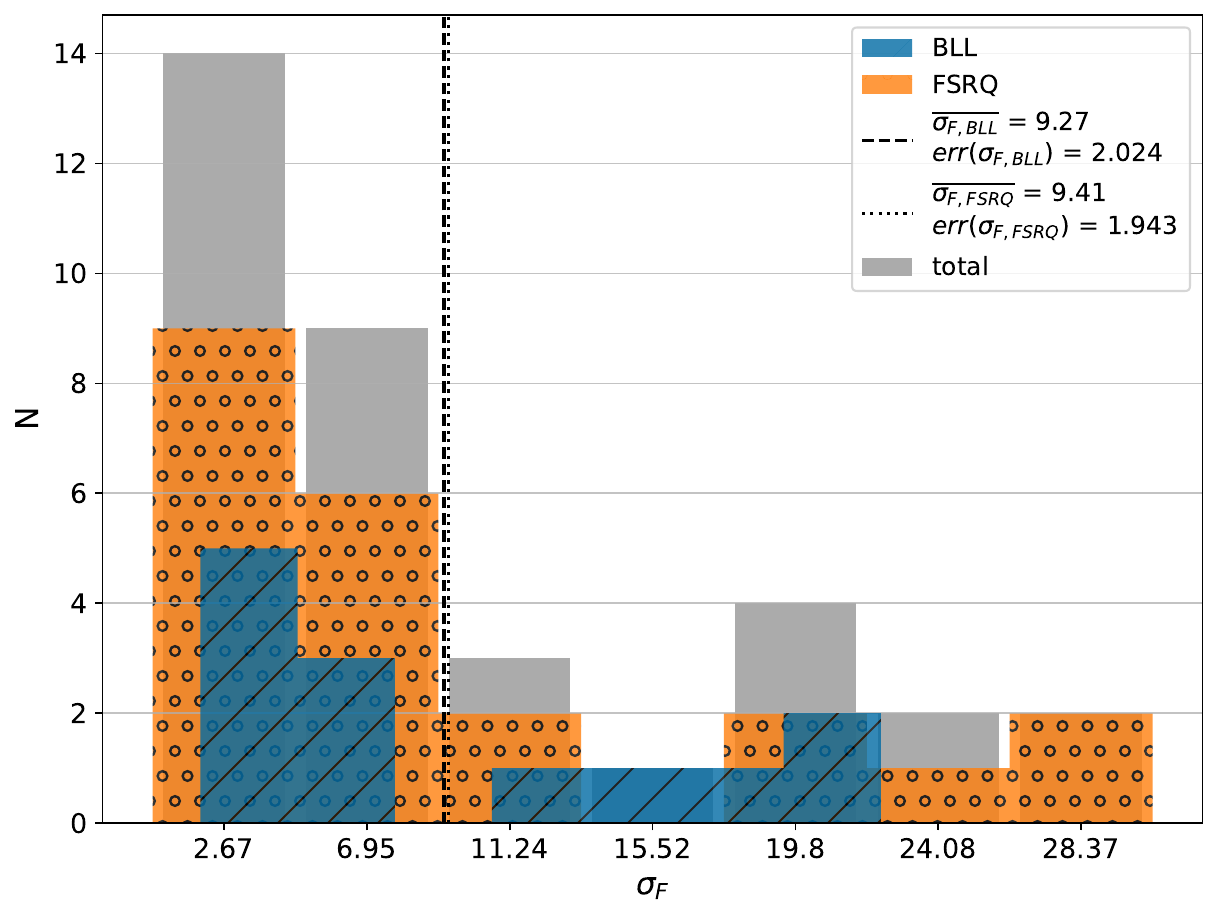}
            \includegraphics[width=0.5\textwidth]{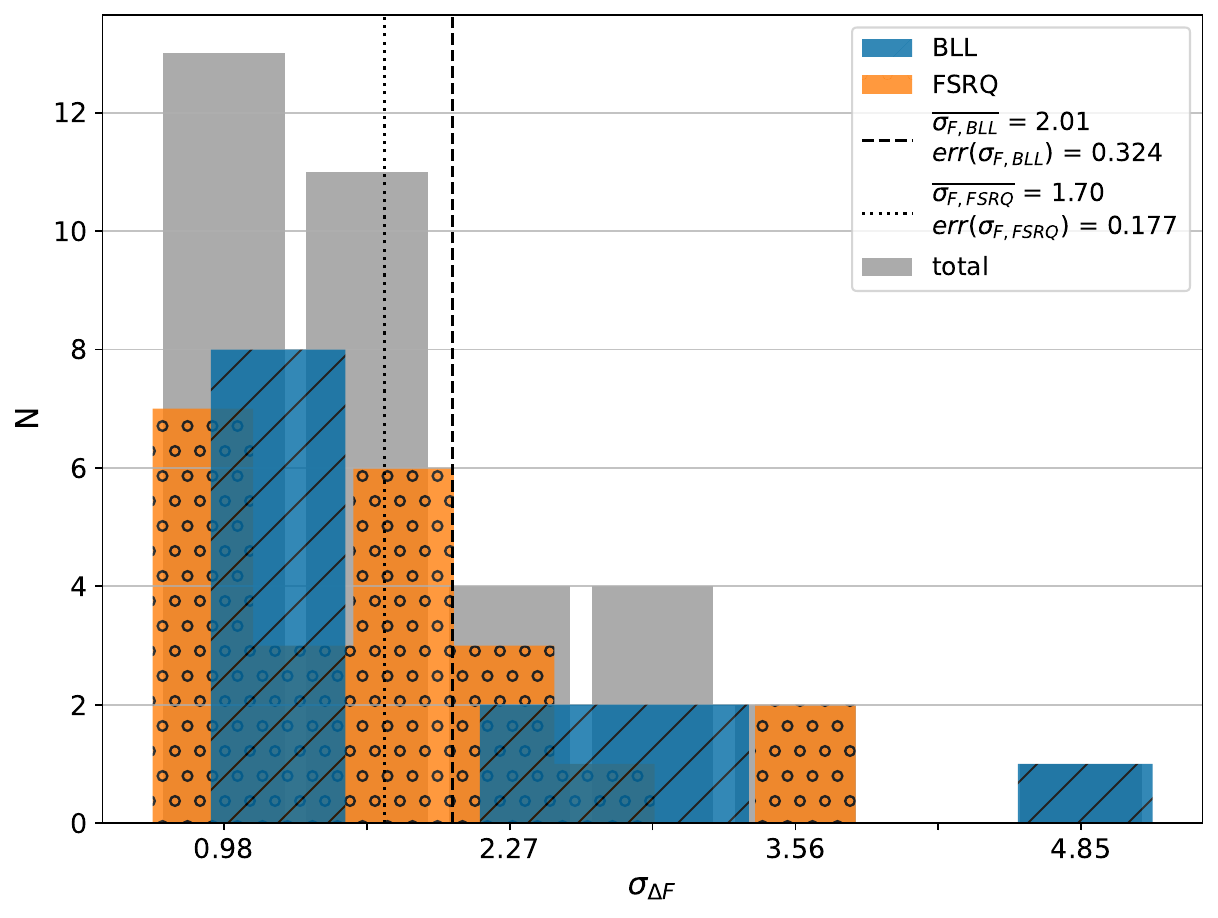}
            \caption{Histograms of the standard deviations for the flux (top) and subsequent flux (bottom) distributions for all variable light curves by activity type. The mean values and standard errors are given in the labels.}
            \label{fig:std_histos}
        \end{figure}

        We next investigate the distributions of both the simple flux and subsequent flux, i.e., the difference between each subsequent flux ($\Delta F_{i,i+1} = F_{i+1} - F_{i}$); these are shown in Figure \ref{fig:1ES2322_fluxdist}. The standard deviation or variance is particularly interesting, directly measuring the flux variation (Figure \ref{fig:std_histos}). We generate histograms for each flux distribution in which the number and spacing of bins are optimized under a combined algorithm following the methodologies of \citet{Shimazaki2007} and \citet{Knuth2013}. We assume the flux distribution to be normal and utilize standard least-squares fitting to estimate the distribution parameters. If this method fails to provide a sufficiently good fit, we make another attempt using an iterative optimization fitting method that utilizes `Broydon-Fletcher-Goldfarb-Shanno' (BFGS) minimization. If this still does not provide a good fit, we use the standard numerical definition of the mean and standard deviation of the flux distribution for further analysis. 
    
        Figure \ref{fig:std_histos} shows the histograms of the flux distributions; we find no difference between BLL and FSRQ sub-populations. We provide the parameters describing the flux and subsequent flux distributions in Tables \ref{tab:variability1} and \ref{tab:variability2}. Examples of flux distributions are provided for 1ES~2322+409 in Figure \ref{fig:1ES2322_fluxdist}, along with statistics and p-values from a Kolmogorov-Smirnov test to show the significance of each Gaussian fit. For each distribution, we conduct a 2-sample Kolmogorov-Smirnov test permutatively with 5000 resamples, comparing random samples of equal size to the original, drawn from each of the above-described distributions \citep{Hodges1958}.
        
        Note, in cases where the light curve is monotonically increasing, the observation gap from TESS light curves may manifest as a false bi-modality in the flux distribution, showing a lack of counts about the approximate ``center''. In such cases, the Gaussian flux statistics are more accurately described by those derived from the binned light curves or by removing the longest-term trend in the light curve while preserving the mean. We find that 32 light curves from our sample are not particularly well-fit by Gaussian distributions, prompting the use of a more complex \textit{a priori} model (Section \ref{sec:simulation}) to generate more accurate simulated light curves in the next phase of our analysis.
        \newline

    \subsubsection{Shortest Timescales of variability}
        \label{sec:shortest_timescales}
        
        We have also calculated the shortest timescale of variability in each light curve using 
        \begin{equation}
            \label{eqn:short_timescale}
            \tau = \frac{\Delta t\ln{2}}{\ln(F_{j}/F_{i})}~,
        \end{equation}
        where $F_{i}$ is a flux value in the light curve and $F_{j}$ is the flux value at later time $F(t+\Delta t)$. Correction into the rest-frame of the target accounting for cosmological redshift and Doppler boosting is provided by multiplication by the factor $\frac{\delta}{(1+z)}$ such that $\tau_{d} = \tau\delta/(1+z)$ in which $\delta$ is the Doppler beaming factor (estimated by \cite{Liodakis2018}) and $z$ is the redshift. This measurement is done only for significant increases and decreases in flux, $\Delta F  = F_{j} - F_{i} > 3\xi_{i}$ in which $\xi_{i}$ is the photometric errors of $F_{i}$ in each measurement \citep{Ruan2012}. In this manner, $\tau_{d}$ provides a characteristic timescale of variability, and $(\tau_{d}/\ln{2})$ may provide the exponential growth or decay timescales of flares \citep{Chatterjee2021}. In Table 2, we provide timescales in the rest frame of the blazar for every object in our sample except for WISE~J105912.43$-$113422.6, which has no redshift measurement. 
        
        To mitigate the selection effects of single cadence flares, we identify a significant change in flux using a Gaussian smoothed light curve \citep{Chatterjee2012}. Implementing such a technique ensures that selection of variability attributed to noise is avoided allowing the real flux values to be used reliably. However, light curves can be over-smoothed, and removing variability over short times must be done systematically; as such, we implement a  light curve smoothing procedure using a 1-hour kernel. If we do not identify a significant change in flux under this smoothing, the kernel is reduced to 30 minutes; likewise, if $\Delta F > 3\xi_{i}$ is still not located, the calculation is made without smoothing. Note that we only utilize this method to ensure the change in flux is significant without only selecting flares; any subsequent analyses, however, are performed on the actual light curve data. Such cases only applied for two light curves, that of NGC128 and PKS0336-177, wherein any smoothing removed significant variability. Examples of this iterative smoothing procedure wherein variability is selected provided in Appendix \ref{sec:timescales}.
        
        The shortest timescales of increasing and decreasing variability, within the rest frame of the target, fall in the range of $0.28 \leq \tau_{d,inc} \leq 160.02$ days and $0.21 \leq \tau_{d,dec} \leq 273.00$ days where the smallest values of $\Delta t$ correspond to our sample spacing. We then select the shortest timescale between $\tau_{d,inc}$ and $\tau_{d,dec}$ to estimate an upper limit to the radii of the emitting region via
        \begin{equation}
            R = \frac{\delta c \tau}{1 + z} = c \tau_{d}.
            \label{eqn:radii_est}
        \end{equation}
        Discussion on the error on $\tau_{d}$ is provided in Appendix \ref{sec:errors}. 
        
\begin{figure}
            \centering
            \includegraphics[width=0.5\textwidth]{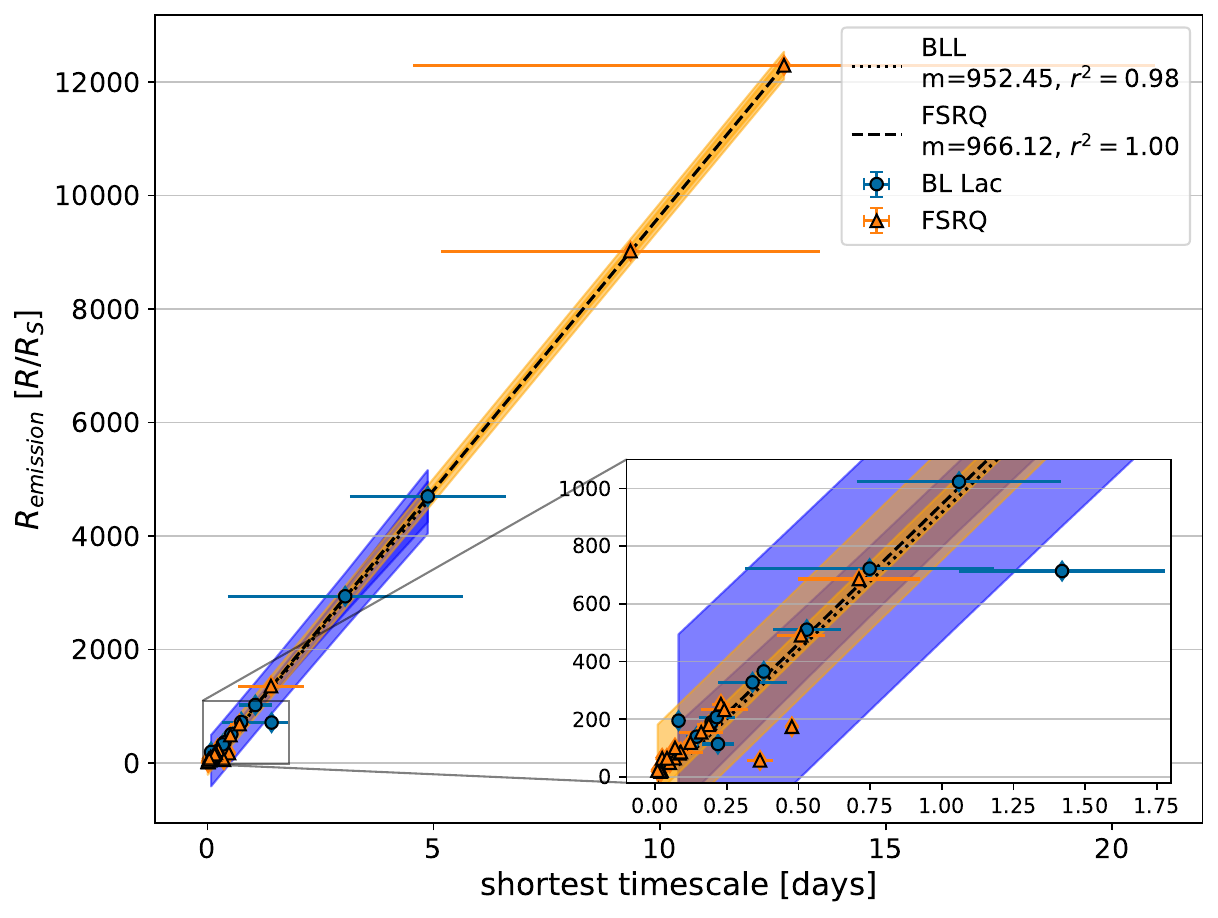}
            \caption{The upper limit radii of the observed emitting region in units of Schwarzchild radii, assuming a black hole mass as estimated by \citet{Paliya2021}, otherwise $M_{BH} = 10^{8.5}~M_{\odot}$, with provided 95\% confidence and prediction intervals for the linear relation of each subpopulation.}
            \label{fig:BHradii}
        \end{figure}

        For objects with published estimates, we utilize the Doppler factors, $\delta$, from a constrained calculation of the limiting brightness temperature. For those that do not have estimates of $\delta$, we perform this calculation under the reasonable assumption of $\delta = 11$ \citep{Sandrinelli2014,Liodakis2017,Liodakis2018}. Figure~\ref{fig:BHradii} shows the ratio of the calculated radii relative to the expected Schwarzchild radius against the shortest timescales. Results are tabulated in Table \ref{tab:variability1} with examples in Appendix \ref{sec:timescales}. 

    \subsubsection{RMS-flux relation}
        \label{sec:rms_flux_rel}
\begin{figure}
            \centering
            \resizebox{0.5\textwidth}{!}{\includegraphics{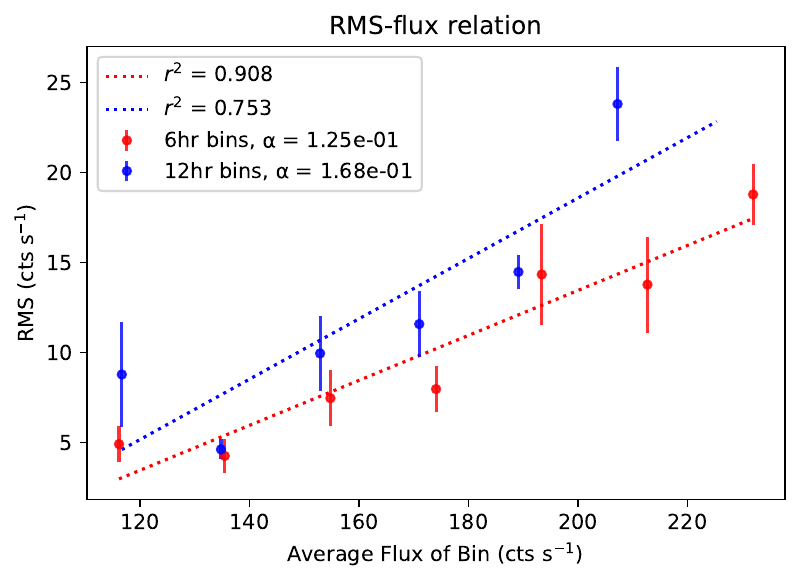}}
            \caption{A linear rms-flux relation is observed in the light curve of PKS~0035$-$252 when binned into either 6- or 12-hour intervals.}
            \label{fig:rms_flux}
        \end{figure}

        X-ray binaries and Seyfert galaxies often appear to have a higher level of rms-scatter when brighter, with a linear trend between flux and variability of a given light curve segment \citep{Uttley2001,Gleissner2004,Alston2019,Bhattacharyya2020,Kundu2021}. Such a relation can, but does not necessarily, imply multiplicative stochastic processes which connect variability between short and long timescales \citep{Scargle2020}. We calculate the average fluxes, the rms variability, and excess variance in 6- and 12-hour bins with error propagation. We then fit the relation between the mean flux and rms for each bin with a linear relation, obtaining a slope and coefficient of determination for goodness-of-fit \citep{Vaughan2003,Sandrinelli2014,Smith2018,Bhattacharyya2020}. Most of our sample shows at least a weak correlation between the rms-scatter and flux over 6- and 12-hour timescales. Figure \ref{fig:rms_flux} shows an example of such a relation. We find that $\sim$74\% of our sample's light curves when binned into 6-hour intervals, show a weak correlation ($r^{2} \geq 0.25$), and $\sim$23\% show a strong correlation ($r^{2} \geq 0.75$). Likewise, when binned into 12-hour intervals, $\sim$69\% of our sample's light curves demonstrate a similarly weak relation with $\sim$20\% showing a strong correlation. These results are summarized in Table \ref{tab:rms_flux_tbl}.
        \begin{deluxetable}{lcccc}
            \tabletypesize{\scriptsize}
            \tablewidth{0pt} 
            \tablenum{4}
            \tablecaption{Slopes (m) and $r^{2}$ coefficients for the best-fit linear trends of the rms-flux relationship\label{tab:rms_flux_tbl}}
            \tablehead{
            \colhead{Target} & \colhead{m (6 hrs)} & \colhead{$r^{2}$} & \colhead{m (12 hrs)} &\colhead{$r^{2}$}\\ 
            \colhead{} & \colhead{$[\times10^{-2}]$} & \colhead{} & \colhead{$[\times10^{-2}]$} &\colhead{} }
            \startdata
            1ES2322-409 & 0.32 & 0.01 & 1.45 & 0.59\\
            1RXSJ054357.3-553206 & 0.73 & 0.73 & 1.42 & 0.81\\
            3C120 & 38.42 & 0.77 & 40.40 & 0.89\\
            NGC1218 & 23.17 & 0.65 & 11.60 & 0.27\\
            PKS0035-252 & 12.49 & 0.91 & 16.77 & 0.75\\
            PKS0130-17 & 2.45 & 0.58 & 4.25 & 0.75\\
            PKS0208-512 & 2.12 & 0.70 & 1.31 & 0.26\\
            PKS0226-559 & 1.60 & 0.81 & 1.96 & 0.46\\
            PKS0235-618 & 1.56 & 0.36 & 0.88 & 0.11\\
            PKS0301-243 & 2.67 & 0.75 & 2.94 & 0.22\\
            PKS0336-177 & 0.73 & 0.00 & 0.43 & 0.00\\
            PKS0420-01 & 1.87 & 0.03 & 3.68 & 0.13\\
            PKS0422+00 & 0.96 & 0.17 & 0.40 & 0.02\\
            PKS0426-380 & 0.56 & 0.44 & 0.71 & 0.09\\
            PKS0454-234 & -3.57 & 0.53 & -1.03 & 0.36\\
            PKS0521-36 & 10.21 & 0.75 & 12.20 & 0.78\\
            PKS0637-75 & ~ & ~ & ~ & ~\\
            ~~sect. 1+2+3 & 1.89 & 0.08 & 4.31 & 0.09\\
            ~~sect. 4+5+6 & 3.84 & 0.18 & 2.57 & 0.12\\
            ~~sect. 7+8+9 & 6.32 & 0.63 & 8.38 & 0.46\\
            PKS0736+01 & 29.76 & 0.63 & 36.09 & 0.83\\
            PKS0829+046 & 4.55 & 0.88 & 8.94 & 0.59\\
            PKS1244-255 & 3.54 & 0.57 & 5.90 & 0.53\\
            PKS2155-304 & -0.53 & 0.07 & -1.57 & 0.30\\
            PKS2155-83 & 4.33 & 0.77 & 7.32 & 0.86\\
            PKS2255-282 & 1.89 & 0.52 & 5.69 & 0.65\\
            PKS2326-502 & 1.49 & 0.28 & 2.99 & 0.43\\
            PKS2345-16 & 3.33 & 0.57 & 4.61 & 0.70\\
            PMNJ0948+0022 & 11.59 & 0.51 & 11.73 & 0.39\\
            PMNJ2141-6411 & 4.47 & 0.45 & 0.91 & 0.02\\
            PMNJ2345-1555 & 7.90 & 0.82 & 11.15 & 0.81\\
            WISEJ031612.72+090443.3 & -2.10 & 0.04 & 8.73 & 0.62\\
            WISEJ085009.64-121335.3 & 5.95 & 0.91 & 10.07 & 0.96\\
            WISEJ095302.70-084018.2 & 1.51 & 0.68 & 2.53 & 0.14\\
            WISEJ105912.43-113422.6 & -1.74 & 0.16 & -4.18 & 0.20
            \enddata
        \end{deluxetable}

    \subsection{Power Spectral Analysis}
    \label{sec:psanalysis}
    
    In this section, we conduct analyses of our target PSDs to test their goodness of fit to 3 power-law variant models: a single power law, broken power law, and bending power law, defined as 

    \begin{itemize}
        \item Single power law
        \begin{equation}
            \label{eqn:p-law}
            \mathcal{P}(\nu) = \mathcal{A} \nu^{-\alpha} + c
        \end{equation}
        \item Broken power law
        \begin{equation}
            \label{eqn:brkn_p-law}
            \mathcal{P}(\nu) = \mathcal{A}
            \begin{cases}
                (\sfrac{\nu}{\nu_{k}})^{-\beta} + c, & \text{$\nu < \nu_{k}$}\\
                (\sfrac{\nu}{\nu_{k}})^{-\alpha} + c, & \text{$\nu > \nu_{k}$}
            \end{cases}
        \end{equation}
        \item Bending power law
        \begin{equation}
            \label{eqn:bend_p-law}
            \mathcal{P}(\nu) = \mathcal{A} \Bigg{(} \frac{ \nu^{-\beta}}{1 + (\sfrac{\nu}{\nu_{b}})^{\alpha-\beta}}\Bigg{)} +c
        \end{equation}

    \end{itemize}
    The spectral index or slope and normalizations of the model provide physical information for the stochastic processes within the jet. Generally, the power spectrum of a blazar is well-modeled by a single power law model. However, it is expected that for the stationary stochastic processes, the slope will tend towards 0 at lower frequencies corresponding to the relaxation timescale. As such, we explore the potential for probing this timescale at which the power spectrum flattens. An identification of a break or bend as the slope changes, would provide timescales related to key physical characteristics about the emission region and cooling timescales \citep{Uttley2002, Vaughan2003, Edelson2013, Kasliwal2015, Goyal2021}.
    
    \subsubsection{Light Curve Simulation}
        \label{sec:simulation}

        A single realization of an AGN's PSD provides no error on any model that attempts to describe it. Consequently, the simulation of long light curves, representative of the underlying PSD model, is essential for performing statistically significant analyses of variability properties. We therefore simulate 500 evenly sampled light curves per object. We discuss the effects and treatment of distortion in Section \ref{sec:distortions}. For an even basis of the comparison, we wish to preserve the potential effects of cadence masking, likely to present as an additional white-noise contribution. We reproduce the ``gaps'' created by our masking procedure in the observed light curve at the exact respective times in each simulated light curve. These gaps are filled by linear interpolation, evenly spaced at the original sampling rate, to allow for direct Fourier analysis.

        We apply the Monte Carlo simulation method of \citet{Emmanoulopoulos2013}, hereafter EMP13, using a modified version of the python code provided by \citep{Connolly2015}\footnote{\url{https://github.com/samconnolly/DELightcurveSimulation}}. This method expands upon that of \citep{Timmer1995} (TK95) to handle the spectral information and has shown successful application to simplistic light curve simulations \citep{Sartori2019}. This method works in conjunction with that of \citep{Schreiber1996} (SS96), which appropriately distributes the Probability Density Function (PDF) properties. Our simulation program and procedure version is publicly available\footnote{DOI:\dataset[10.5281/zenodo.10957741]{https://doi.org/10.5281/zenodo.10957741}}.

    \subsubsection{Modified Power Spectral RESPonse (PSRESP) Method}
        \label{sec:PSRESP}
        Our analysis technique closely follows the PSRESP method as described in \citet{Uttley2002} and \citet{Smith2018} except for the PSD fitting and simulation as prescribed by the methods of EMP13. 

        Initially, we fit each light curve, observed and simulated, with a linear trend, which is then subtracted, excluding the offset, from the flux values. This subtraction effectively flattens our light curve about the mean. The mean is then subtracted from these light curves so that the light curves vary around zero. Appendix \ref{sec:lightcurves} provides our light curves prior to flattening and mean subtraction. If there are multiple TESS sectors, we stitch them together, and then lineraly interpolate over gaps. We then generate and normalize a periodogram using standard Discrete Fourier Transformation (DFT) \citep{Vaughan2003}, and determine the best-fit parameters for single, broken, and bending power-laws of this periodogram by iterative, maximum-likelihood minimization methods as discussed in \citet{Barret2012} and \citet{Emmanoulopoulos2013}. 
        
\begin{figure}
            \centering
            \subfloat{
                \includegraphics[width=0.5\textwidth]{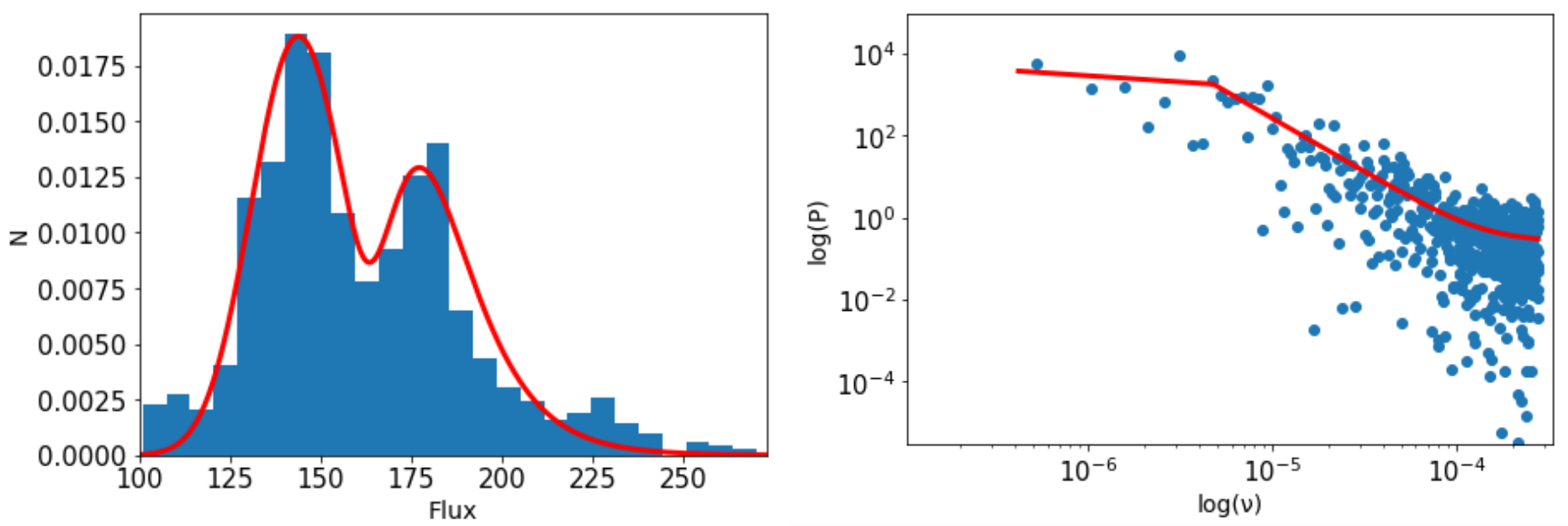}}
                \caption{PKS~0035-252, Eq. \ref{eqn:brkn_p-law}}
                \label{subfig:PKS0035_PDF+PSDfit}
            
            \subfloat{
                \includegraphics[width=0.5\textwidth]{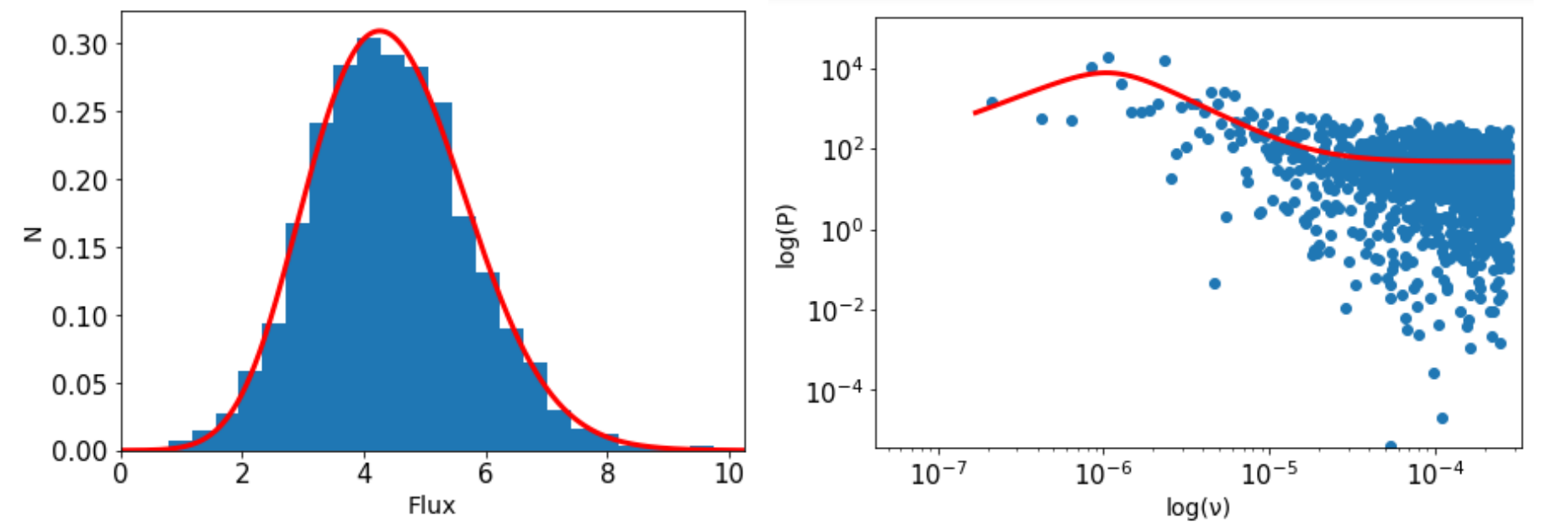}}
                \caption{PKS~2326-502, Eq. \ref{eqn:bend_p-law}}
                \label{subfig:PKS2326_PDF+PSDfit}
            
            \subfloat{
                \includegraphics[width=0.5\textwidth]{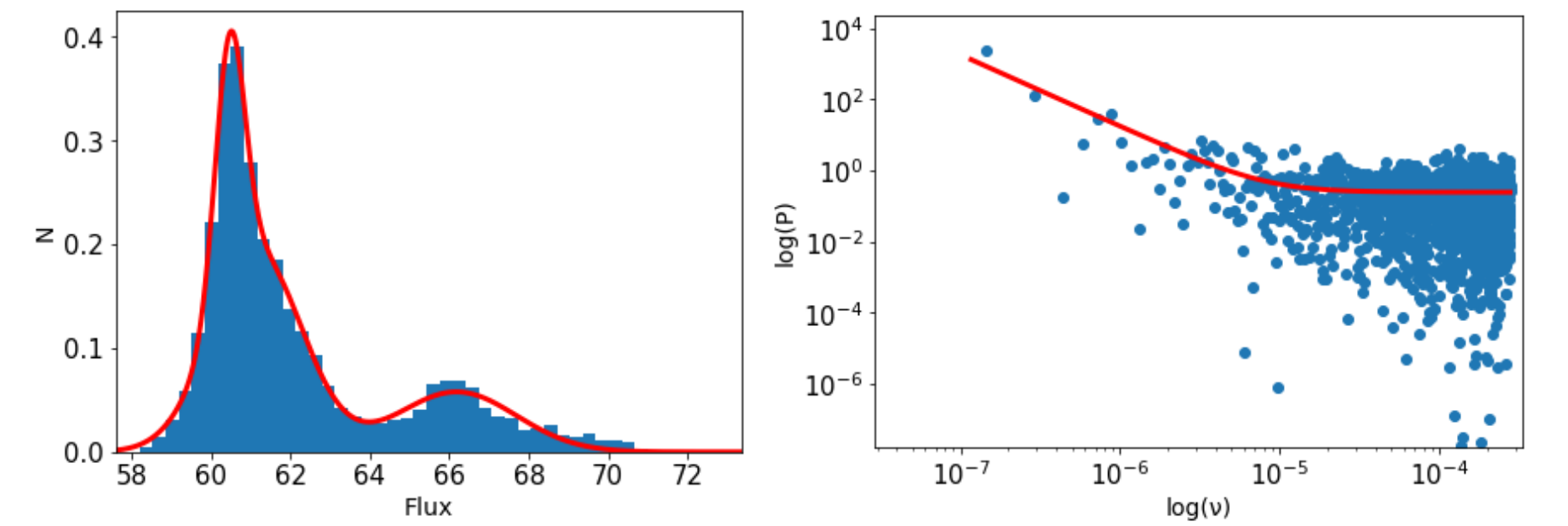}}
                \caption{PKS~0235-618, Eq. \ref{eqn:p-law}}
                \label{subfig:PKS0235_PDF+PSDfit}
            
            \caption{PDFs (left) and PSDs (right) with associated model fits (red).\label{fig:PDF+PSDfits}}
        \end{figure}

        The best-fitting model will then serve as the underlying PSD (Fig. \ref{fig:PDF+PSDfits}), renormalized by a factor, $A_{rms^{2}} = 2\Delta t_{samp}/(\overline{F}^{2}N)$, according to the van der Klis convention as defined in \citet{Babu1997},  where $\Delta t_{samp}$ is the sampling interval, $\overline{F}$ is the mean flux (cts~s$^{-1}$) after flattening but before mean subtraction, and $N$ is the number of unbinned flux values. Using this model, our simulated, normally distributed time series are generated using the TK95 algorithm, except for the noise component $c$, from Equations \ref{eqn:p-law} -- \ref{eqn:bend_p-law}, which is zero. We then apply the white-noise sampling, spectral adjustment, and amplitude adjustment techniques from \citet{Emmanoulopoulos2013} iteratively until the synthetic time series converges with consistent statistical properties, and  generate power spectra for each of the 500 simulated light curves. It is important to note that this process intends to simulate light curves with the same underlying power-spectral model. In most cases, these PSDs will require renormalization to minimize goodness-of-fit statistics.

        We then segment the power spectra into 25 log-normal frequency bins and take the mean, per bin, of the synthetic periodograms to act as our `model average spectrum,' $\overline{\mathcal{P}_{sim}(\nu)}$. All spectra follow the same binning scheme as determined by the actual PSD. The error bars on this mean, $\Delta\overline{\mathcal{P}_{sim}(\nu)}$, represent the spread of the simulated spectra in each bin. While it is non-standard to assign the error bars to the model rather than the data, it is valid in this context where we base our statistic on the variance of the population from the model \citep{Vaughan2003,Emmanoulopoulos2013,Connolly2015}. Binning methods for periodogram estimation require that each bin is Gaussian about the mean, which is not always observed. \citet{Lawrence1993} suggests a correction for such non-Gaussian bins using logarithmic periodograms, in which the normally distributed spread of periodogram estimates centers about the geometric mean instead of the arithmetic mean. However, this requires at least 10 points per frequency bin, which significantly limits its usefulness in low-frequency regions and so, while potentially useful in longer baseline investigations, we omit this correction in our analysis.

    \subsubsection{Red Noise Spectral Distortions}
        \label{sec:distortions}

        Light curves of AGN tend to have PSDs biased to high-frequencies, known as ``red-noise leak'', as well as a fold-back to low-frequencies due to a low time resolution of observations \citep{Uttley2002,Vaughan2003,Emmanoulopoulos2013}.  We create surrogate light curves 1000$\times$ longer than the observed light curve and select random samples of the observed light curve length to serve as the simulated sample to ensure we reduce red-noise leakage as much as possible.
        Utilization of the EMP13 method and code provided by \citet{Connolly2015} provide a simple manner of addressing and correcting such distortions due to aliasing. However, since our TESS observations have excellent time resolutions, $\Delta t \lesssim 30$ min, this effect is expected to be negligible.

        \subsubsection{Treatment and expected level of Poisson Noise}
        \label{sec:poisson}
        The observed light curve is affected by Poisson Noise, reflected as a constant offset in the observed PSD. To account for this in our simulations, each simulated light curve point, $F_{sim,k}(t)$, is replaced with a Poisson random variate, $F_{sim,Pois,k}(t)$ from a Poisson distribution with mean $F_{sim,1}(t) \Delta t_{samp}$ as defined in EMP13.  
        
        We observed that most of our sample did not consistently have near-Gaussian error distributions and frequently displayed log-normal error distributions. Thereby, with our convention of normalization, $A_{rms^{2}}$, the expected level of Poisson noise  for our unbinned (i.e., $\Delta t_{bin}$ = $\Delta t_{samp}$) data set is provided by, 
        \begin{equation}
            \label{eqn:Poisson_noise}
            P_{noise} = 2(\overline{F}+B) /\overline{F}^{2} ,
        \end{equation}
        where $\overline{F}$ is the mean flux (cts s$^{-1}$) after flattening but before mean subtraction, and $B$ is the average background count as found from the TESS Target Pixel Files.        %
    \subsubsection{Goodness-of-fit testing}
        \label{sec:goodfit}
        We define our statistical test as $\chi^{2}_{dist}$, following \citet{Uttley2002,Chatterjee2008,Smith2018,Goyal2022}. This test compares the PSDs of our simulated and real spectra to our model spectrum:
        \begin{equation}
            \label{eqn:chi-sqr_obs}
            \chi^{2}_{obs} = \sum_{\nu_{n}}^{\nu_{N_{bin}}}\frac{[\overline{\mathcal{P}_{sim}(\nu_{n})} - \mathcal{P}_{obs}(\nu_{n})]^{2}}{\Delta\overline{\mathcal{P}_{sim}(\nu_{n})}^{2}}~,
        \end{equation}
        \begin{equation}
            \label{eqn:chi-sqr_sim}
            \chi^{2}_{dist,i} = \sum_{\nu_{n}}^{\nu_{N_{bin}}}\frac{[\overline{\mathcal{P}_{sim}(\nu_{n})} - \mathcal{P}_{sim,i}(\nu_{n})]^{2}}{\Delta\overline{\mathcal{P}_{sim}(\nu_{n})}^{2}}~.
        \end{equation}
        The above sums represent the values used in testing the goodness-of-fit of the observed power spectra to the best-fit model PSD where $\mathcal{P}_{obs}(\nu_{n})$ is the binned, observed PSD and $\mathcal{P}_{sim,i}(\nu_{n})$ are the binned simulated power spectra. $\overline{\mathcal{P}_{sim}(\nu_{n})}$ and $\Delta\overline{\mathcal{P}_{sim}(\nu_{n})}$ are the binned model average and bin error, as discussed in Section \ref{sec:PSRESP}, which remain constant in every test. Because each $\mathcal{P}_{sim,i}(\nu)$ is simulated such that their overall, underlying PSD, $\mathcal{P}_{sim}(\nu)$, matches the input model PSD, a comparison of each simulated PSD to that of the ensemble average, Eq. \ref{eqn:chi-sqr_sim}, forms a distribution of 500 values which all intrinsically represent a reasonably good fit to the best-fit model PSD.

        We adopt the convention defined by \citet{Chatterjee2008} in which we present our goodness-of-fit to the model PSD as the success fraction, $p_{\alpha}$ $\epsilon$ [0.0,1.0] in which a higher value of $p_{\alpha}$ represents a fit of lower rejection confidence. Under this definition, \textit{the probability of model rejection is provided by the percentile above which $\chi^{2}_{dist}$ for the real PSD exceeds that of the simulated PSD.} For example, if $\chi^{2}_{obs}$  is smaller than 75\% of the simulated $\chi^{2}_{dist,i}$, the model shows a success fraction of $p_{\alpha}$ = 0.75 and has a rejection probability of 0.25. This statistical method is described further in \citet{Uttley2002} and \citet{Press1992}. This process ultimately aims to minimize the $\chi^{2}$ statistic for the real PSD relative to those simulated.

        For our study, of $p_{\alpha} $ > 0.25 represents a good fit to the model, i.e., a model which we cannot reject with at least 75$\%$ confidence, adequately describes the PSD of the observed AGN \citep{Press1992,Chatterjee2008,Goyal2022}.

    \subsubsection{Results}
        \label{sec:pow_spec_results}

\begin{figure}
            \centering
            \resizebox{0.5\textwidth}{!}{\includegraphics{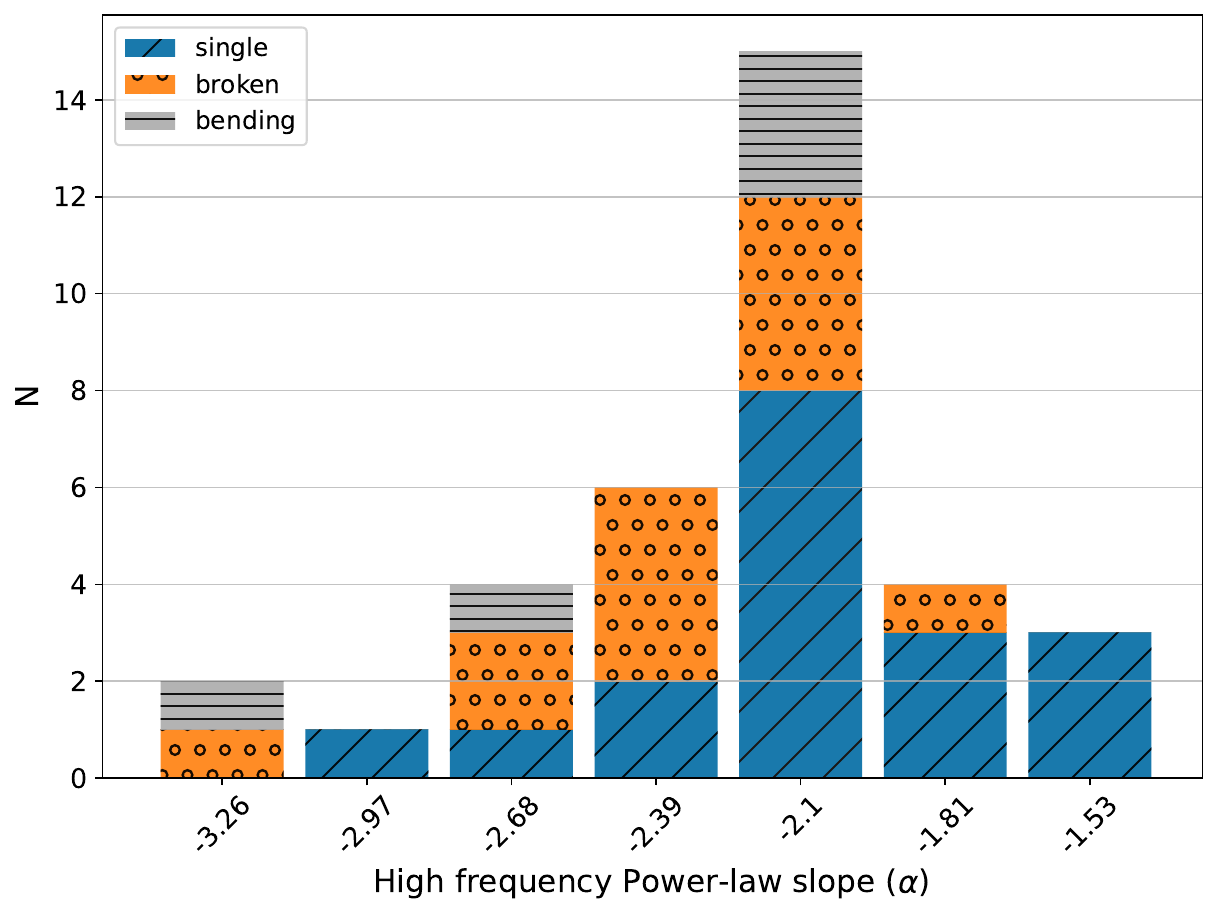}}
            \caption{Histogram of $\alpha$, showing counts per PSD model type per bin.}
            \label{fig:p-law_type}
        \end{figure}

        From our sample of 35 light curves from 33 blazars exhibiting statistically significant variability, we find that 18 power spectra are best fit by a single power-law model, 12 are best fit by a broken power-law and 5 by a bending power-law. These results are presented in Table \ref{tab:psd_results} and illustrated by Figure \ref{fig:p-law_type}. 

        \begin{figure}
            \centering
            \resizebox{0.5\textwidth}{!}{\includegraphics{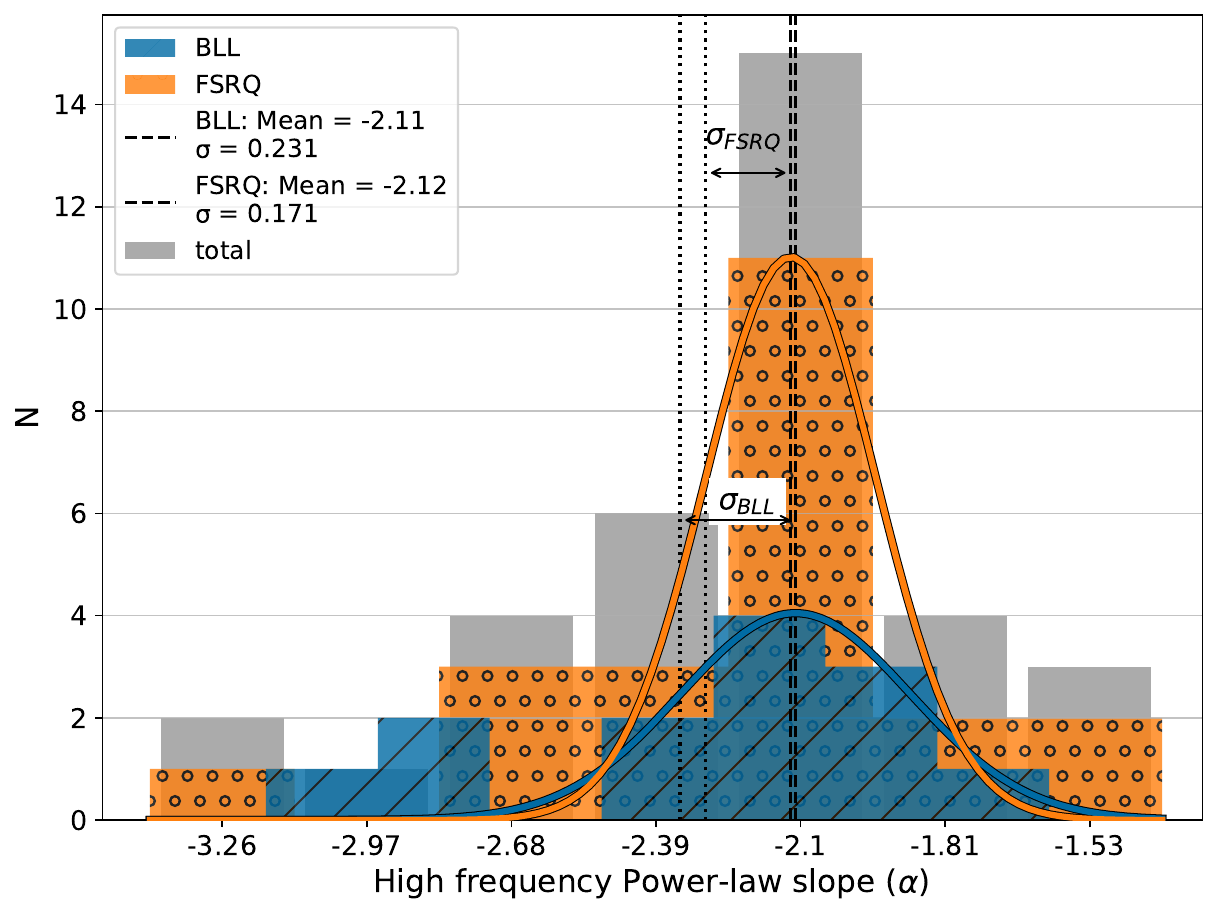}}
            \caption{Histogram of power-law slopes by activity type and associated Gaussian fits for BLL and FSRQ populations.}
            \label{fig:p-law_slopes_all}
        \end{figure}

\begin{figure}
            \includegraphics[width=.5\textwidth]{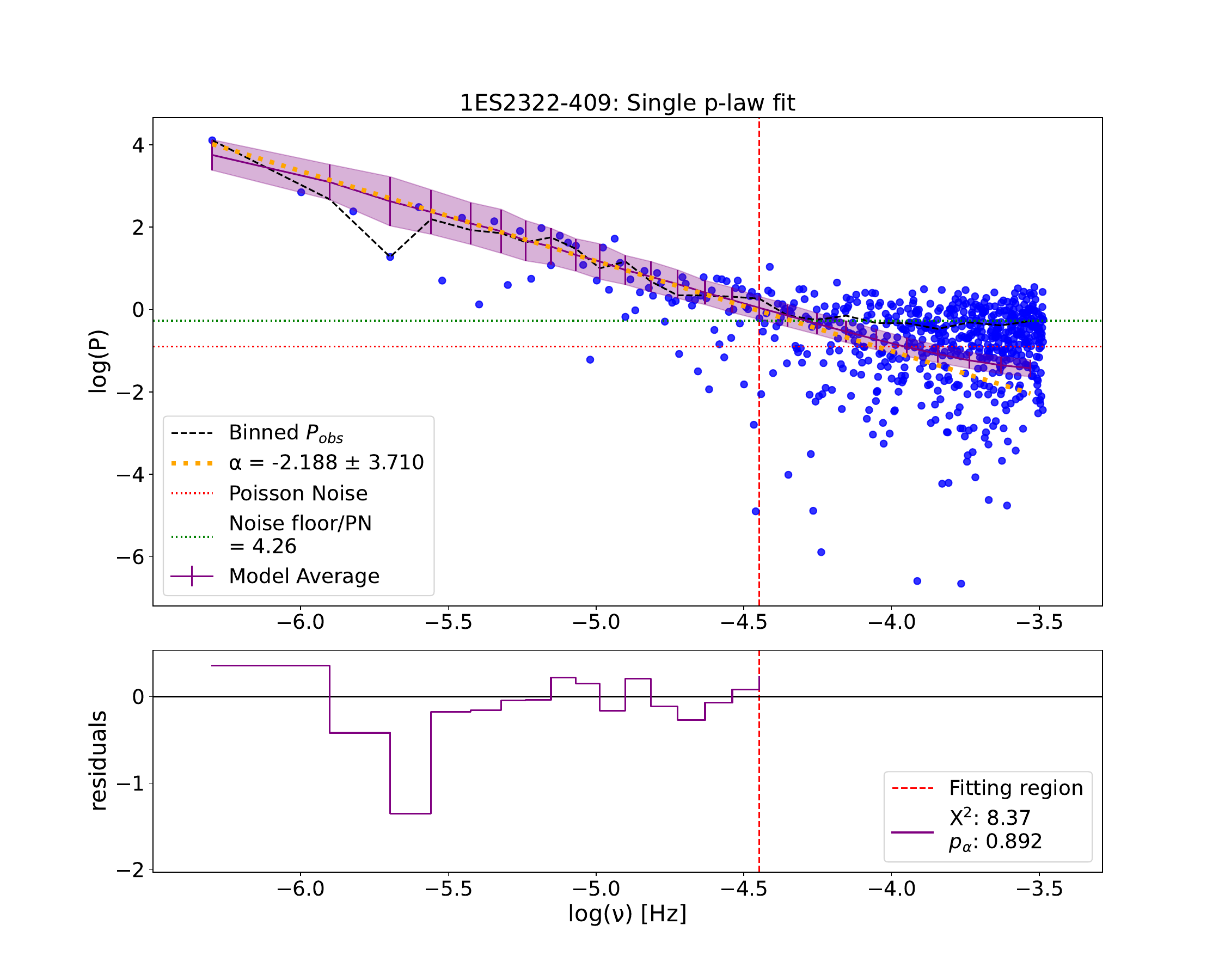}\\
            \includegraphics[width=.5\textwidth]{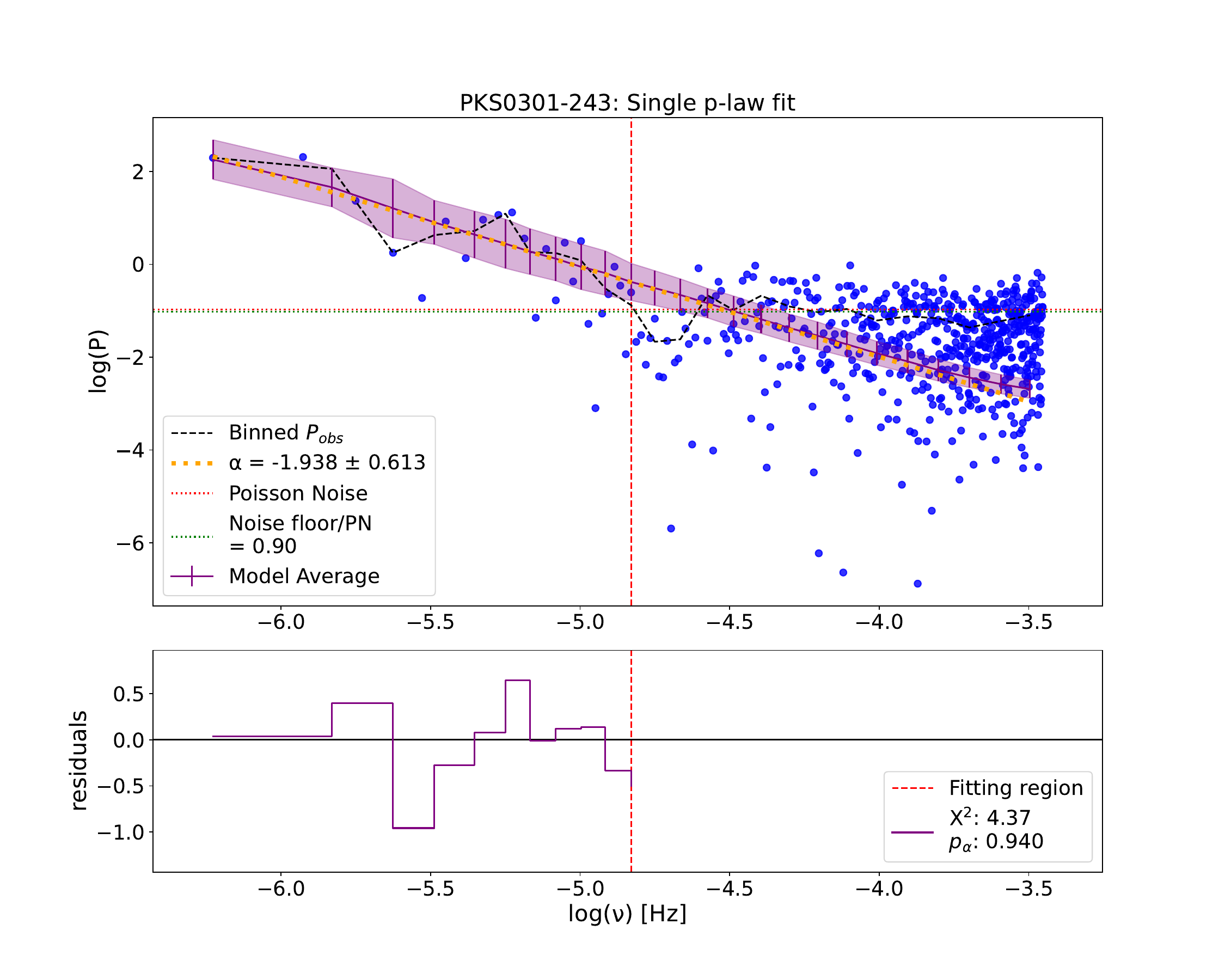}\\
            \includegraphics[width=.5\textwidth]{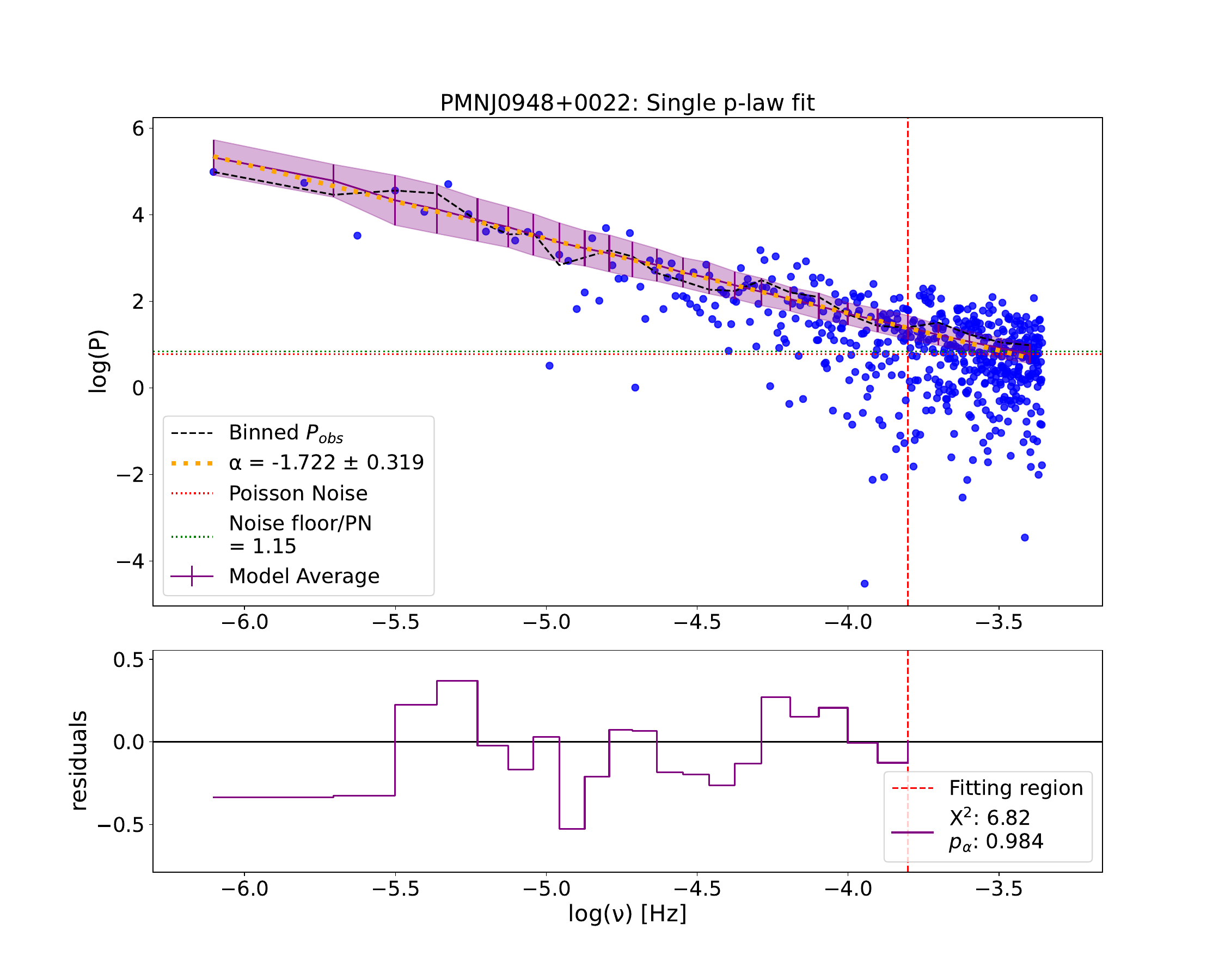 }
            \caption{Examples of objects best-fit by a single power-law model. The binned, target  (black) and model average (purple) PSDs, are shown overlaid on the full target PSD (blue) with a projection of the best-fit $\alpha$ (yellow). Expected and measure levels of Poisson noise are given by the red, and green lines respectively. The lower section displays the object and model average PSD residuals in the fitting region (red, vertical line) along with $\chi^{2}$ value and success fraction.}
            \label{fig:PSD_results_goodfit_plaw}
        \end{figure}

\begin{figure}
            \includegraphics[width=.5\textwidth]{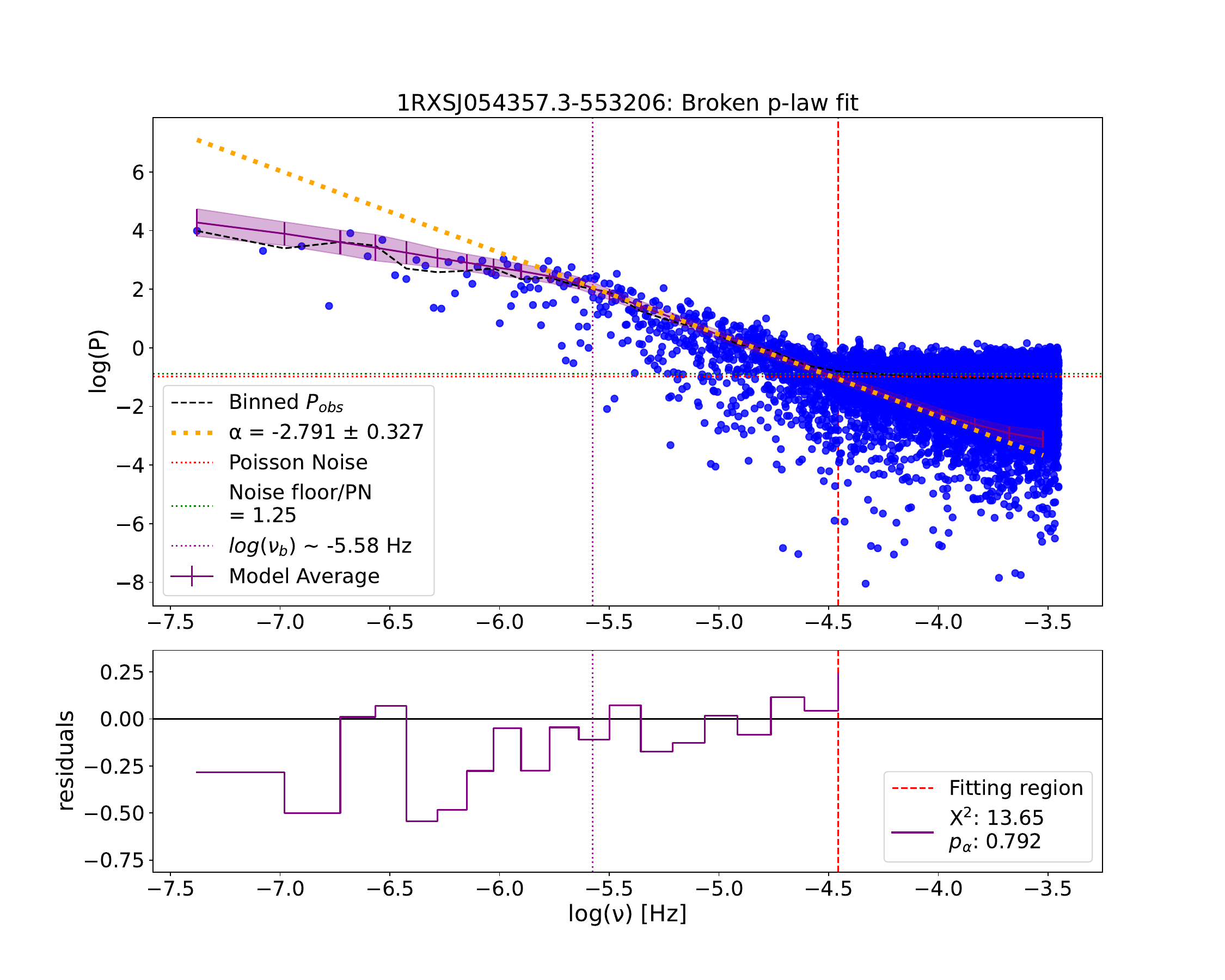}\\
            \includegraphics[width=.5\textwidth]{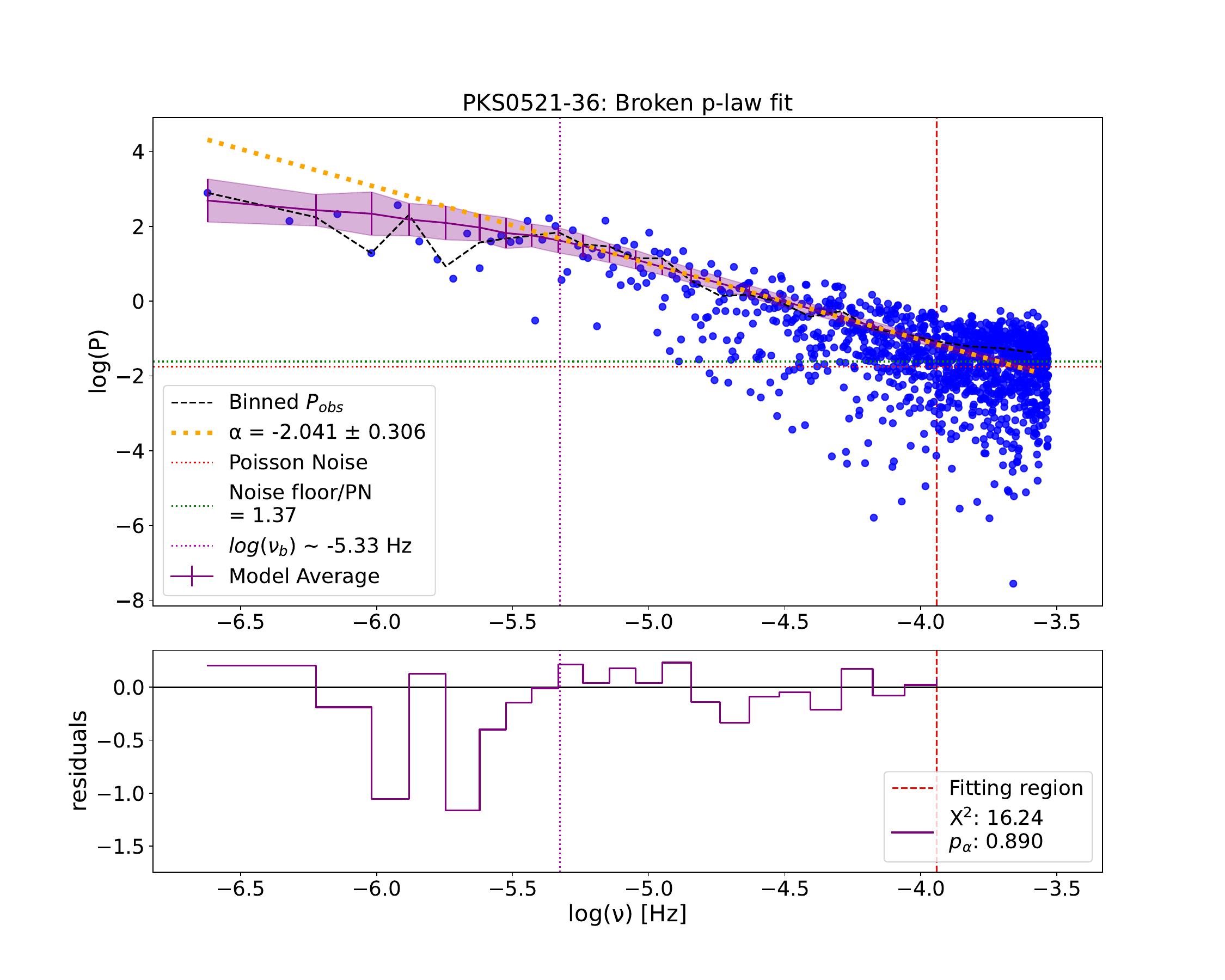}\\
            \includegraphics[width=.5\textwidth]{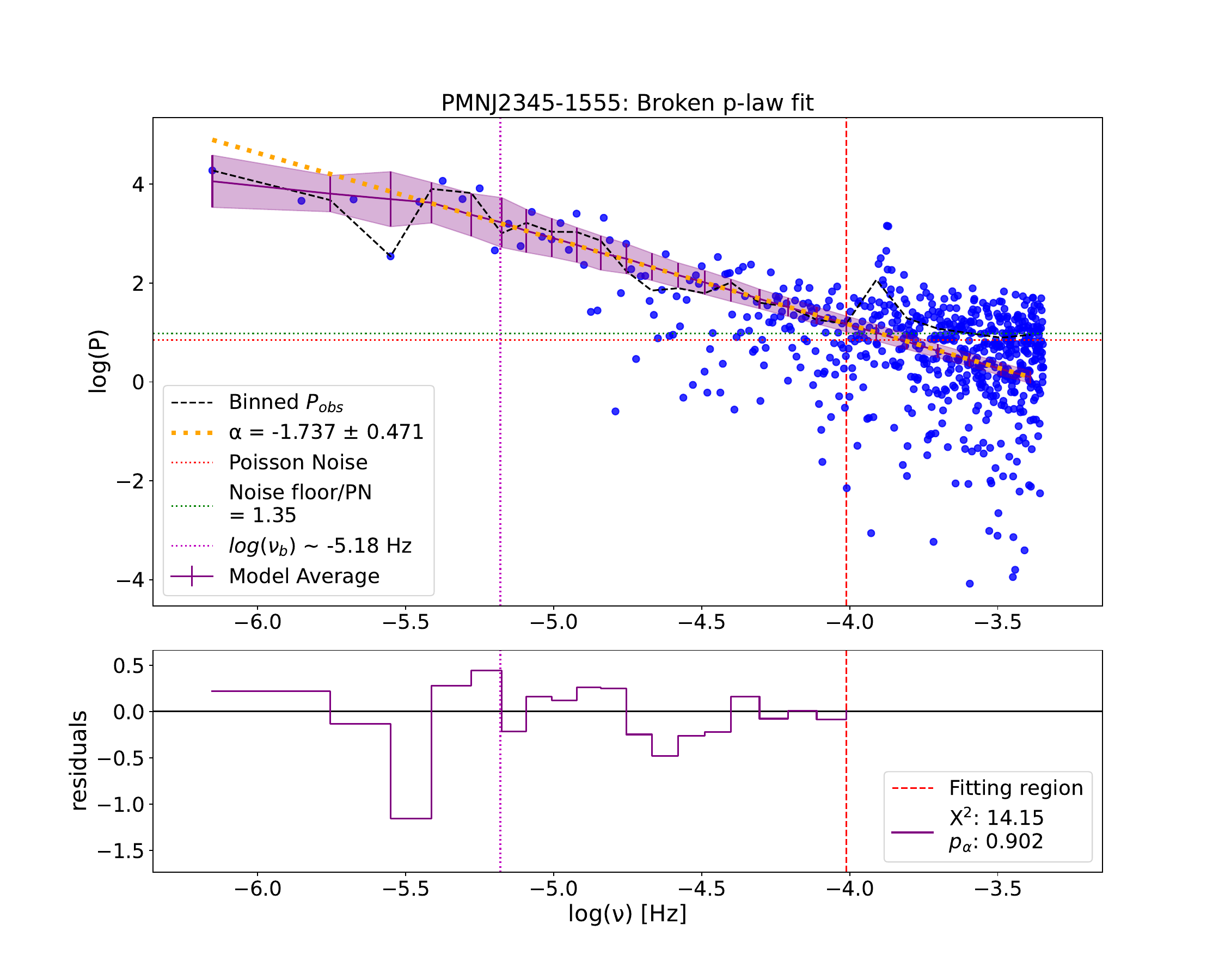}
            \caption{Examples of objects best-fit by a broken power-law model. The binned, target  (black) and model average (purple) PSDs, are shown overlaid on the full target PSD (blue) with a projection of the best-fit $\alpha$ (yellow). The vertical purple line shows $\nu_{k}$. Expected and measure levels of Poisson noise are given by the red, and green lines respectively. The lower section displays the object and model average PSD residuals in the fitting region (red, vertical line) along with $\chi^{2}$ value and success fraction.}
            \label{fig:PSD_results_goodfit_brkn}
        \end{figure}

\begin{figure}
            \includegraphics[width=.5\textwidth]{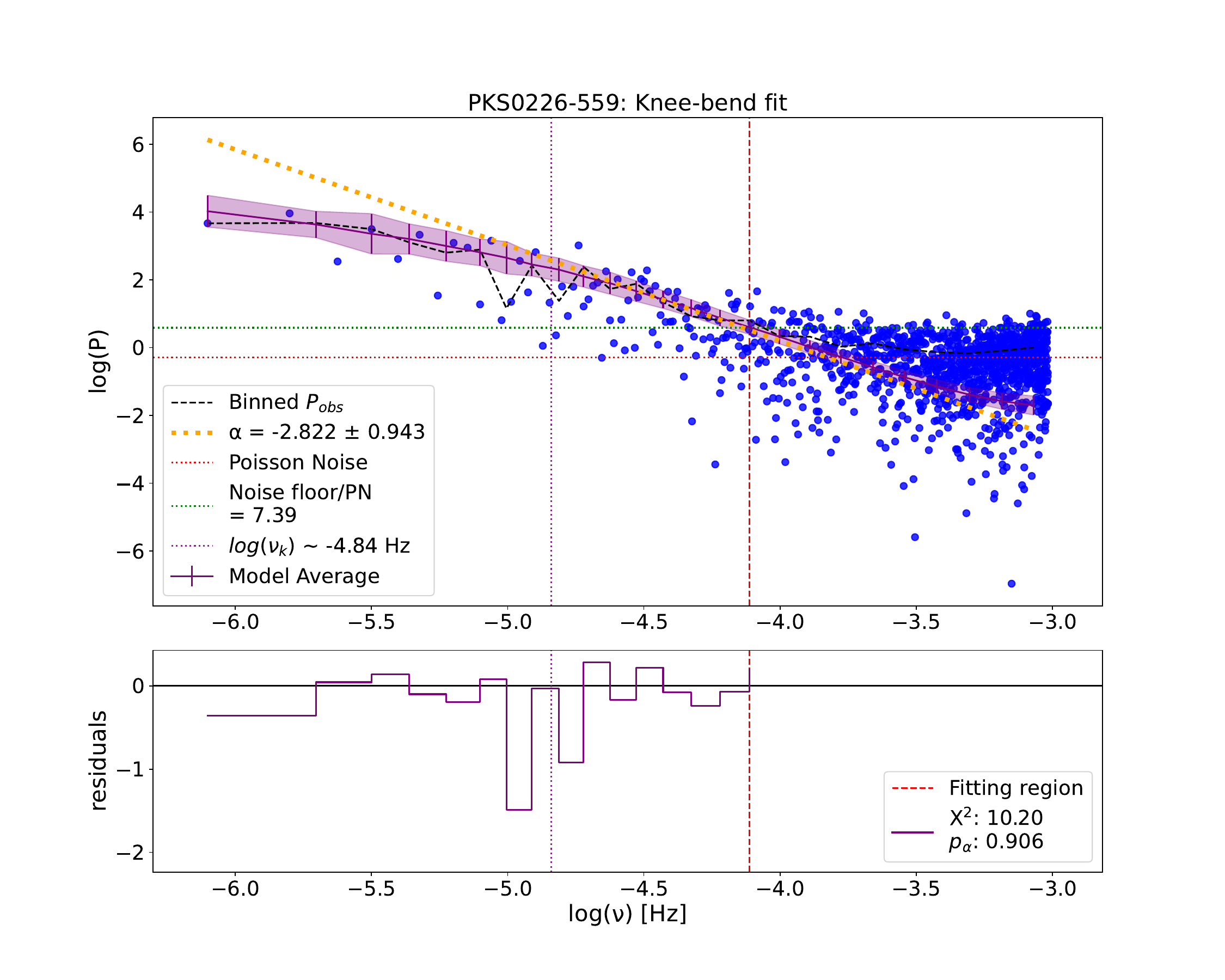}\\
            \includegraphics[width=.5\textwidth]{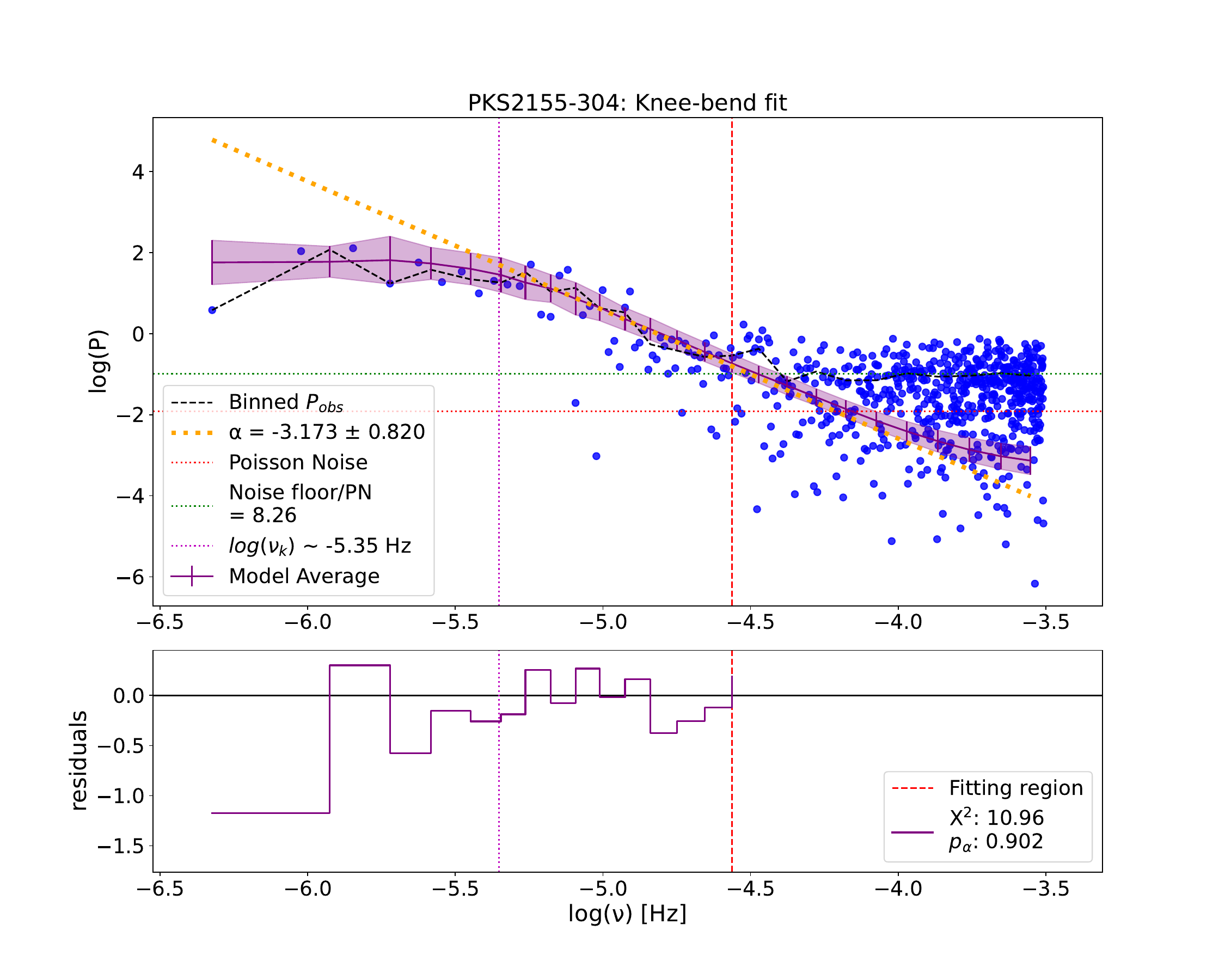}\\
            \includegraphics[width=.5\textwidth]{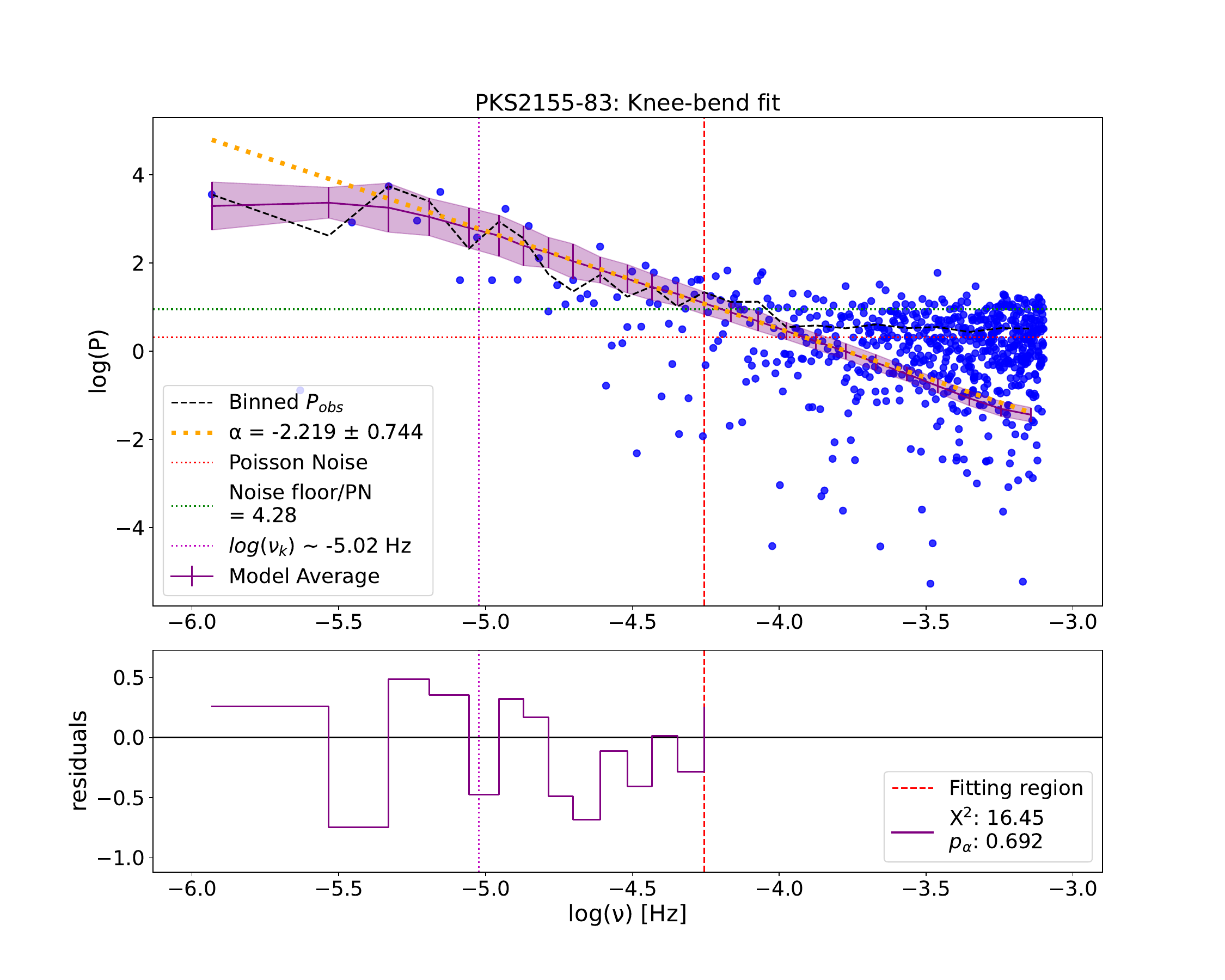}
            \caption{Examples of objects best-fit by a bending power-law model. The binned, target  (black) and model average (purple) PSDs, are shown overlaid on the full target PSD (blue) with a projection of the best-fit $\alpha$ (yellow). The vertical purple line shows $\nu_{b}$. Expected and measure levels of Poisson noise are given by the red, and green lines respectively. The lower section displays the object and model average PSD residuals in the fitting region (red, vertical line) along with $\chi^{2}$ value and success fraction.}
            \label{fig:PSD_results_goodfit_bend}
        \end{figure}

        As seen in Figure \ref{fig:p-law_slopes_all}, the high-frequency power-spectral slopes for all data, as well as those for the BLL and FSRQ populations individually, are approximately normally distributed with an overall mean of $\alpha = -2.14 \pm 0.28$. The isolated BLL and FSRQ populations display respective means of $ \alpha_{BLL} = -2.11 \pm 0.23$ and $\alpha_{FSRQ} = -2.12 \pm 0.17$. Break or bend frequencies identified range between $\log{_{10}(\nu_{k/b})} \approx [-5.83,-4.84]$ Hz. These correspond to timescales, in the rest-frame of the target, between $\approx$ [0.801, 7.825] days. It is not clear what contribution the accretion disk provides to the power spectrum. As such, we do not correct these timescales for doppler-beaming. If the contribution were to be considered negligible, such timescales would be increased by approximately an order of magnitude. In Figures \ref{fig:PSD_results_goodfit_plaw} -- \ref{fig:PSD_results_goodfit_bend}, we provide three examples of objects that best fit each model. While not included in this paper, we provide a publicly available, complete collection of figures for all power-spectral fits\footnote{DOI:\dataset[10.5281/zenodo.10632989]{https://zenodo.org/doi/10.5281/zenodo.10632988}}. To check the consistency of our results with other standard methods, we perform an additional CARMA analysis to find the best fitting damped-random-walk (DRW, CARMA(1,0)) and damped harmonic oscillator model fits, see Appendix \ref{sec:CARMA}.


\section{Discussion}
    \label{sec:discussion}
    
    \subsection{Temporal Variability}
    \label{sec:var_prop_discuss}
    Our main measure for the variability of our sample is the excess variance, $\sigma^{2}_{XS}$. This value can be very small; however, as long as $\sigma^{2}_{XS} - 5~err(\sigma^{2}_{XS}) >0 $, there is some level of variability observed, provided that the standard deviation of the light curve exceeds that of the background region after correction for systematics. We found that sources can have falsely high excess variances due to single-cadence events like cosmic ray hits or long-term trends insufficiently removed by correction techniques. In such cases, the light curves produced very poor and noisy power spectra, which were unsuitable for analysis. We then re-extracted the light curve under the ``full'' hybrid method of \texttt{quaver} to correct such incidents and recalculate our variability metrics.  We then usually found lower levels of variability, if any. We find that 20 objects in our original sample, as indicated by an asterisk in Table \ref{tab:sample}, fall into this category. From these 20 objects, we recovered light curves with significant variability from only PKS~0637-75 and NGC~1218; we excluded the rest from further analysis.
    
    We find the level of variability above the photometric error to be statistically identical between the BLL and FSRQ subpopulations. Due to the small sample size and visually obvious non-normality, we perform a permutative 2-sample Anderson-Darling test \citep{Scholz1987} with 9999 resamples. This test returns a statistic of $\sim-0.816$ and p-value of p$\sim0.845$ implying no significant difference in the level of variability observed between BLLs and FSRQs with $\sim85\%$ likelihood that the samples are drawn from the same distribution.      
    
\begin{figure}
        \centering
        \includegraphics[width=0.5\textwidth]{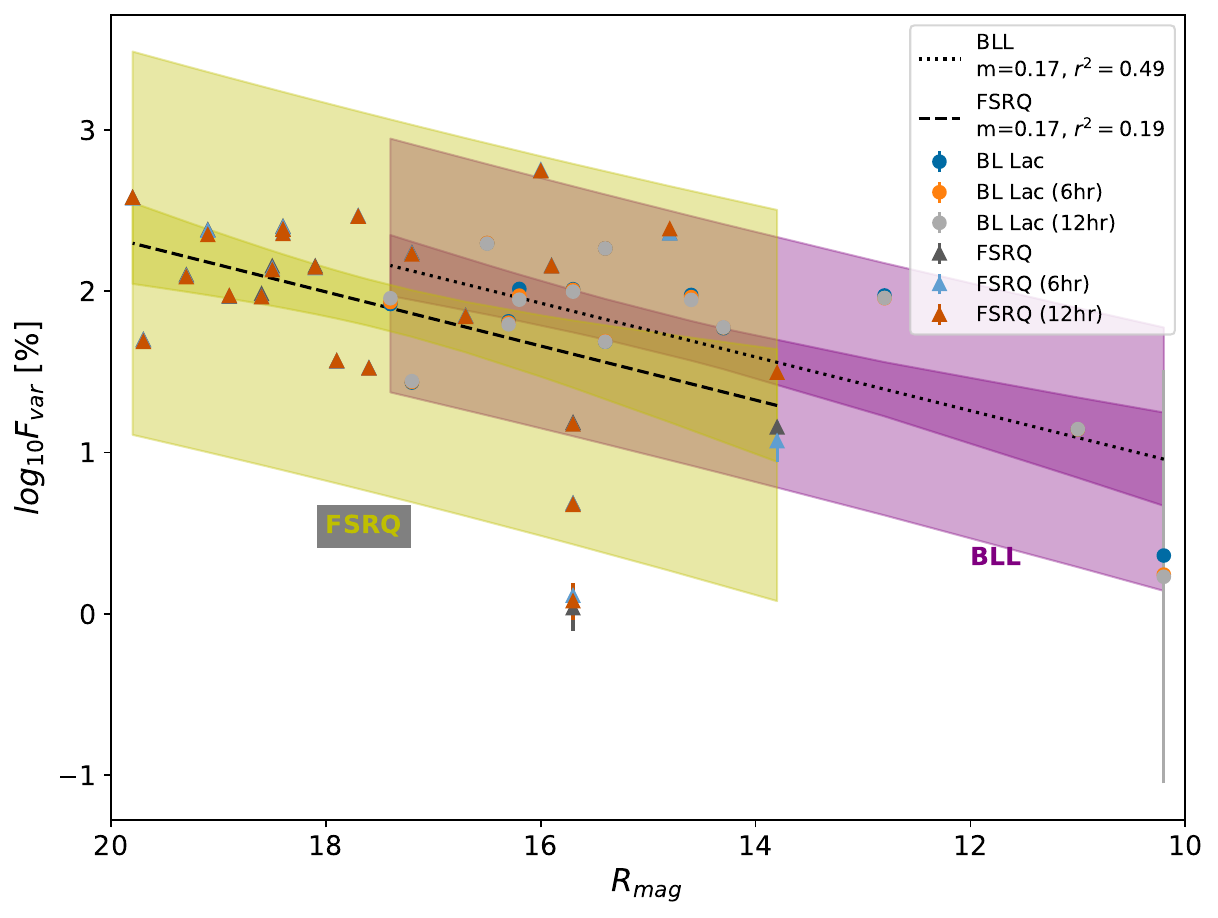}
        \includegraphics[width=0.5\textwidth]{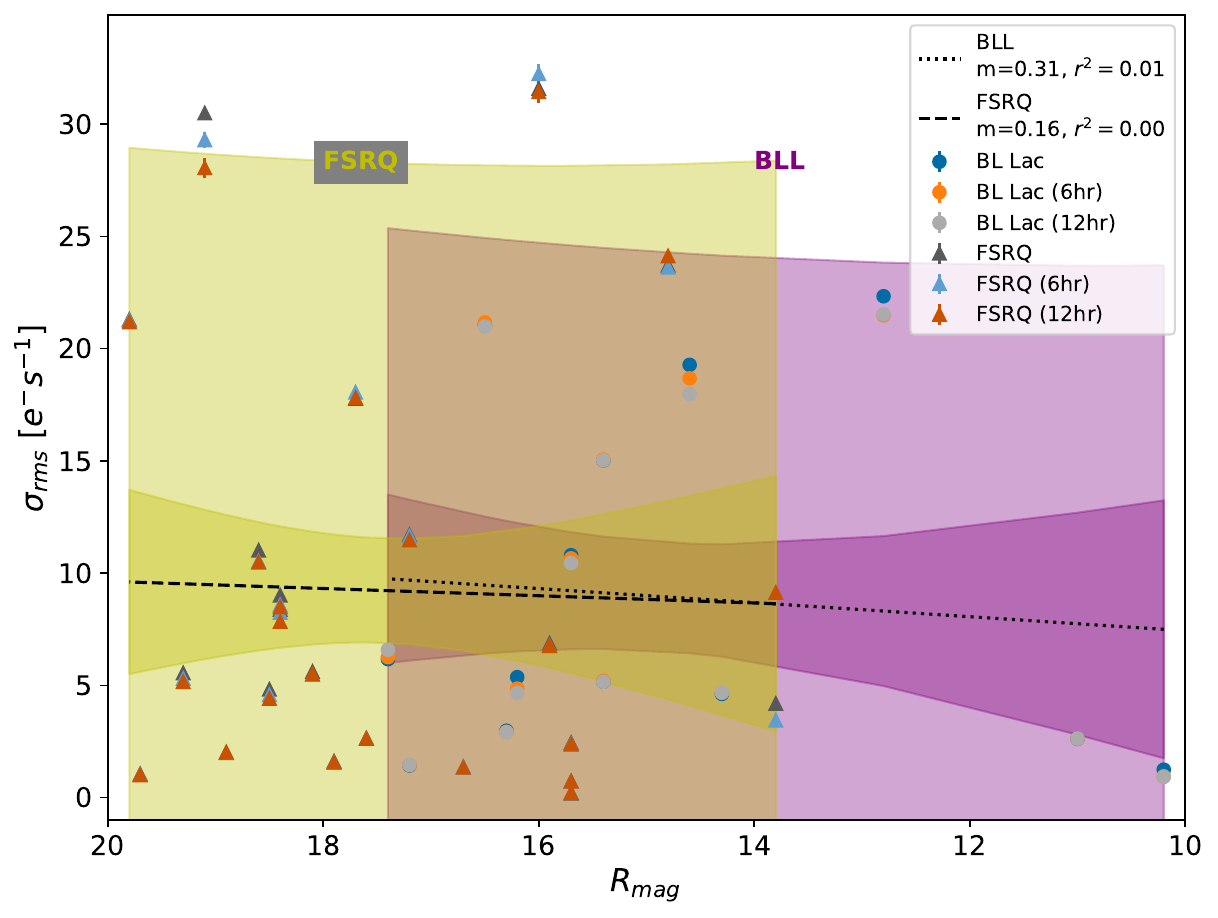}
        \caption{Logarithmic values of $F_{var}$ (top) and $\sigma_{rms}$ (bottom) plotted against the apparent R-band magnitude with 95\% confidence and prediction intervals for each sub-population. The provided coefficients of determination show the fractional variance to have a mildly strong dependence on apparent magnitude, which is not present in the unnormalized observable.}
        \label{fig:Fvar_v_Rmag}
    \end{figure}

    Note that the results of our flux distributions are affected by our regression method and our aperture (Table \ref{tab:variability2}, Figure \ref{fig:Fvar_v_Rmag}). Mean fluxes following background subtraction may not be representative of the true mean. Objects of fainter magnitude are subject to stronger background signals and lower resultant mean fluxes. Also, the photometric errors on the flux of TESS light curves are small and generally do not vary much within their distribution. The effect of the photometric errors on our variability metrics is much smaller than the effects of our reduction methods. Thereby, a normalized measure of variations such as the fractional variance ($F_{var} = \sqrt{\sfrac{\sigma^{2}_{XS}}{\overline{F}^{2}}} = \sigma_{rms}/{\overline{F}}$) is highly dependent on the mean flux and its use is not recommended. When the excess variance is normalized (equation \ref{eqn:sigmaXS}), fainter objects may show a falsely higher value of $F_{var}$ as evidenced by Figure \ref{fig:Fvar_v_Rmag}. Additionally, the effects of fluctuations in the point-spread function of our targets may become more prevalent (producing larger mean flux and/or deviation) in narrower aperture selections \citep{Smith2023}. 

\begin{figure}
        \centering
        \includegraphics[width=0.5\textwidth]{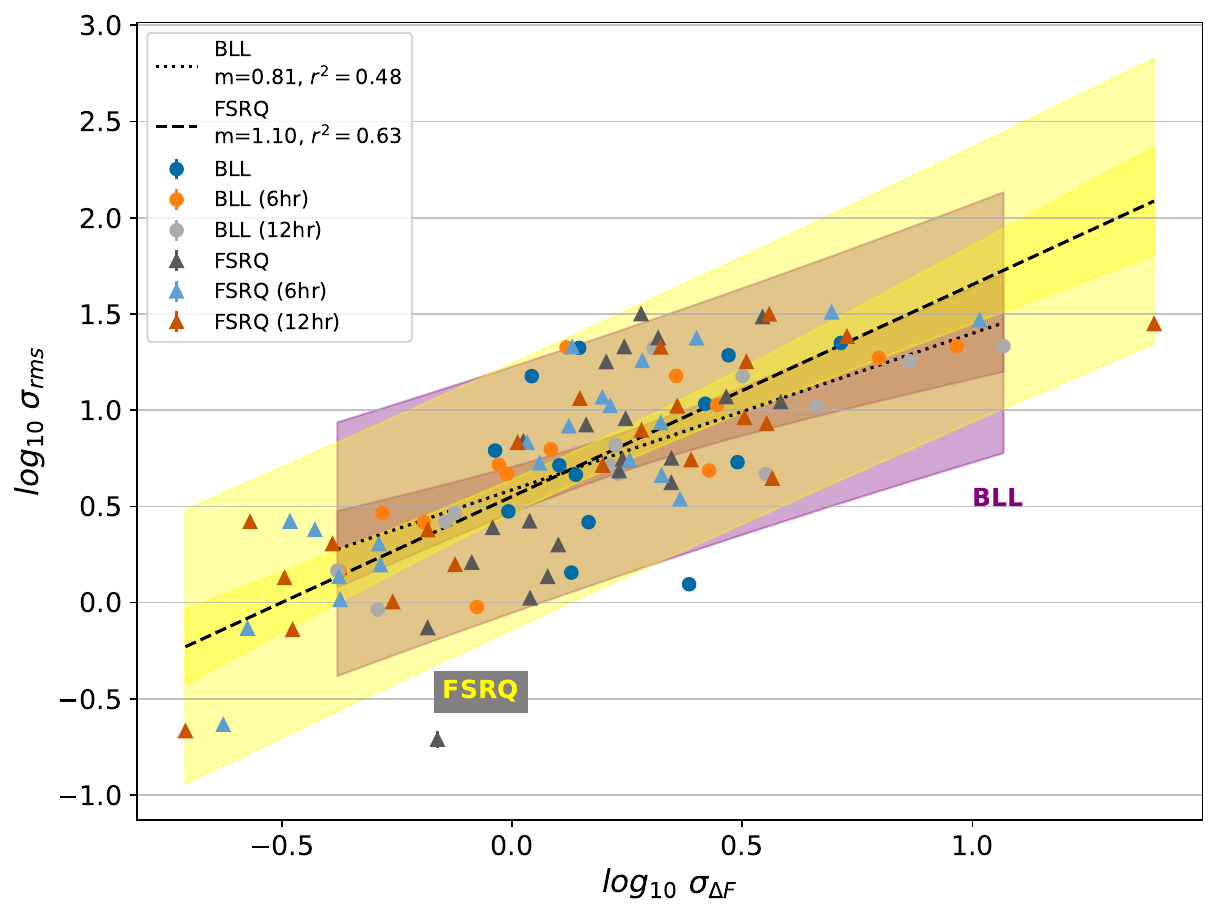}
        \includegraphics[width=0.5\textwidth]{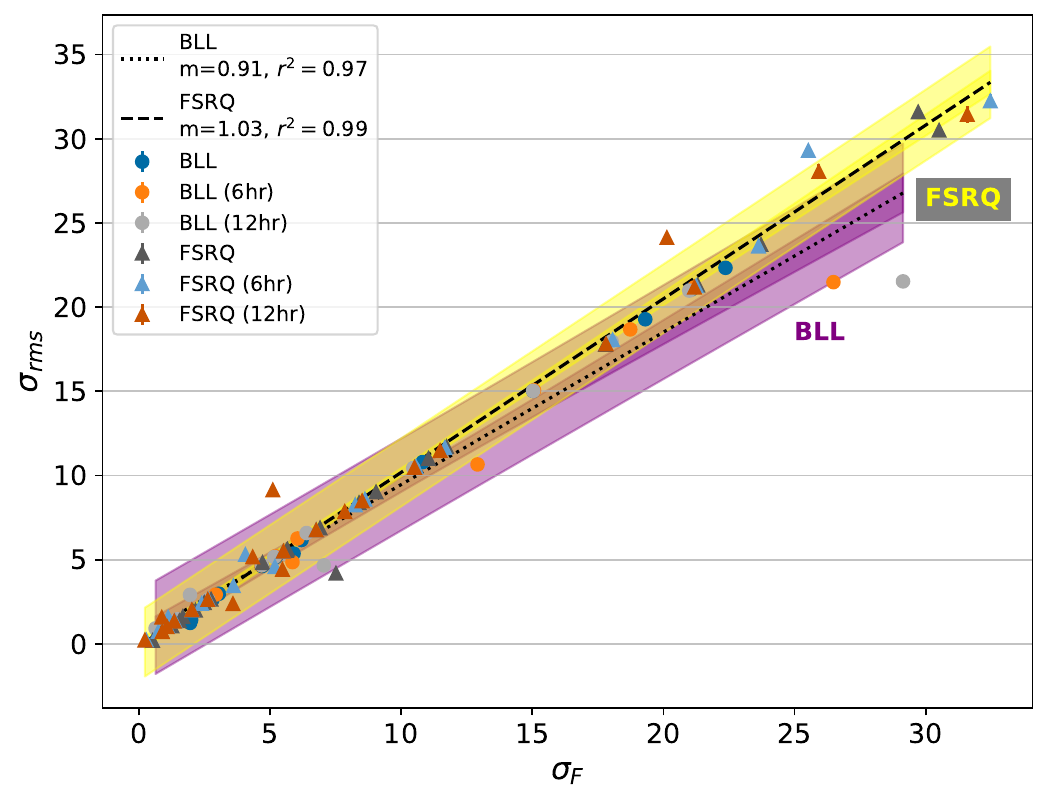}
        \caption{Log-log plot of $\sigma_{rms}$ against $\sigma_{\Delta F}$ (top) and plot of $\sigma_{rms}$ against $\sigma_{\Delta F}$ (bottom) with 95\% confidence and prediction intervals for each sub-population. The provided coefficients of determination show a robust correlation between parameters.}
        \label{fig:xs_v_std}
    \end{figure}

\begin{figure}
        \centering
        \includegraphics[width=0.5\textwidth]{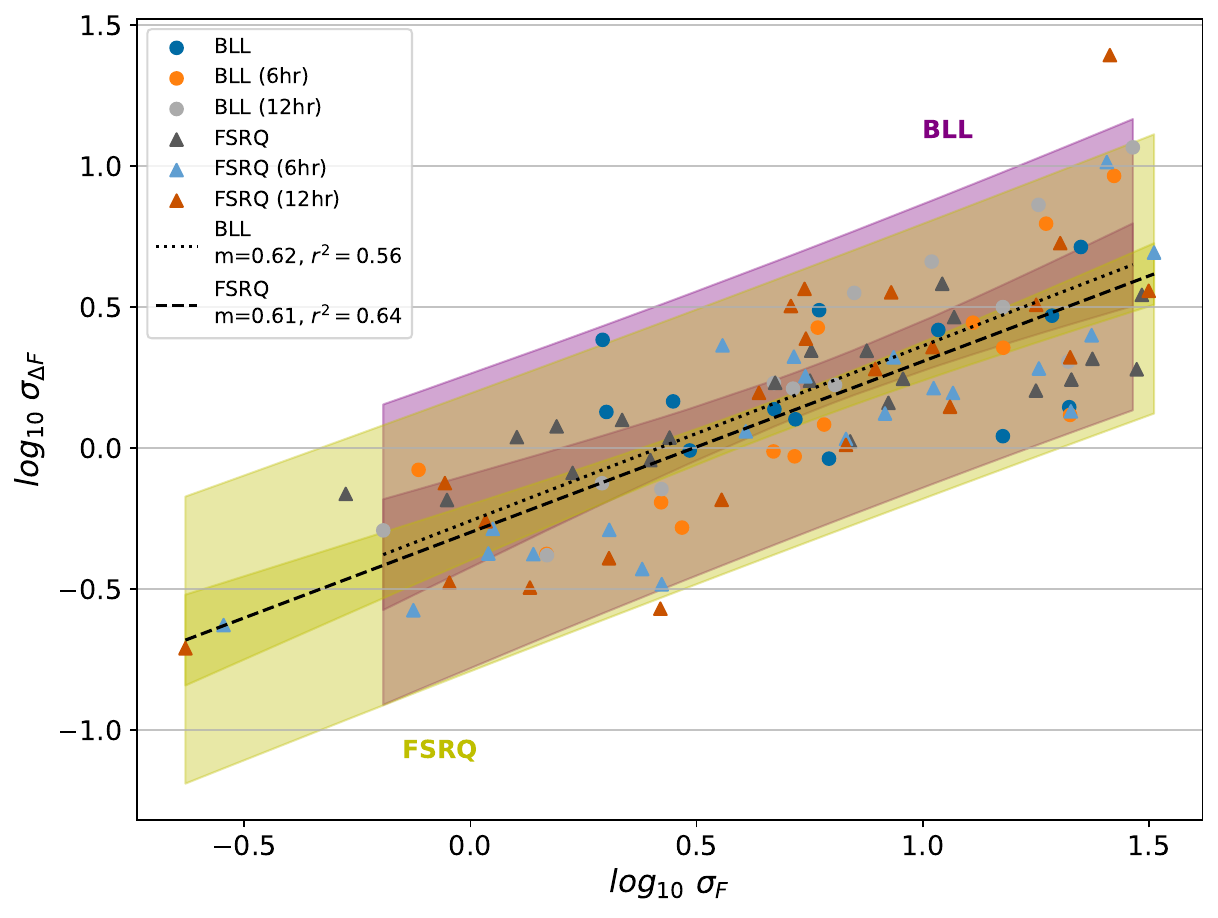}
        \caption{Log-log plot of $\sigma_{\Delta F}$ against $\sigma_{F}$with 95\% confidence and prediction intervals for each sub-population. The provided coefficients of determination show a mildly strong correlation between these parameters.}
        \label{fig:std_v_delstd}
    \end{figure}

    It is more informative to consider the spread of each distribution such that the continuous deviation from the mean characterizes variability. We find a strong, positive correlation between the excess variance and rms-scatter against the flux standard deviation (Figure \ref{fig:xs_v_std}), demonstrating the standard deviation is a standardized measure of variability which should generally be less dependent on extraction methods \citep{Smith2018}. Likewise, the standard deviation of the subsequent flux distribution demonstrates a similar, though less significant, trend (Figure \ref{fig:std_v_delstd}). This measure can characterize the significance of the variability level, whereas a more considerable average change in flux displays a more significant level of variability. 

    Similarly to the subclass comparison of $\sigma^{2}_{XS}$ and $\sigma_{rms}$, we perform 2-sample Anderson-Darling tests on these distributions \citep{Scholz1987}. These tests returns a statistics of $\sim-0.500$ and $\sim-0.351$ with p-values of $p\sim0.621$ and $p\sim0.524$ for $\sigma_{F}$ and $\sigma_{\Delta F}$, respectively. Again, this implies no significant difference in the flux and intra-flux spreads between subclasses.    

    \subsection{Power Spectral Fitting}
    \label{sec:pow_spec_discuss}
    The determination criteria of the best-fit power-law variation are defined below: 
    \begin{enumerate}
        \item For each target, which of the three power-spectral models has the highest relative success fraction? 
        
        \item The high-frequency slope of the best-fit model must be estimated with less than 50\% error; otherwise, the model defaults to the following highest success fraction with less than 50\% error on the slope.
        
        \item In some cases, a broken or bending power-law model may show a higher success fraction than a single power-law; however, the break/bend frequency often hits or comes close to the lower limit that can be probed in the frequency space,  as limited by the light curve baseline. This results in 3 or fewer data points in the low-frequency regime \textit{without a notable drop in power in that regime}, such that $\beta \geq 0$ (Eqs. \ref{eqn:brkn_p-law} and \ref{eqn:bend_p-law}). Since the bending or broken models are, in this case, potentially due to poor low-frequency statistics, we default to the single power law model.
        
        \item As in (3), in some cases, a broken or bending power-law model may show a higher success fraction than a single power-law. The break/bend frequency can come close to the lower limit of the frequency space, as discussed above. In such cases, the single power-law can be solely within the high-frequency regime as determined by the broken/bending power-law models. Suppose the single power-law model in the new regime fails to produce a success fraction higher than its broken/bending counterpart(s). In that case, we consider the broken or bending models the better fit if $\nu_{b/k}$ does not fall below the lower limit frequency.
    \end{enumerate}

    Under the above mentioned criteria, we determine the best fit high-frequency spectral index for comparison. The distributions of $\alpha$ between activity type sub-populations show an $\sim85\%$ overlap. An Anderson-Darling test, performed similarly as in the previous section \citep{Hodges1958}, demonstrates, with a statistic of $\sim$ -0.93, ($p\sim0.92)$ that the BLL \& FSRQ high-frequency slope distributions are equivalent, see Figure \ref{fig:p-law_slopes_all}.
    
\begin{figure}[H]
        \centering
        \includegraphics[width=0.5\textwidth]{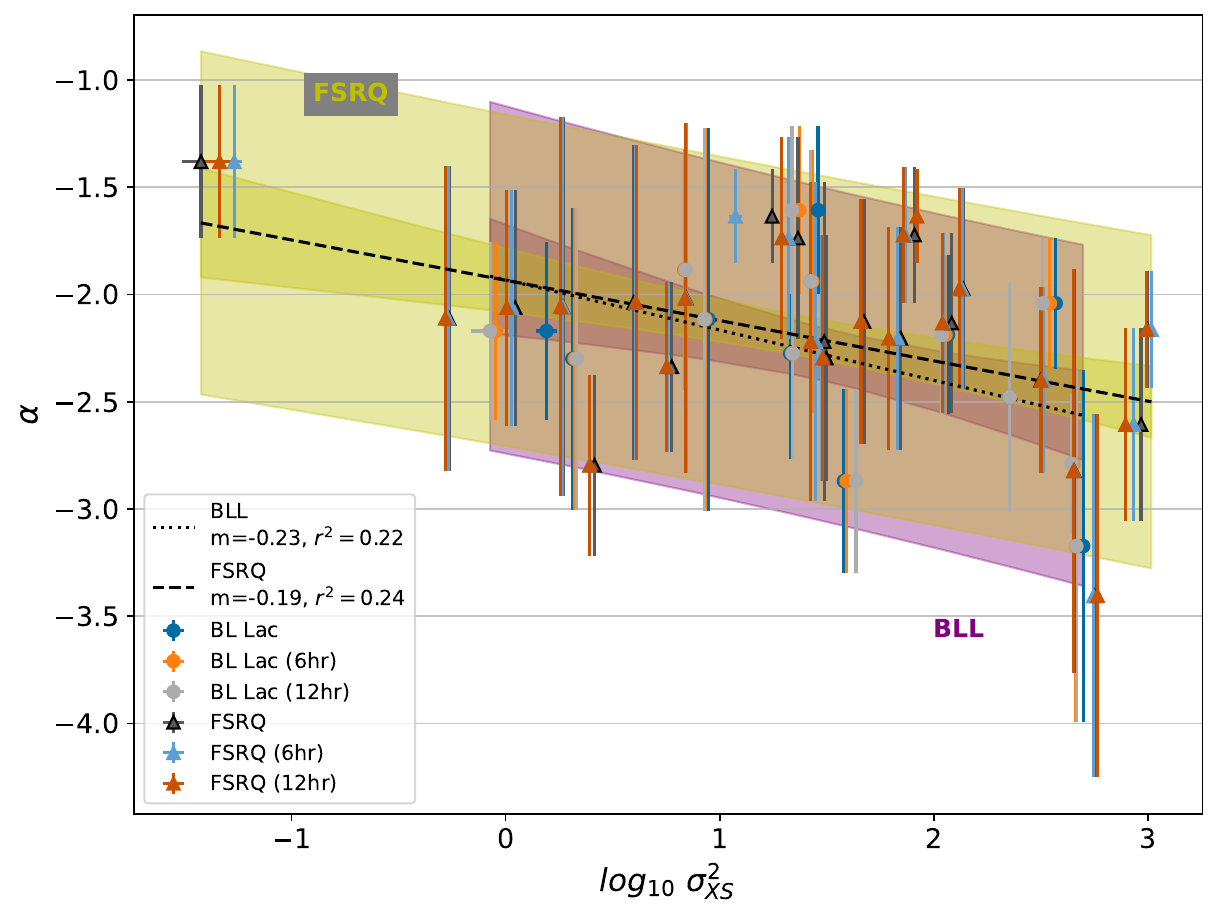}
        \caption{Plot of spectral indices $\alpha$ against $log_{10}~\sigma^{2}_{XS}$ showing a weak to mild correlation.}
        \label{fig:slope_vs_XS}
    \end{figure}

    A comparison of the best-fit spectral indices against the objects' excess variances shows a weak to mild anti-correlation, Figure \ref{fig:slope_vs_XS}, implying that increased variability will manifest as a steeper slope in the power spectrum.

    \subsection{Physical implications}
    \label{sec:physical_implications}
    Investigations of optical variability at a rapid cadence of many blazars are rare. We compare our results to a few key studies and discuss their physical ramifications.
    
    \subsubsection{Detection Fraction of Optical Variability}
        \label{sec:detection_frction}
        We found that all targets in our sample with statistically significant variability in these TESS observations are also detected in $\gamma$-rays by \textit{Fermi}-LAT.

        Out of our 67 targets, 33 have excess variances inconsistent with zero. While non-variable optical light curves from blazars seem counter-intuitive, it is not unexpected based on previous research: \citet{Guar2012} searched for ``short-term'' and intra-day variability in 11 and 4 blazars, respectively, and found that 9 sources varied on weeks-to-months timescales, and only 2 of the 4 blazars with intra-night searches exhibited significant variations; the objects that were statistically flat on hours-to-days timescales included objects with well-known longer-term optical variability (e.g., OJ~287). \citet{Bonning2012} analyzed the SMARTS light curves of six blazars and found that one object (3C~273) barely varied, despite highly variable $\gamma$-ray emission. \citet{Abrahamyan2019} reports that 60\% of the BZCAT blazars exhibited significant variability in Palomar Sky Survey multi-epoch data, and \citet{Kurtandize2005} found that only 7 out of 9 X-ray selected blazars showed any long-term ($\sim$weeks-months) variability in the Abastumani Observatory Blazar Monitoring Program, and only 2 of them showed significant intra-night variability. So, at the timescales probed by the TESS light curves, the percentage of objects with detectable optical variations is similar to that of optical surveys on longer timescales.
        
        However, our other results differ from previous studies, indicating that the optical behavior at short timescales behaves differently than at longer timescales. 
        
    \subsubsection{FSRQs vs BLLs}
        \label{sec:FSRQ_v_BLL}
        
        We find no statistical distinction between our sample's optical variability properties, excess variances, and RMS-scatter of FSRQs and BLL objects. Likewise, we find that the high-frequency power spectral slopes are drawn from statistically identical distributions, and there is no preference for single, bending, or broken power-law models by blazar type. There is also no apparent difference in the characteristic timescales (for those 17 blazars well-fit with a bending or broken model) or shortest timescales of variability between the two groups. This result differs from the modest differences between BLLs and FSRQs as seen in \citet{Wehrle2023}.
        
        The physical difference between blazar subclasses is currently unclear. \citet{Ghisellini2009} suggests that these distinctions arise due to differing accretion disk contributions and emission mechanisms. In this postulate, the $\gamma$-ray emission mechanism for FSRQs is primarily External Compton emission while Synchrotron Self-Compton processes are primary in BLLs. Such differences would manifest as statistically distinct PSD slopes and break-frequencies \citep{Ghisellini2009,Finke2014,Ghisellini2017}. It is also true that sources can transition between subclasses when the jet or disk becomes more dominant, so while the classes are intrinsic, they are not necessarily stable \citep{Ghisellini2011}. This may confuse attempts to differentiate their variability properties, if light curves are not obtained simultaneously with the optical spectra used for classification.
        
        Previous studies using ground-based data over timescales of days to months, with sampling intervals > 1 day, have found various results. In a study of optical blazar light curves from the Palomar-QUEST survey, \cite{Bauer2009} found that the highest amplitude optical flares ($\Delta V > 1$~mag) are more common in FSRQs than in BLLs at all timescales. \cite{Hovatta2014} analyze the PTF and CRTS data of a large sample of blazars and find decisively that FSRQs are more optically variable than BLLs and that $\gamma$-detected objects are more variable than $\gamma$-non-detected objects. However, the significance of the FSRQ/BLL distinction decreases to $\sim2\sigma$ when only low-redshift objects are considered; additionally, when only $\gamma$-detected LSP blazars are considered, there is no statistically significant difference between the populations. Those authors suggest that this supports the picture of \cite{Giommi2013} that a large number of \textit{Fermi} BLLs are FSRQs; that is, they possess a standard accretion disk but are misclassified as BLLs because of the dominance of the beamed non-thermal emission.
            
        The short-term optical variability of these source groups showing no significant difference indicates that the forces driving the differences at $\sim$month timescales are potentially irrelevant at the flaring timescales of the jet: the dominance of the accretion disk in blazar low states, for example. Even when probed with TESS-like cadence, typical quasar variability timescales are on the order of weeks to months \citep{Edelson2014,Kasliwal2015,Smith2018}, so these trends would only manifest in our much shorter light curves as quasi-linear increases or decreases. 
    
    \subsubsection{RMS-Flux Relation}
        \label{sec:rms-flux_implication}
        
        The so-called RMS-flux relation is commonly observed in the X-ray light curves of Seyfert galaxies. While \citet{Sandrinelli2014} found no evidence for it in the optical-NIR light curves of six \emph{Fermi} blazars, this relation was observed in the $\sim180$~day \emph{Kepler} light curve of the blazar W2R1926+42 \citep{Uttley2002,Edelson2013}. Additionally,  \citet{Bhattacharyya2020} observed this relation for blazar Mrk~421 in the X-rays over four epochs from short baseline (5 days) \textit{AstroSat}-SXT and ASCA-GIS3 light curves, as well as in long-term ($\sim$10 years) observations by \textit{Swift}-XRT and simultaneously in $\gamma$-rays by \citet{Bhatta2021}. \citet{Bhatta2021} observes this relation in an additional 11 blazars. 
        
        Samples from both \citet{Sandrinelli2014}, \citet{Bhattacharyya2020}, and \citet{Bhatta2021} include PKS~2155-304 in common with ours; however, \citet{Bhattacharyya2020} does not analyze this object for a potential RMS-flux relation. Our results generally agree with the behavior observed in X-ray observations of blazars such as Mrk~421 \citep{Bhattacharyya2020}, simulations of X-ray blazar light curves, and a previous optical study by \citet{Bhatta2021} of 12 $\gamma$-ray bright blazars. This common feature over multiple wavebands may imply that this is the case for all blazars, in agreement with the unified model of AGN and blazars \citep{Uttley2002,Ghisellini2017,Hovatta2019,Bhattacharyya2020,Kundu2021}. Additionally, this may suggest that the physical mechanism responsible for the X-ray and optical variability in blazars may be the same, or at least, connected processes. 

        Furthermore, intra-day timescales suggest that variability originates in the jet in which timescales are significantly truncated by relativistic Doppler boosting. However, a correlation between flux and the RMS-scatter indicates multiplicative process from additional contributions. \citet{Biteau2012} have shown that the statistical flux properties, i.e. lognormal flux distributions and RMS-flux relation, and rapid variability of blazars can be reproduced by models of `minijets-in jets'.
                    
    \subsubsection{PSD models and Characteristic Timescales}
        \label{sec:timescale_implication}
              
        We find that 17 of our power spectra are best fit by bending or broken power law models. The location of the bend or break corresponds to characteristic timescales of 0.8 -- 7.8 days. The average slope of the PSDs, $\alpha \sim2$ as identified in this and other studies is consistent with present variability models to support leptonic emission and 'minijet-in-jet' models and magnetohydrodynamic simulations of a turbulent disk \citep{Pollack2016,Curd2019,Thiersen2019}. These timescales are in good agreement with those identified in $\gamma$-ray modelling by \citet{Ryan2019} and may correspond to dynamical fluctuations or physical processes in the jet due to fluid dynamics, such as the kink instability \citep{Jorstad2022}. Such timescales suggest leptonic emission mechanisms and superimposed stocastic variabilty mechanism which may coincide with light-crossing, electron escape, or cooling timescales \citep{Finke2014,Goyal2021}. If this study were able to reach even lower frequency regimes, it is suggested by \citet{Ryan2019} that we would also be able to recover maximum correlation times. Furthermore, the agreement between optical and $\gamma$-ray break frequencies coincide with Synchrotron Self-Compton emission relaxation time \citep{Finke2014,Fan2023}.
    
        Utilizing the same relation as defined in equation \ref{eqn:radii_est}, excluding the Doppler factor since the disk contribution is unknown, we provide radii of emitting regions, ranging from $R_{em}\sim(0.021 - 0.203)\times 10^{12}$ km ($0.67 - 6.57~\mathrm{mpc}; \sim22 - 217~R_{s}$ for $M_{BH} = 10^{8.5} M_\odot$), which may correspond to characteristic timescales as derived from the break/bend frequencies (see Tables \ref{tab:psd_results} and \ref{tab:psd_timescales}) \citep{Sandrinelli2014,Liodakis2017,Liodakis2018}. These results show that the optical emission likely originates from regions in the jets and the accretion disk exceeding $\sim7-3800\times$ the size of the central supermassive black hole.

        Timescales of variability from this study may provide important insights into the physical structure and processes within the central engine. Based on the disk model of \citet{Shakura1973}, if these timescales are expected to scale with the size of the emitting region, shorter timescales and multiple timescales identified may suggest jet substructures or `jets-in-jets'.  Such a result is further supported in conjunction with an RMS-flux relation as we have found for many objects in our sample as well as symmetric rise and fall timescales \citep{Biteau2012,Carini2020}. It has been modelled that, while some "short-term" may be the result of orbiting "hot-clumps" in the accretion disk, purely disk emission provides much longer timescales than those identified in this work \citep{Zhang1991,Carini2020}. Such a result may imply that shorter timescales in our study are due to highly jet-dominated emission, while longer timescales have a more significant disk contribution.

        \begin{deluxetable}{lccc}
            \tabletypesize{\scriptsize}
            \tablewidth{0pt} 
            \tablenum{6}
            \tablecaption{Emission region estimates for characteristic timescales\label{tab:psd_timescales}}
            \tablehead{
            \colhead{Target} & \colhead{$log{(\nu_{b,k})}$ [$10 ^{x}$ Hz]} &
            \colhead{$\tau_{char}$ [days]} & \colhead{$R_{em}~[\times10^{12}$ km]} }
            \startdata
            1RXSJ054357.3-553206 & -5.68 & 5.54 & 0.143\\
            PKS0035-252 & -5.14 & 1.60 & 0.041\\
            PKS0208-512 & -4.99 & 1.13 & 0.029\\
            PKS0226-559 & -4.84 & 0.80 & 0.021\\
            PKS0235-618 & -5.83 & 7.83 & 0.203\\
            PKS0346-27 & -5.69 & 5.67 & 0.147\\
            PKS0426-380 & -5.43 & 3.12 & 0.081\\
            PKS0521-36 & -5.35 & 2.59 & 0.067\\
            PKS0637-75 & ~ & ~ & ~\\
            ~~sect. 4+5+6 & -5.71 & 5.94 & 0.154\\
            ~~sect. 7+8+9 & -5.56 & 4.20 & 0.109\\
            PKS0829+046 & -5.25 & 2.06 & 0.053\\
            PKS1244-255 & -5.62 & 4.82 & 0.125\\
            PKS2155-304 & -5.4 & 2.91 & 0.075\\
            PKS2155-83 & -5.48 & 3.50 & 0.091\\
            PKS2326-502 & -5.78 & 6.97 & 0.181\\
            PKS2345-16 & -5.46 & 3.34 & 0.086\\
            PMNJ2345-1555 & -5.39 & 2.84 & 0.074
            \enddata
        \end{deluxetable}

\section{Summary}
    \label{sec:summary}

We have analyzed the TESS, Cycle 1, 30-minute cadence light curves of 67 blazars, including 41 FSRQs and 26 BLLs. We have found the following:
\begin{enumerate}
    \item{Eighteen FSRQs and 15 BLLs, or $49.3\%$ of our original sample meeting our selection criterion, were found to be statistically variable.}
    \item{Our sample demonstrates no significant difference in the level of excess variance and RMS-scatter between FSRQs and BLLs, and the level of RMS-scatter provides for an intrinsic variability level of $\sigma_{rms}\sim0.19-35.31~e^{-}s^{-1}$ above the photometric error.}
     \item{Approximately half of the sample have PSDs best fit by a single power law, while the other half are best fit by bending or broken power laws.}
    \item{We find no statistically significant difference in either the power-spectral slope distributions nor amongst the identified $\nu_{k}$ or $\nu_{b}$ between FSRQs and BLLs.}
    \item{When selecting the minimum time between $\tau_{d,inc}$ and $\tau_{d,dec}$, we find that the shortest timescales of variability within our sample fall within the range $\tau_{d}~\epsilon~$[0.21,140.22] days in the objects' rest frames. These timescales correspond to emission region upper-limits of $R_{em}\sim0.17-117.70$ mpc ($\sim 6 - 3900~R_{s}$ for $M_{BH} = 10^{8.5} M_\odot$).}
    \item{Characteristic timescales suggested by frequencies $\nu_{k}$ or $\nu_{b}$ range between $\tau_{char}\sim$ 0.8--7.8 days. Such timescales would correlate to emission region upper limits of $R_{em}\sim0.67-6.57$ mpc $\sim(0.021 - 0.203)\times 10^{12}$ km ($\sim22 - 217~R_{s}$ for $M_{BH} = 10^{8.5} M_\odot$).}
    \item{We identify a linear rms-flux relation in $69 - 74\%$ of our sample light curves with some level of correlation ($r^{2} \geq 0.25$) and strong correlations ($r^{2} \geq 0.75$) are found in $20-23\%$.}
    \item{Spectral indices display a weak anti-correlation to the excess variance such that greater variability in the time-domain manifests as a steeper power-spectral slope in the frequency domain.}
\end{enumerate}


\renewcommand\refname{References Cited}
\bibliography{blazar_references}

\appendix{
    \section{Light Curves}
    \label{sec:lightcurves}

    We present here, in Figure set \ref{fig:lightcurves}, the full sample of light curves (BLACK) after regression by the \texttt{quaver} pipeline, prior to any correction. Provided in red are Gaussianly smoothed light curves for a kernel of 1-hour to be representative as the continuous average while, in blue, we provide the longest linear trend identified to be removed before power spectral analysis. We provide the target name and the Cycle 1 sector(s) of observation. The axes show the TESS Julian Date (TJD = JD - 2457000) and the flux (counts per second) on the x- and y- axis, respectively. For reference to the corresponding Gregorian dates date, we provide, in Table \ref{tab:dates}, a full list of start and end dates for each sector and orbit in Cycle 1. A complete collection of figures for all light curves after extraction, including the aperture selection and identified TESS systematics can be found at {DOI:\dataset[10.5281/zenodo.10982328]{https://doi.org/10.5281/zenodo.10982328}} while all figures for the regressed light curves and linear trends can be found at {DOI:\dataset[110.5281/zenodo.11490590]{https://doi.org/10.5281/zenodo.11490590}}.

    \begin{figure}[H]
        \includegraphics[]{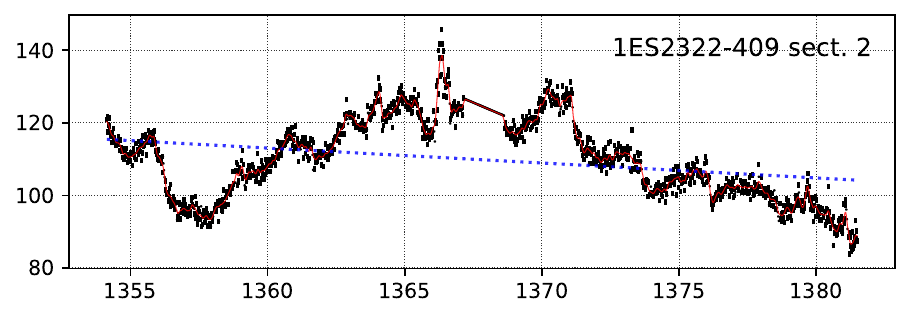}
        \includegraphics[]{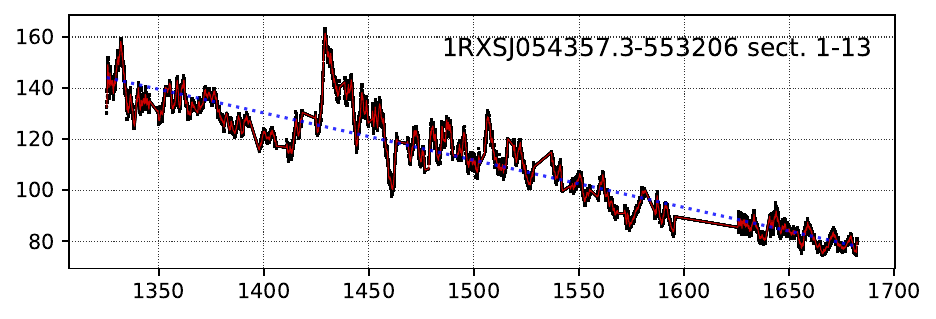}
        \includegraphics[]{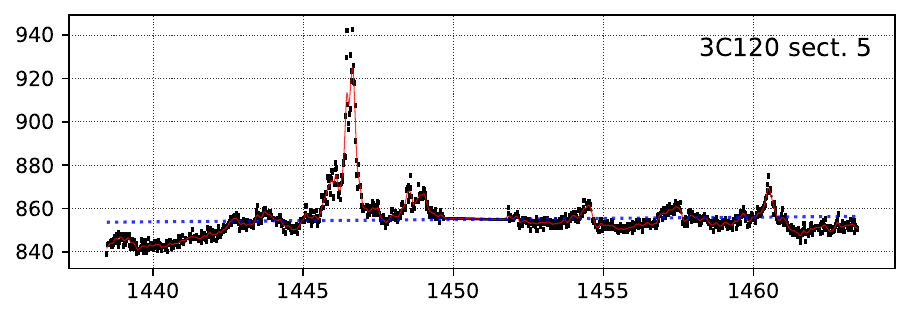}
        \includegraphics[]{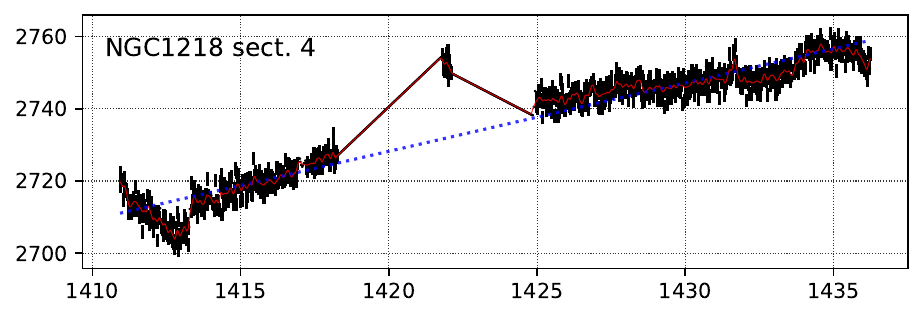}
        \centering
        \caption{}
        \label{fig:lc_group1}
    \end{figure}
    \begin{figure}[H]
        \includegraphics[]{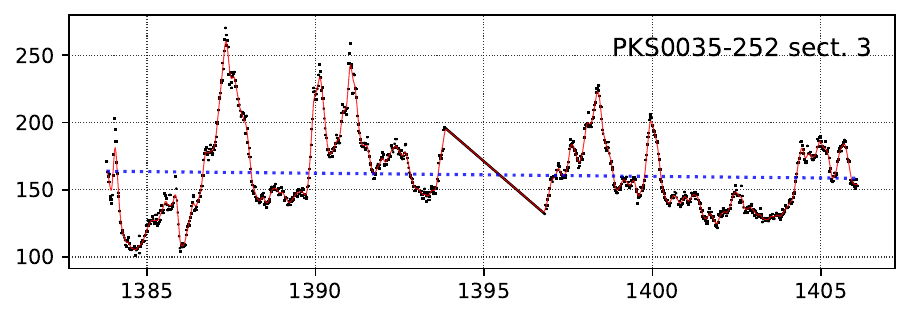}
        \includegraphics[]{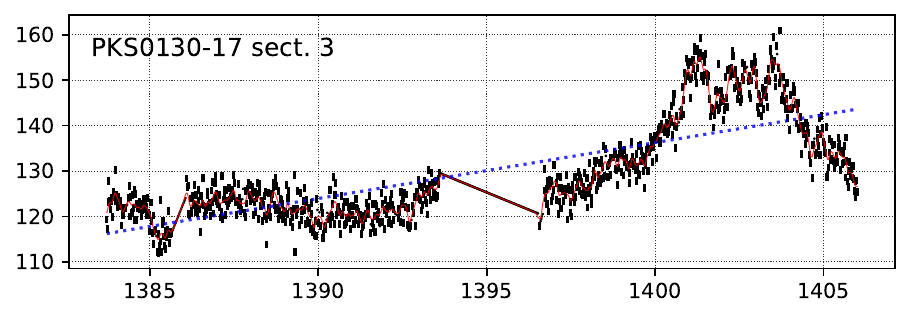}
        \includegraphics[]{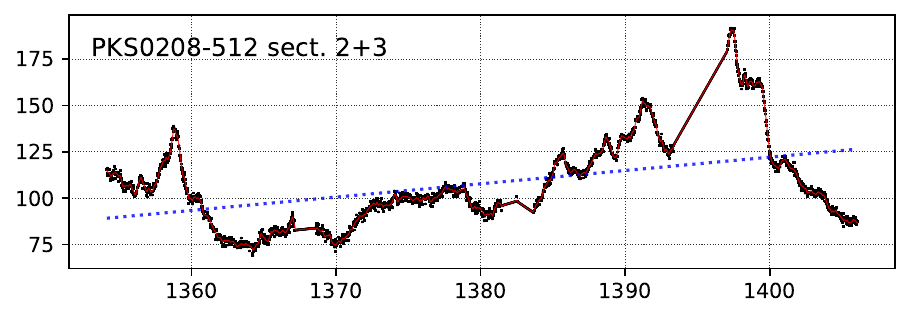}
        \includegraphics[]{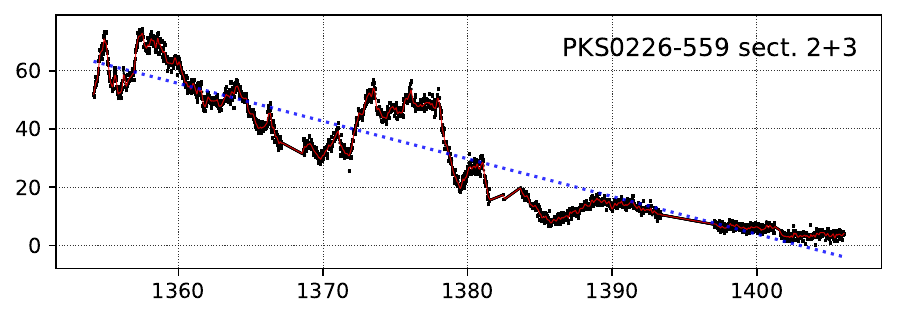}
        \centering
        \caption{}
        \label{fig:lc_group2}
    \end{figure}
    \begin{figure}[H]
        \includegraphics[]{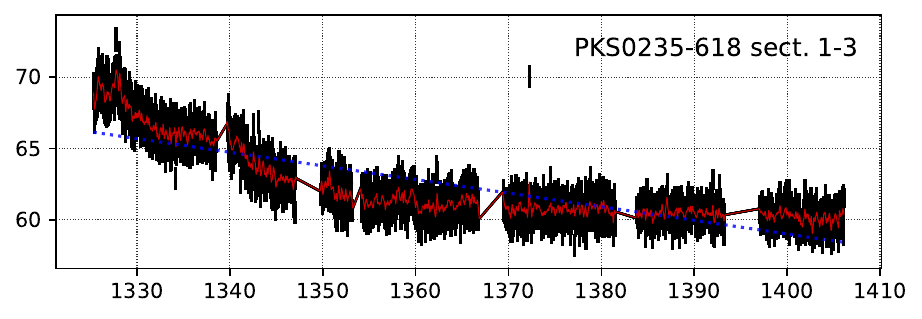}
        \includegraphics[]{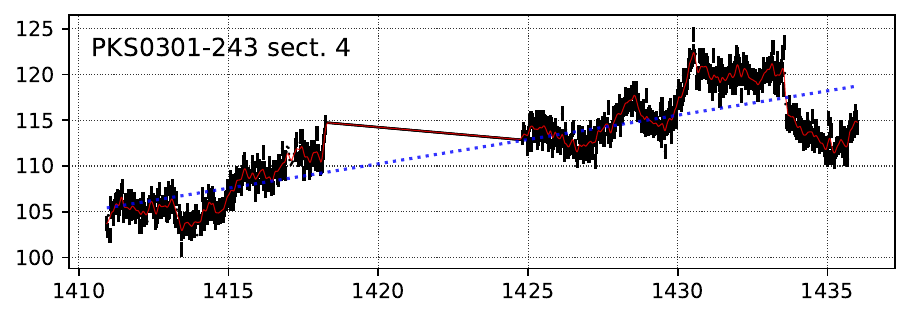}
        \includegraphics[]{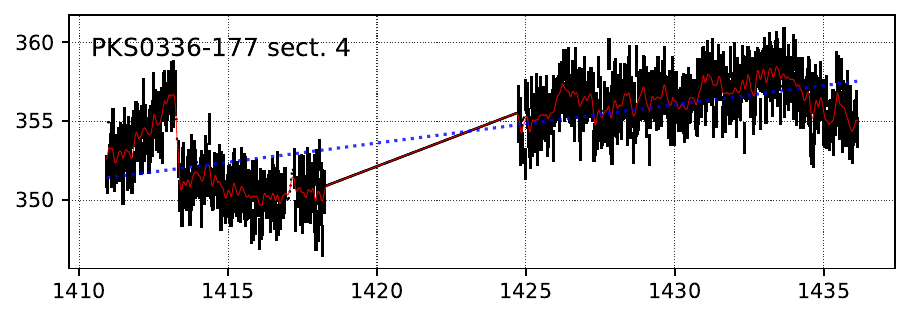}
        \includegraphics[]{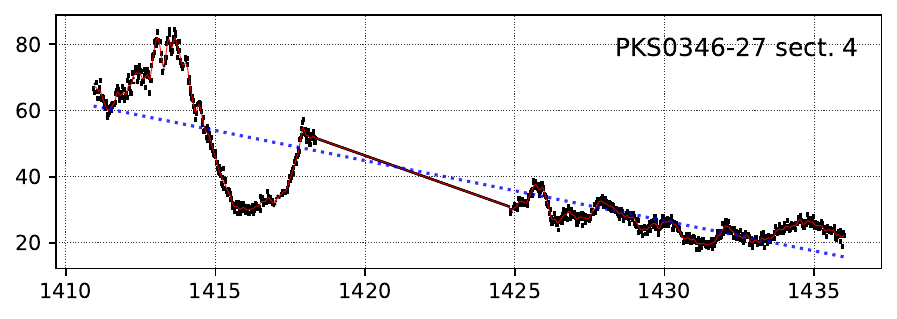}
        \centering
        \caption{}
        \label{fig:lc_group3}
    \end{figure}
    \begin{figure}[H]
        \includegraphics[]{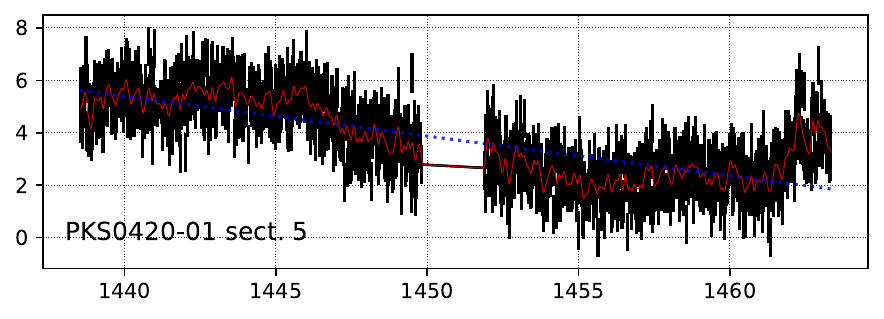}
        \includegraphics[]{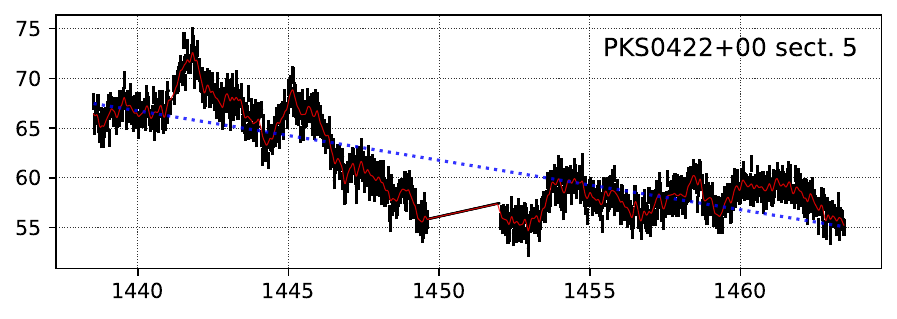}
        \includegraphics[]{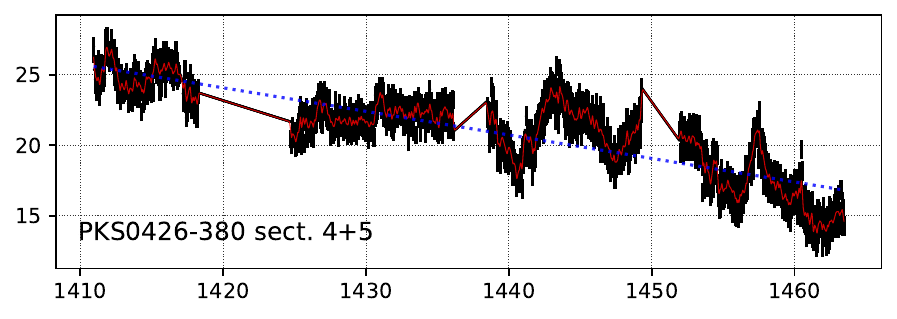}
        \includegraphics[]{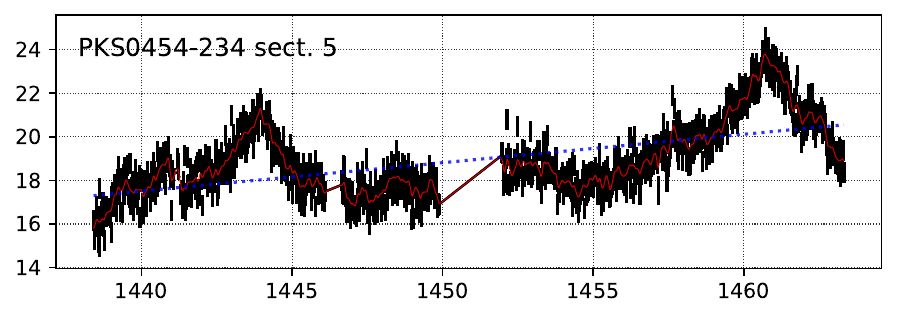}
        \centering
        \caption{}
        \label{fig:lc_group4}
    \end{figure}
    \begin{figure}[H]
        \includegraphics[]{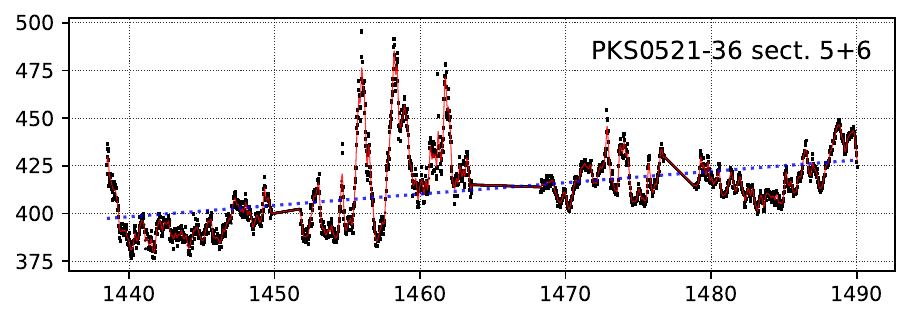}
        \includegraphics[]{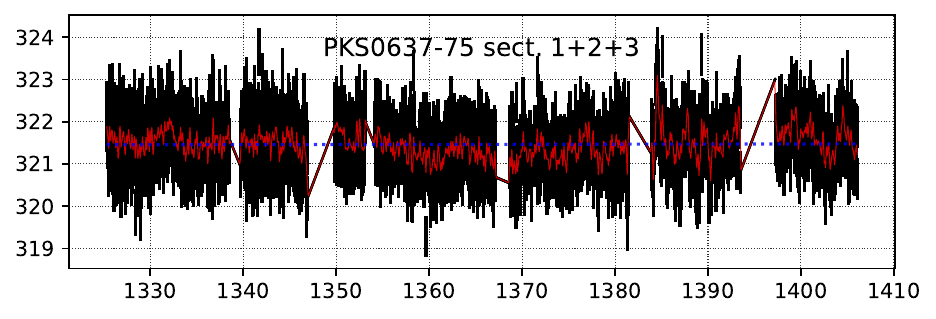}
        \includegraphics[]{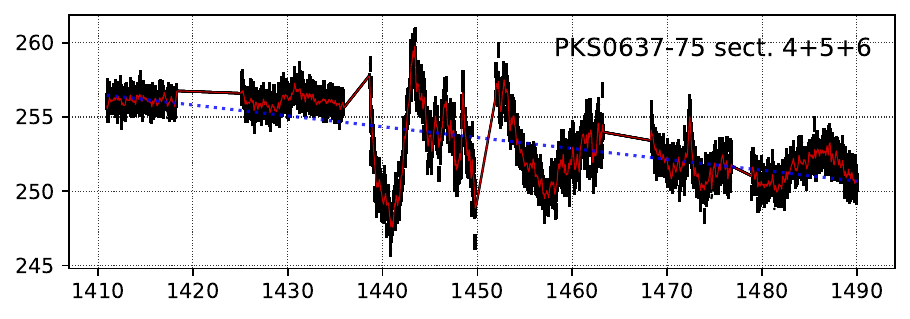}
        \includegraphics[]{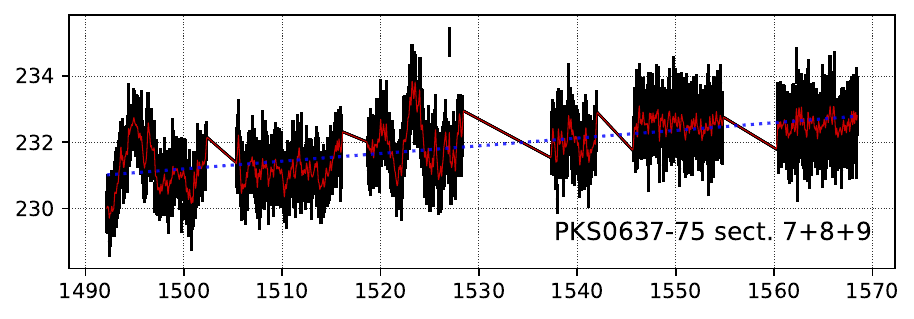}
        \centering
        \caption{}
        \label{fig:lc_group5}
    \end{figure}
    \begin{figure}[H]
        \includegraphics[]{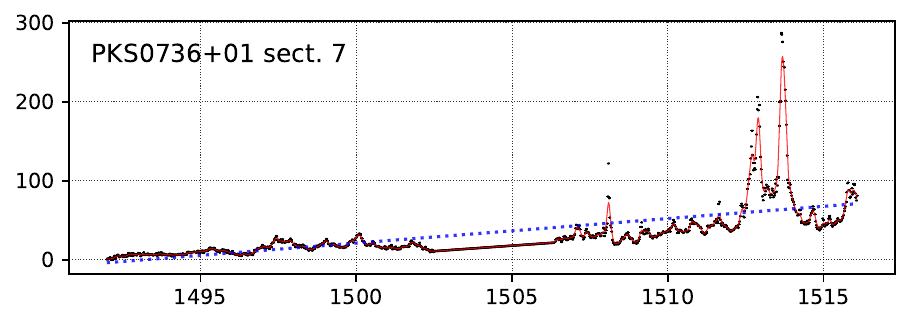}
        \includegraphics[]{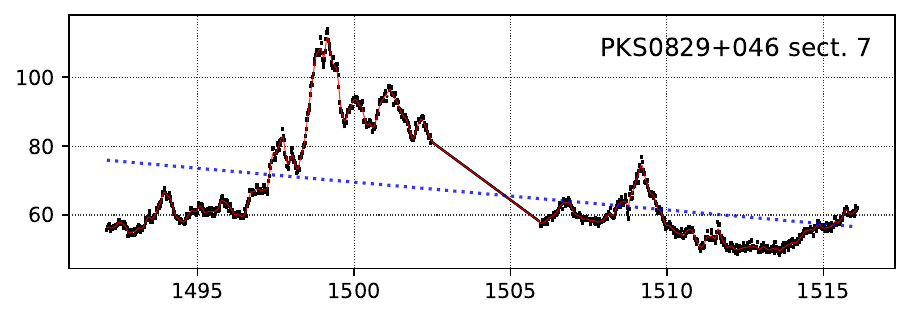}
        \includegraphics[]{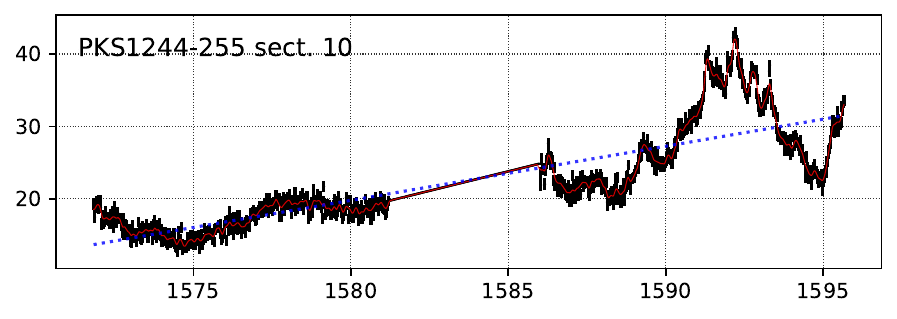}
        \includegraphics[]{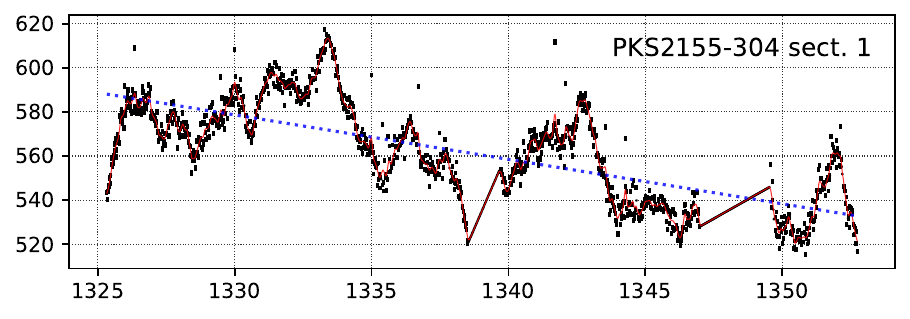}
        \centering
        \caption{}
        \label{fig:lc_group6}
    \end{figure}
    \begin{figure}[H]
        \includegraphics[]{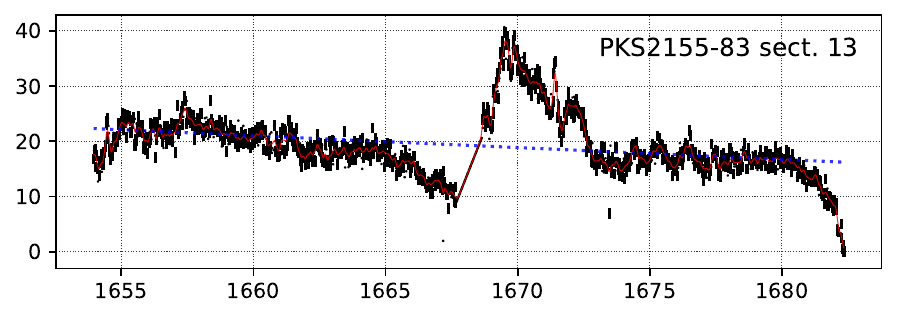}
        \includegraphics[]{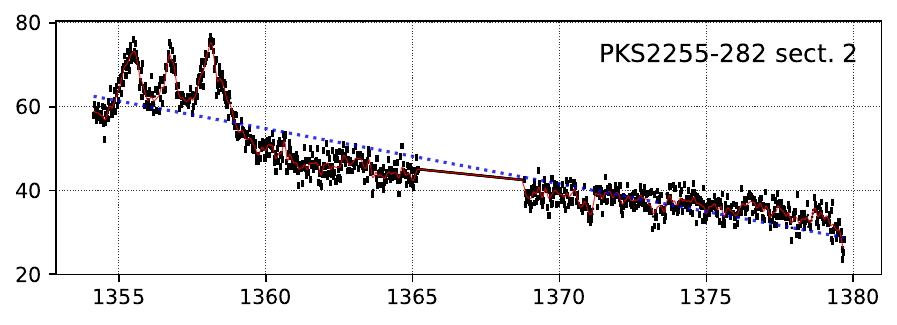}
        \includegraphics[]{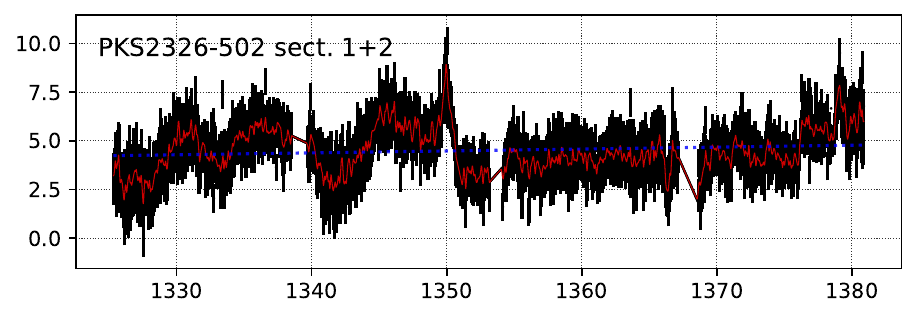}
        \includegraphics[]{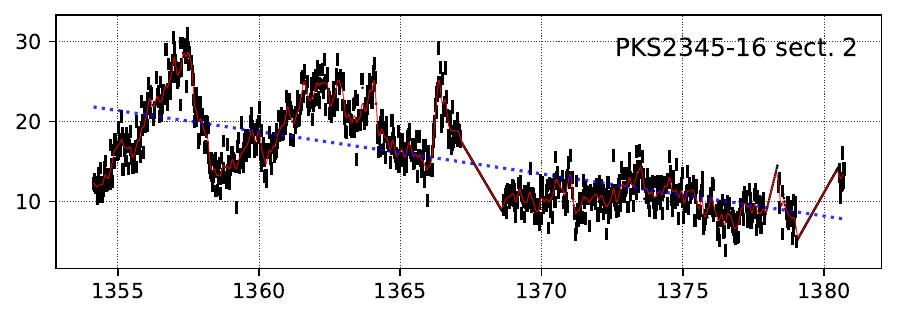}       
        \centering
        \caption{}
        \label{fig:lc_group7}
    \end{figure}
    \begin{figure}[H]
        \includegraphics[]{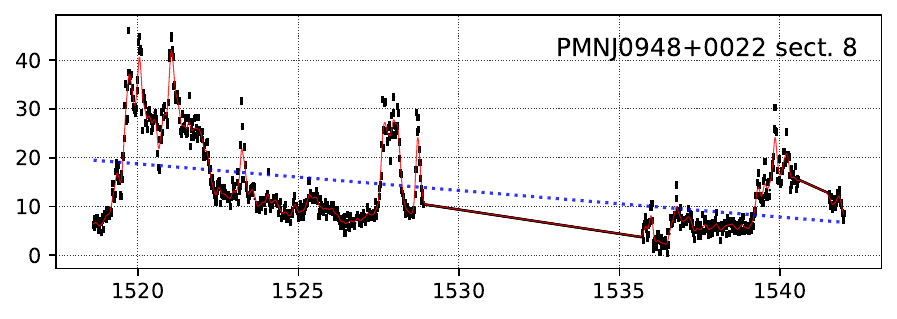}
        \includegraphics[]{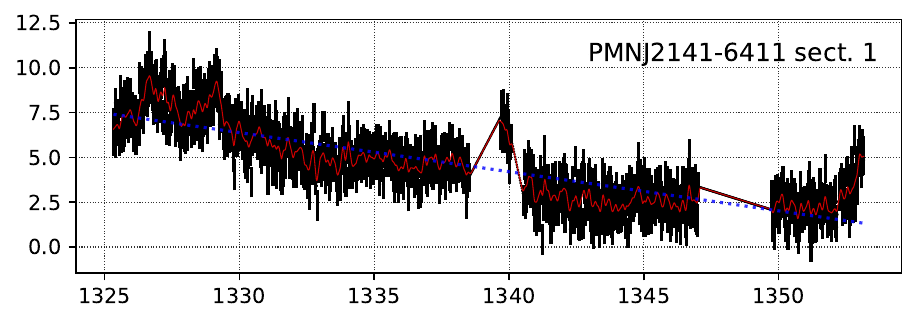}
        \includegraphics[]{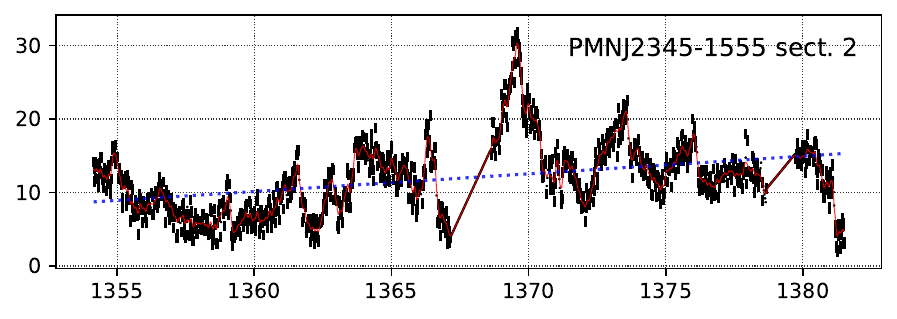}
        \includegraphics[]{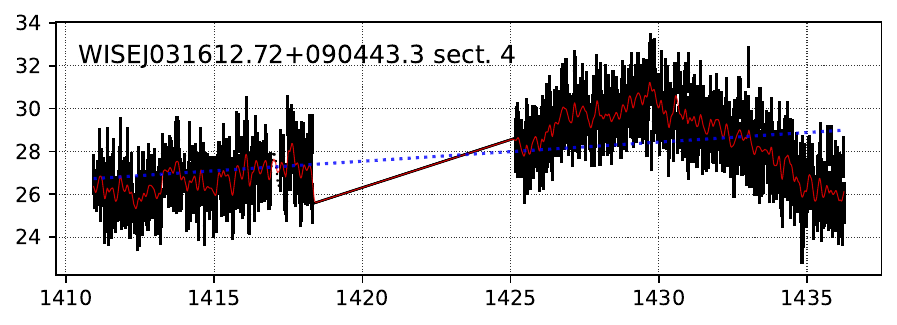}        
        \centering
        \caption{}
        \label{fig:lc_group8}
    \end{figure}
    \begin{figure}[H]
        \includegraphics[]{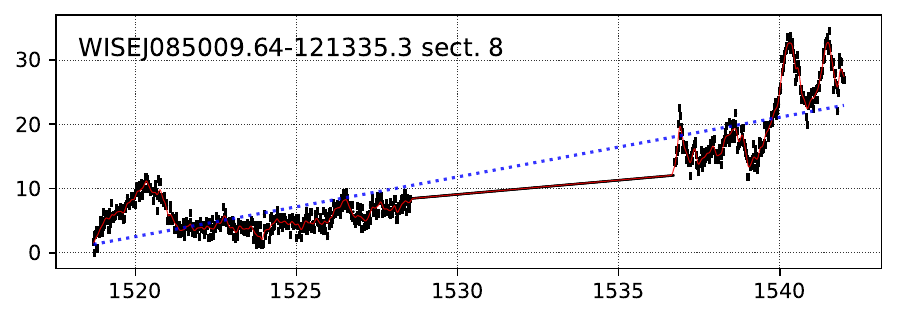}
        \includegraphics[]{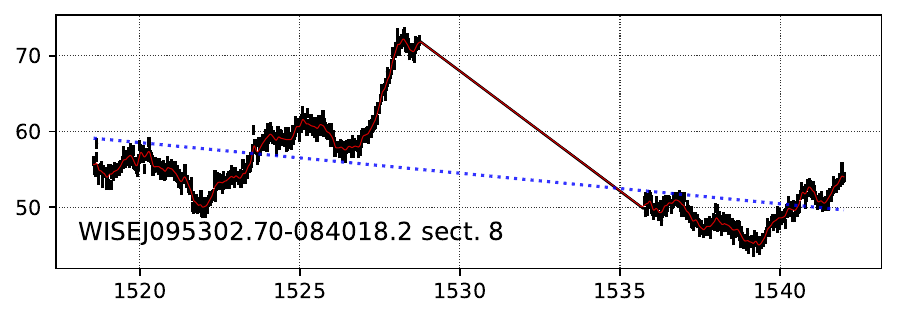}
        \includegraphics[]{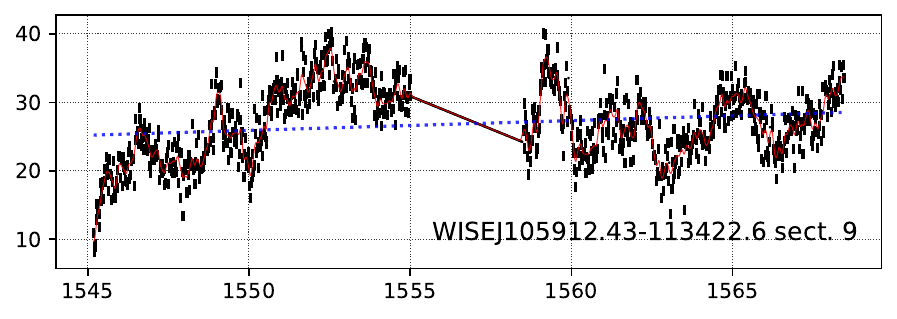}
        \centering
        \caption{}
        \label{fig:lc_group9}
    \end{figure}

    \begin{longdeluxetable}{cccccc}
        \centering
        \tabletypesize{\scriptsize}
        \tablewidth{0pt} 
        \tablenum{7}
        \tablecaption{\label{tab:dates}Start and end dates for each sector and orbit in TESS Cycle 1. These dates are provided in both UTC and TESS Julian Date (TJD) where TJD = JD - 2457000.}
        \tablehead{
        \colhead{Sector} & \colhead{Orbit} & \multicolumn{2}{c}{Start} & \multicolumn{2}{c}{Stop}\\
        \colhead{} & \colhead{} & \colhead{UTC} & \colhead{TJD} & \colhead{UTC} & \colhead{TJD}
        }
        \startdata 
        \multirow{2}{*}{1} & 9 & 2018-07-25 19:00:27 & 1325.29 & 2018-08-08 00:29:51 & 1338.52\\
        ~ & 10 & 2018-08-09 03:39:19 & 1339.65 & 2018-08-22 16:14:51 & 1353.18\\
        \hline
        \multirow{2}{*}{2} & 11 & 2018-08-23 14:24:19 & 1354.10 & 2018-09-05 15:39:51 & 1367.15\\
        ~ & 12 & 2018-09-07 02:14:18 & 1368.59 & 2018-09-20 00:19:52 & 1381.51\\
        \hline
        \multirow{2}{*}{3} & 13 & 2018-09-20 12:56:15 & 1382.04 & 2018-10-03 23:30:02 & 1395.48\\
        ~ & 14 & 2018-10-05 02:30:02 & 1396.60 & 2018-10-17 21:17:58 & 1409.39\\
        \hline
        \multirow{2}{*}{4} & 15 & 2018-10-19 09:34:28 & 1410.90 & 2018-11-01 00:11:40 & 1423.51\\
        ~ & 16 & 2018-11-02 01:09:21 & 1424.55 & 2018-11-14 08:21:39 & 1436.85\\
        \hline
        \multirow{2}{*}{5} & 17 & 2018-11-15 07:47:48 & 1437.83 & 2018-11-27 16:44:46 & 1450.20\\
        ~ & 18 &  2018-11-29 01:09:22 & 1451.55 & 2018-12-11 21:35:39 & 1464.40\\
        \hline
        \multirow{2}{*}{6} & 19 & 2018-12-12 17:05:01 & 1465.21 & 2018-12-12 17:05:01 & 1477.02\\
        ~ & 20 & 2018-12-25 14:41:37 & 1478.11 & 2019-01-06 13:01:37 & 1490.04\\
        \hline
        \multirow{2}{*}{7} & 21 & 2019-01-08 02:59:37 & 1491.63 & 2019-01-19 12:53:30 & 1503.04\\
        ~ & 22 & 2019-01-21 04:43:36 & 1504.70 & 2019-02-01 14:01:30 & 1516.09\\
        \hline
        \multirow{2}{*}{8} & 23 & 2019-02-02 20:09:35 & 1517.34 & 2019-02-14 13:31:35 & 1529.07\\
        ~ & 24 & 2019-02-15 18:09:35 & 1530.26 & 2019-02-27 11:57:34 & 1542.00\\
        \hline
        \multirow{2}{*}{9} & 25 & 2019-02-28 17:09:34 & 1543.22 & 2019-03-13 00:57:34 & 1555.54\\
        ~ & 26 & 2019-03-14 05:19:34 & 1556.72 & 2019-03-25 23:21:33 & 1568.47\\
        \hline
        \multirow{2}{*}{10} & 27 & 2019-03-26 22:19:33 & 1569.43 & 2019-04-08 06:47:33 & 1581.78\\
        ~ & 28 & 2019-04-09 06:15:33 & 1582.76 & 2019-04-22 04:17:32 & 1595.68\\
        \hline
        \multirow{2}{*}{11} & 29 & 2019-04-23 06:29:32 & 1596.77 & 2019-05-06 04:37:32 & 1609.69\\
        ~ & 30 & 2019-05-07 06:35:32 & 1610.78 & 2019-05-20 09:21:31 & 1623.89\\
        \hline
        \multirow{2}{*}{12} & 31 & 2019-05-21 10:45:31 & 1624.95 & 2019-06-04 11:51:31 & 1639.00\\
        ~ & 32 & 2019-06-05 12:45:31 & 1640.03 & 2019-06-18 09:21:30 & 1652.89\\
        \hline
        \multirow{2}{*}{13} & 33 & 2019-06-19 09:55:30 & 1653.92 & 2019-07-03 04:31:30 & 1667.69\\
        ~ & 34 & 2019-07-04 02:49:30 & 1668.62 & 2019-07-17 20:31:29 & 1682.36
        \enddata
    \end{longdeluxetable}

    \section{Discussion of errors and error propagation}
    \label{sec:errors}

    When binning our light curves, we bin to the average flux, therefore the photometric error per bin is provided by the standard error of the flux:
    \begin{equation}
        \label{eqn:binerr}
        \xi_{bin} = \frac{\sigma_{bin}}{\sqrt{n}} = \frac{1}{n}\sqrt{\sum_{i=1}^{n}(F_{i}-\overline{F}_{bin})^{2}}
    \end{equation}
    where n is the number of data points per bin. Binning and error propagation is only be performed on uninterpolated data. To maintain the proper binning schema and uneven time separation caused by cadence masking, n, from equation \ref{eqn:binerr}, will vary between bins and may have less than the expected number if the sample were evenly spaced resulting in larger errors.\\
    
    A definition for the error on the normalized excess variance has been provided by previous articles (see Equation 11 of \cite{Vaughan2003}). From this definition and the general form for the propagation of error, wherein $\sigma_{NXS}^{2} = \sfrac{\sigma_{XS}^{2}}{\overline{F}^2}$ and $\overline{F}^{2}$ is a normalization constant, we find
    \begin{equation}
        \label{eqn:sigmaNXSerr}
        err(\sigma_{NXS}^{2}) = \sqrt{\bigg{\{}\Delta q_{1}\frac{\partial\sigma_{NXS}^{2}}{\partial q_{1}}\bigg{\}}^{2} + \cdot\cdot\cdot+ \bigg{\{}\Delta q_{N} \frac{\partial\sigma_{NXS}^{2}}{\partial q_{N}} \bigg{\}}^{2} } = \frac{err(\sigma_{XS}^{2})}{\overline{F}^2}
    \end{equation}
    Thereby, from equation \ref{eqn:sigmaNXSerr}, the error on the excess variance is given by
    \begin{equation}
        \label{eqn:sigmaXSerr}
        err(\sigma_{XS}^{2}) = \begin{cases}\sqrt{\bigg{\{}\sqrt{\frac{2}{N}}\overline{\xi^{2}}\bigg{\}}^{2} + \bigg{\{}2\sqrt{\frac{\overline{\xi^{2}}\sigma_{XS}^{2}}{N}}\bigg{\}}^{2}};~~~~~~~~\sigma_{XS}^{2} > 0\\
        ~~~~~~~~~~~~~~~~~~~0~~~~~~~~~~~~~~~~~~~;~~~~~~\text{ otherwise}
        \end{cases}
    \end{equation}
    The error on the rms-scatter is also given in the same manner as in Appendix B of \cite{Vaughan2003} such that
    \begin{equation}
        \label{eqn:sigmarmserr}
        err(\sigma_{rms}) = \frac{err(\sigma_{XS}^{2})}{2 \sigma_{rms}} = \begin{cases}
        \sqrt{\bigg{\{}\sqrt{\frac{1}{2N}}\cdot\frac{\overline{\xi^{2}}}{\sigma_{rms}}\bigg{\}}^{2} + \bigg{\{}\sqrt{\frac{\overline{\xi^{2}}}{N}}\bigg{\}}^{2}};~~~~~~~~\sigma_{rms} > 0\\
        ~~~~~~~~~~~~~~~~~~~0~~~~~~~~~~~~~~~~~~~;~~~~~~\text{ otherwise}
        \end{cases}
    \end{equation}
    If no significant variability is observed, due to small amplitude or poor signal to noise, then $\sigma_{NXS}$ approaches zero. In such cases, the left term  of equation \ref{eqn:sigmaXSerr} dominates and vice versa in cases when the sample variance greatly exceeds the mean square error. This also applies to the rms-scatter in equation \ref{eqn:sigmarmserr}) \citep{Vaughan2003}.\\

    Error for the shortest timescales of exponential rise/decay is provided by:
    \begin{equation}
        \label{eqn:tauerr}
        err(\tau_{d}) = \frac{\ln{2}\cdot{\Delta t}}{\ln{\big{(}\frac{F_{2}}{F_{1}}}\big{)}^{2}}\cdot\bigg{[}\bigg{(}\frac{\xi_{1}}{F_{1}}\bigg{)}^{2} + \bigg{(}\frac{\xi_{2}}{F_{2}}\bigg{)}^{2}\bigg{]} = \frac{\tau_{d}}{\ln{\big{(}\frac{F_{2}}{F_{1}}}\big{)}}\cdot\bigg{[}\bigg{(}\frac{\xi_{1}}{F_{1}}\bigg{)}^{2} + \bigg{(}\frac{\xi_{2}}{F_{2}}\bigg{)}^{2}\bigg{]}
    \end{equation}
    in which the error shows dependency on both the flux and its associated photometric error. It is from this error that we provide our range of upper-limit estimations for the emission region radii.\\

    \section{Shortest timescales of variability}
\label{sec:timescales}

    Shown in the section by Figures \ref{fig:timescales_PKS0422} -- \ref{fig:timescales_NGC1218} are examples of our light curve data selection as used to calculate the shortest observed timescales of variability (Equation \ref{eqn:short_timescale}) in the rest frame of the target as discussed in Section \ref{sec:shortest_timescales}. Shown are the full, uncorrected (black) and Gaussianly smoothed (light blue) light curves (TOP) with zoomed-in segments of the selected time intervals (BOTTOM) in which a significant change in flux ($|\Delta F| > 3\sigma$) provides the shortest timescale of variability. In each case, $F_{i}$ and $F_{j}$ (the prior and latter flux values, respectively) are indicated by vertical lines at their associated times and $\Delta t$ is represented by the difference in these marked times. Gaussianly smoothed light curves are either smoothed over 1-hour ($\sigma = 2$) or 30-minute ($\sigma = 1$) kernels and are used only to identify a significant change in flux while mitigating the selection of single cadence flares. Calculation of $\tau_{d}$ is done using the real flux values at these times. If no significant change is found in the smoothed light curve, the significant change is identified by the non-smoothed light curve. Figures \ref{fig:timescales_PKS0422} and \ref{fig:timescales_WISEJ03} demonstrate examples of the Gaussian smoothing procedure in which variability has been selected by smoothing over a 1-hour kernel. Figure \ref{fig:timescales_NGC1218} provides an example for one of the two light curves in which variability is lost by smoothing and selection is made by the real flux values.
    
    \begin{figure}[H]
        \centering
        \begin{tabular}{cc}
             \includegraphics[width=1\textwidth]{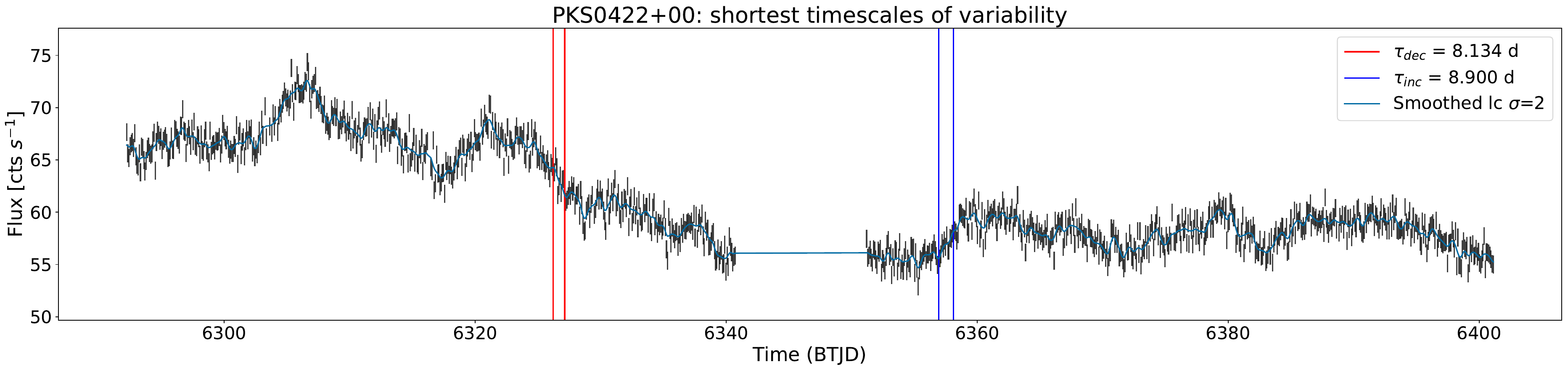}\\
             \includegraphics[width=.5\textwidth]{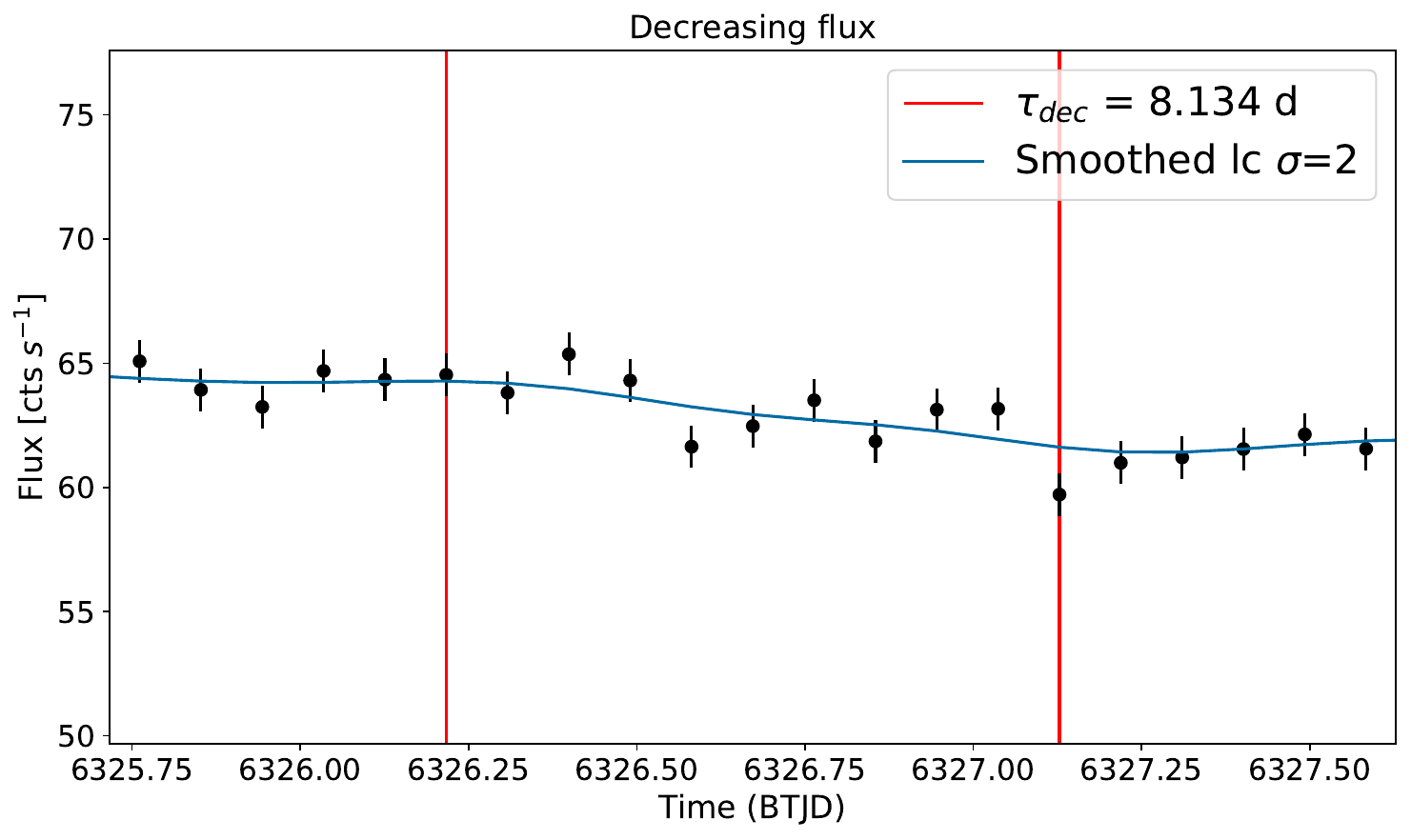}
             \includegraphics[width=.5\textwidth]{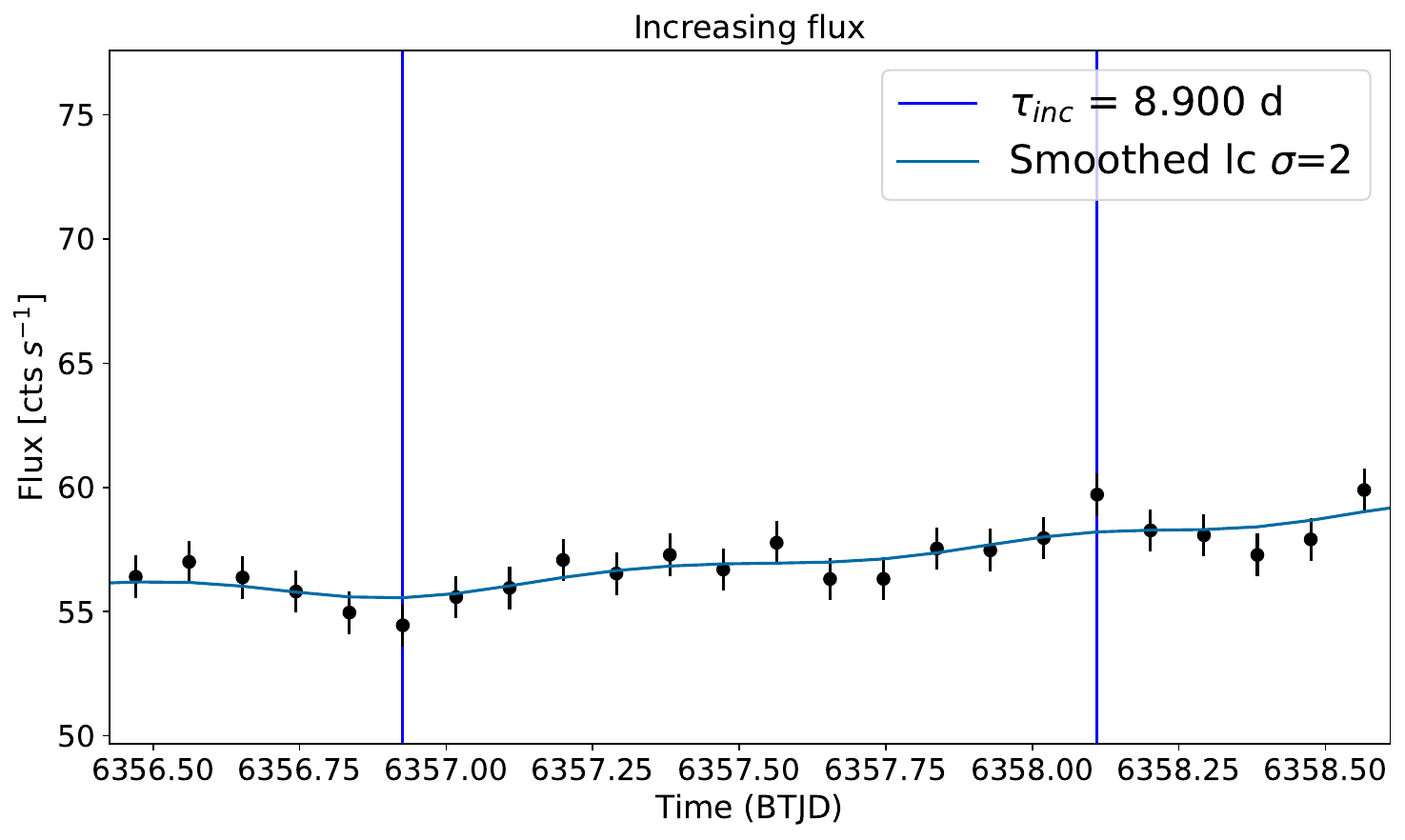}
        \end{tabular}
            \caption{Light curve for for BLL PKS~0422+00 showing the shortest timescales of variability for significant flux change as calculated for increasing (BLUE) and decreasing (RED) flux.}
            \label{fig:timescales_PKS0422}
    \end{figure}
    \begin{figure}[H]
        \centering
        \begin{tabular}{cc}
             \includegraphics[width=1\textwidth]{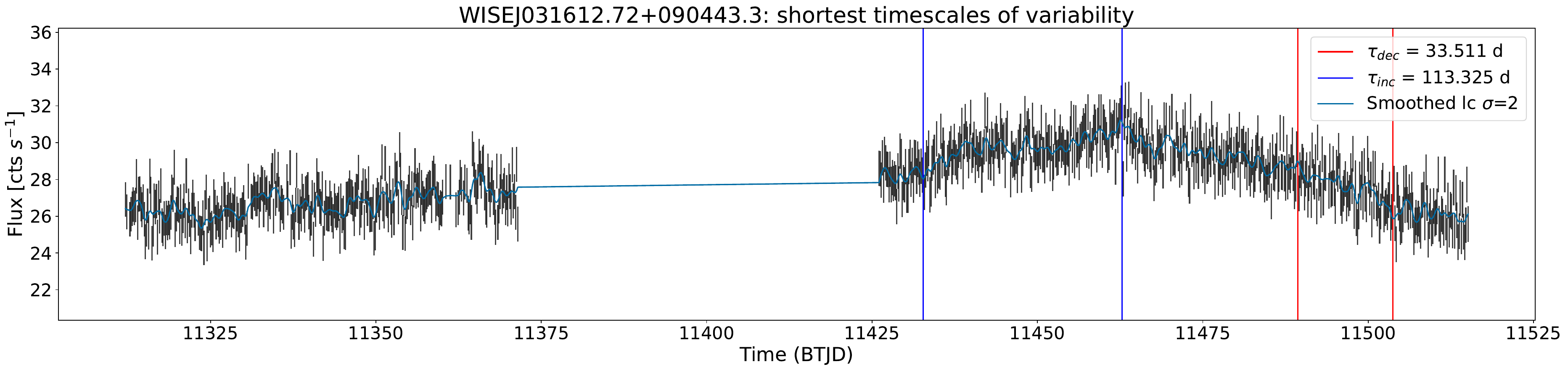}\\
             \includegraphics[width=.5\textwidth]{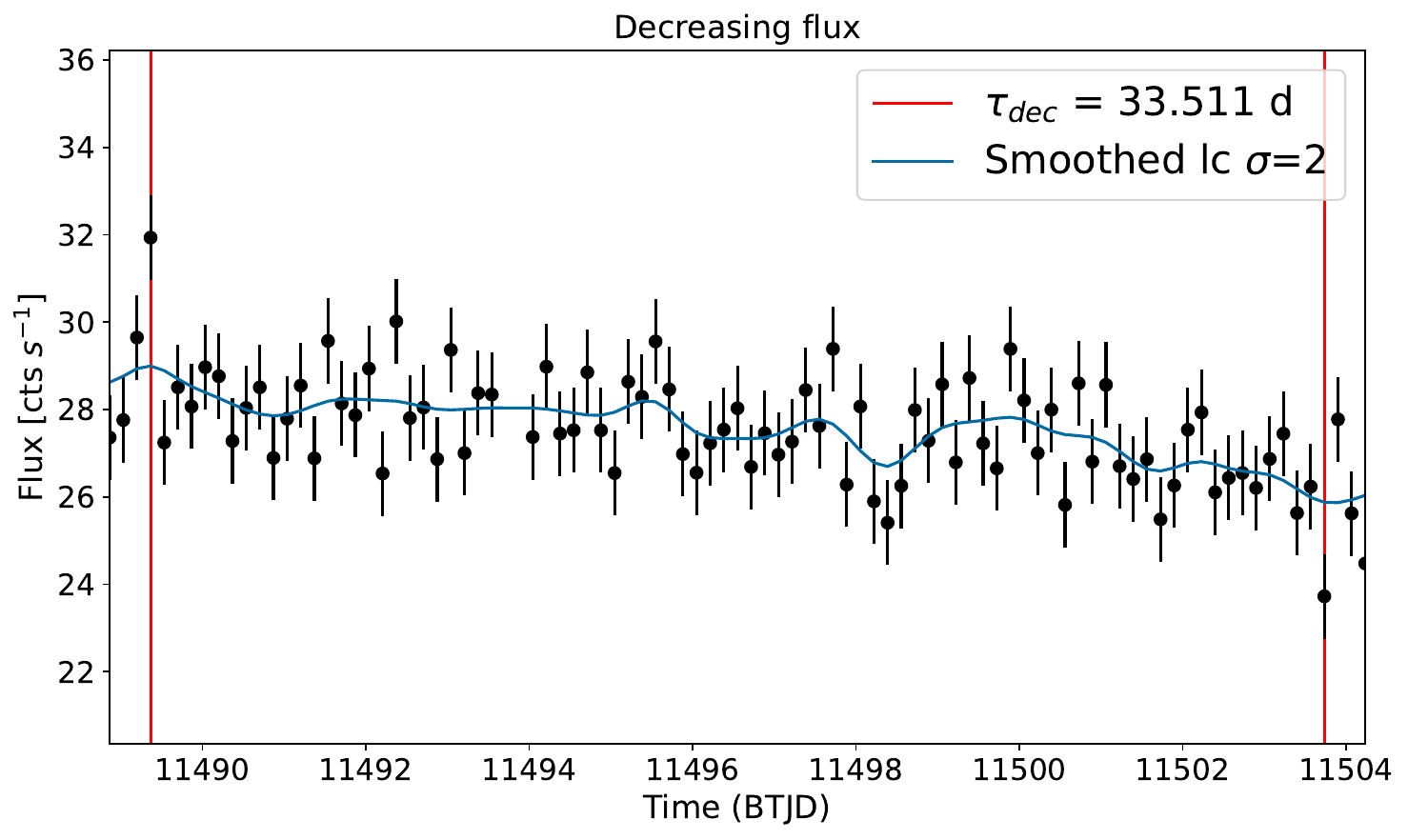}
             \includegraphics[width=.5\textwidth]{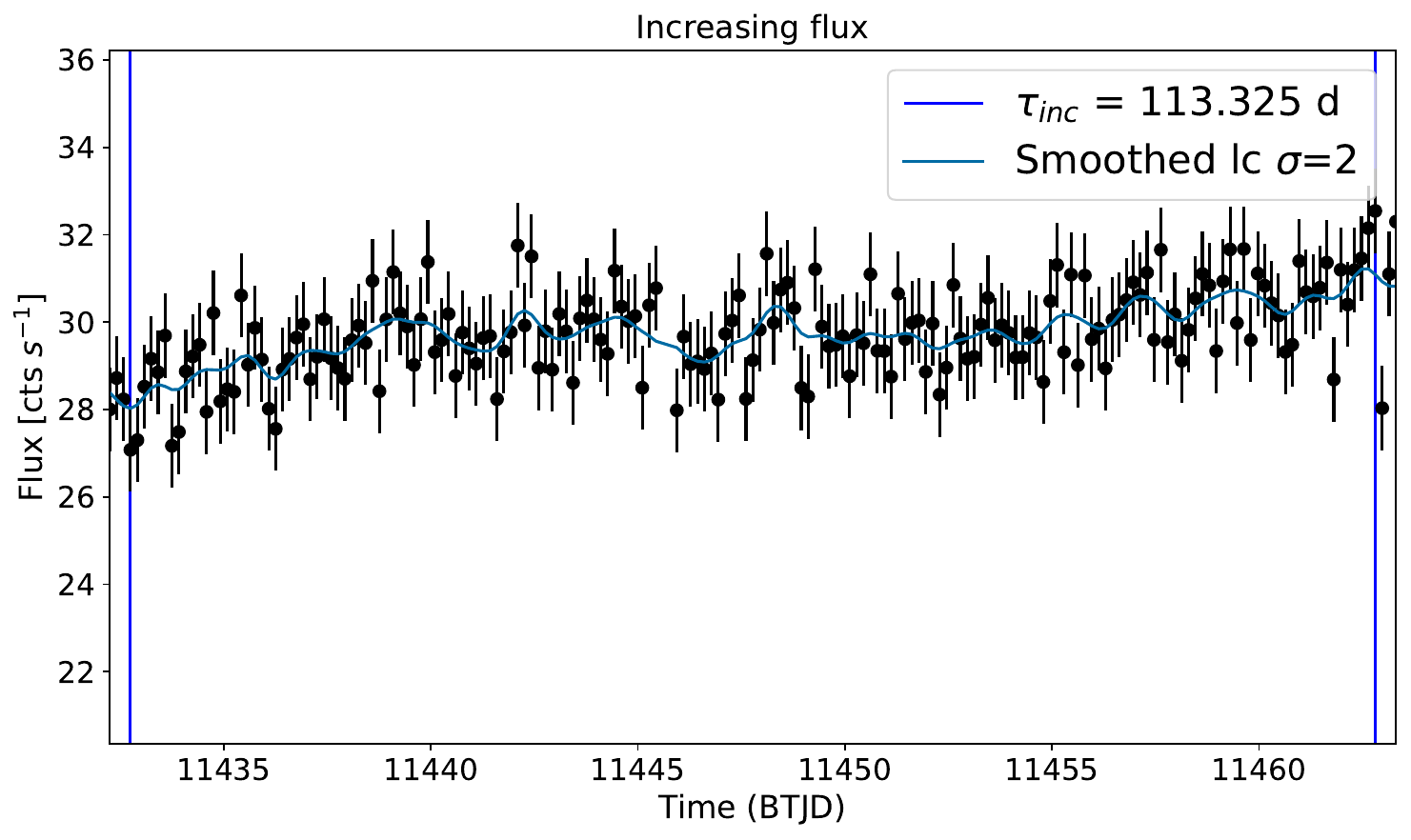} 
        \end{tabular}
            \caption{Light curve for BLL PKS~2155-304 showing the shortest timescales of variability for significant flux change as calculated for increasing (BLUE) and decreasing (RED) flux.}
            \label{fig:timescales_WISEJ03}
    \end{figure}
    \begin{figure}[H]
        \centering
        \begin{tabular}{cc}
             \includegraphics[width=1\textwidth]{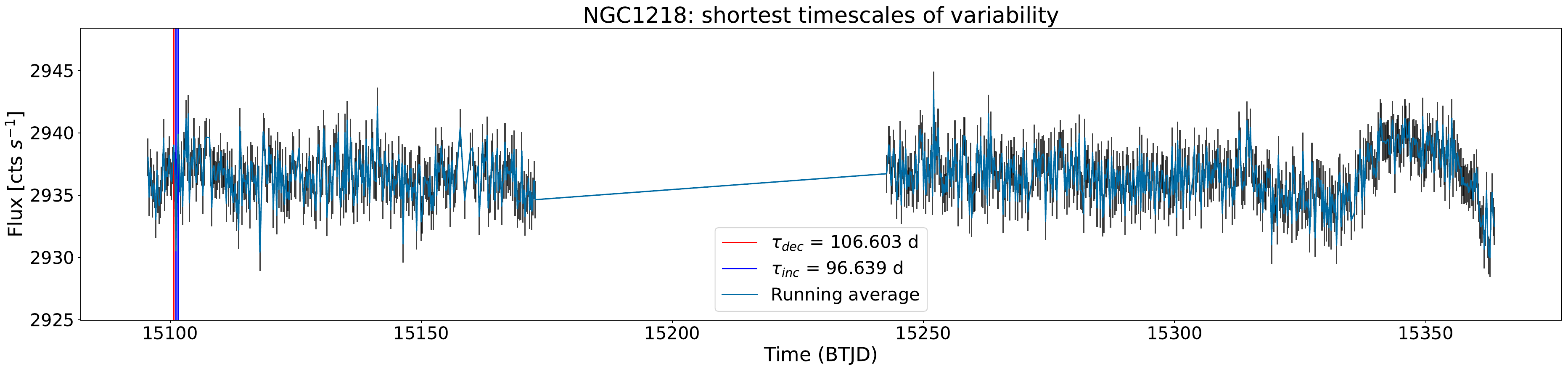}\\
             \includegraphics[width=.5\textwidth]{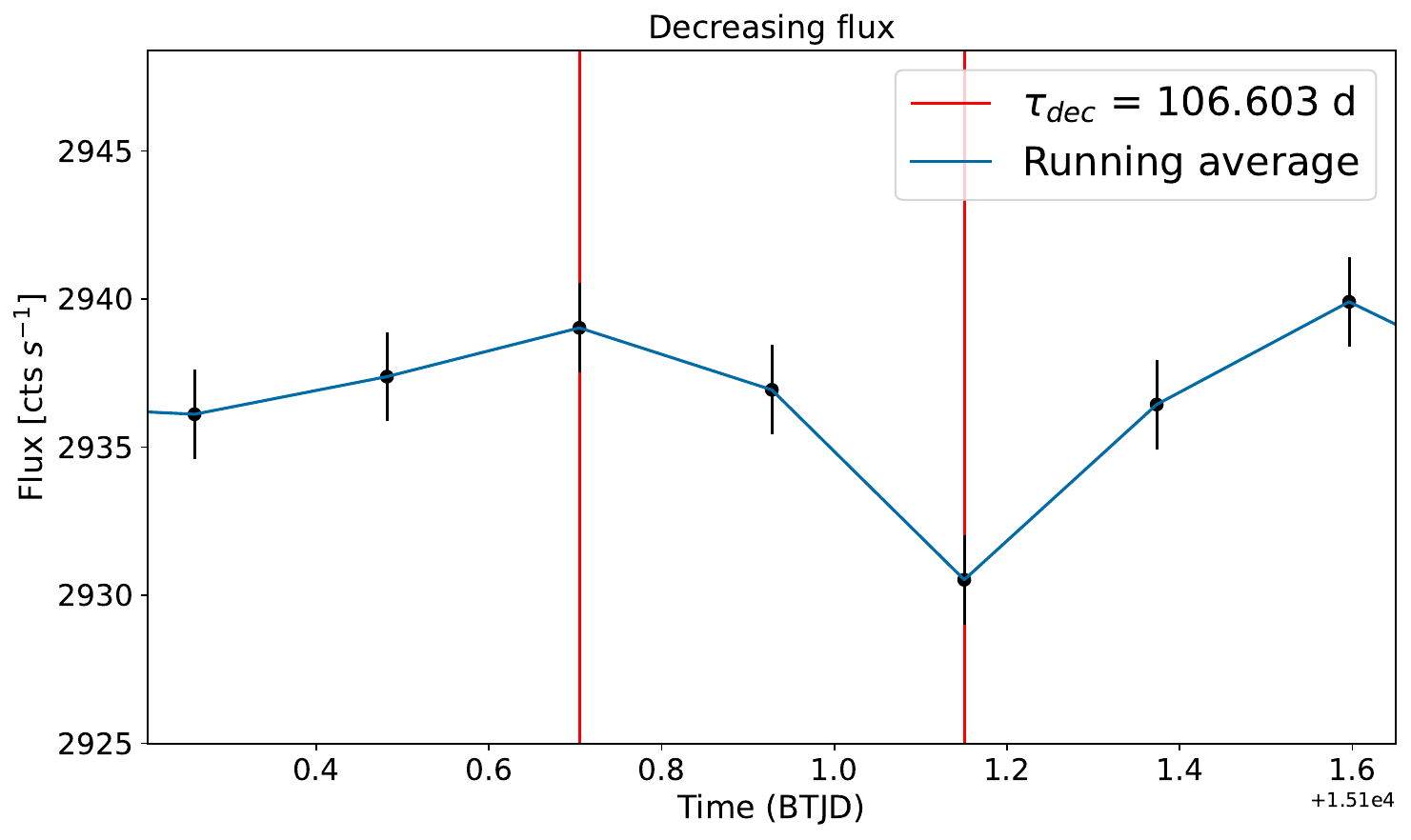}
             \includegraphics[width=.5\textwidth]{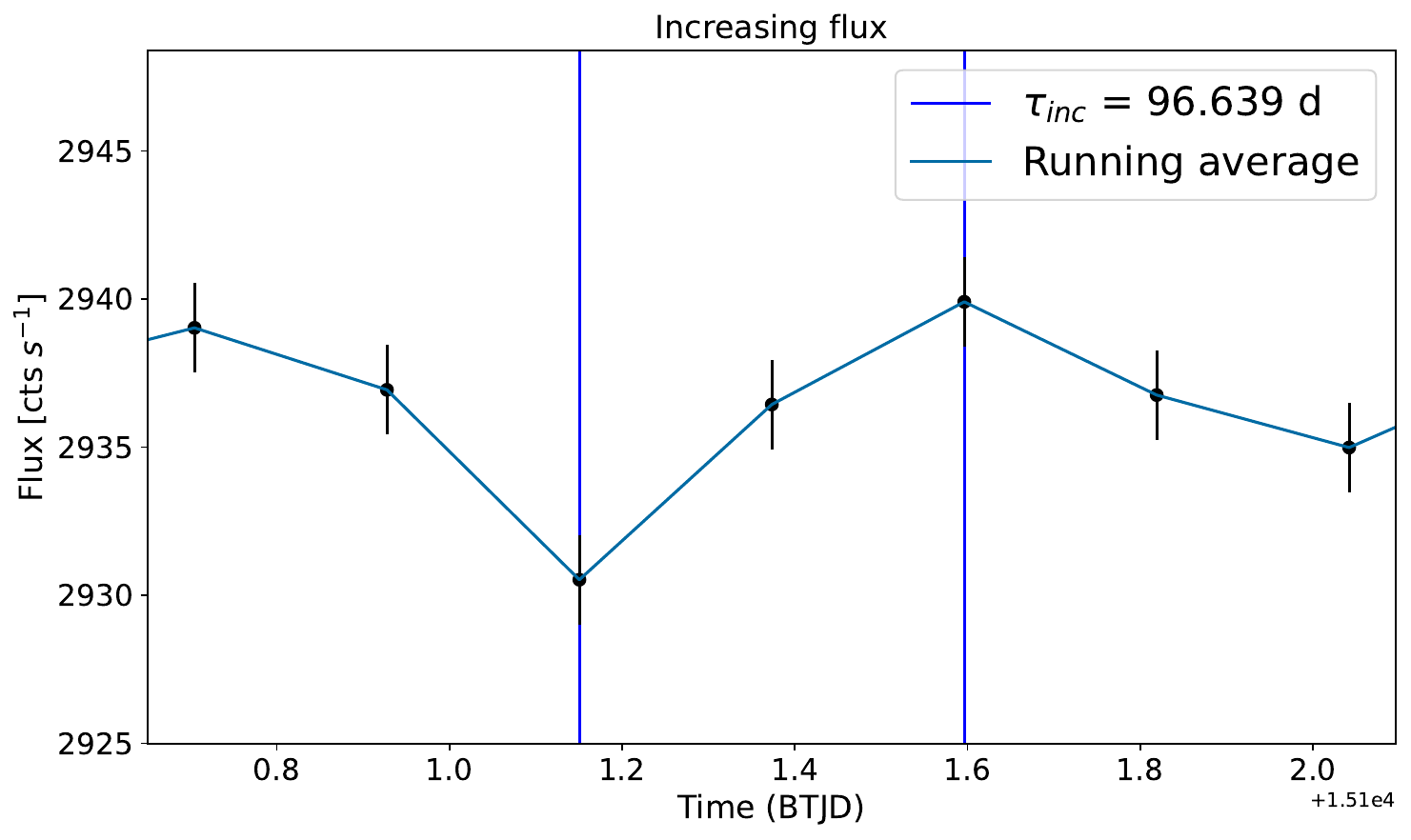} 
        \end{tabular}
            \caption{Light curve for BLL NGC1218 showing the shortest timescales of variability for significant flux change as calculated for increasing (BLUE) and decreasing (RED) flux.}
            \label{fig:timescales_NGC1218}
    \end{figure}

    \section{Consistency with CARMA analysis}
    \label{sec:CARMA}

    We present here the full sample of light curves (BLACK) after regression by the \texttt{quaver} pipeline, after correction in comparison with light curves replicated by best fit damped random walk (DRW, CARMA(1,0), orange) an damped harmonic oscillator (DHO, CARMA(2,1), green) autoregressive models. Excluding one source, PKS0235-618 which demonstrated variability on the timescale of the full observation, the decorrelation timescales observed in this sample from DRW models [break frequencies can be found by dividing by 2$\pi$ \citep{Kelly2014}] are generally consistent with our other results showing variability on times between 0.1 and 10.7 days, see Figure \ref{fig:CARMAhist}.  We find that 13 light curves are best fit by a DRW model while 22 are better fit by a DHO model. However, while a DHO model better describes the latter light curves, the $\chi^{2}$ value is only improved by at most by 45\% to which the magnitude difference never exceeds 10. For example, Figure \ref{fig:CARMAlcs+psd1}, shows that 1ES2322+409 is better modeled by a DHO, but this shows little improvement over a DRW model. As such, we state that all objects in our sample can still be well described by DRW model.  Out of all DRW model fits, the identified timescale is less than a week for all but 4 light curves and intraday for 17 light curves. Provided are relative $\chi^{2}$ statistics such to established which CARMA model best fits the time series. On the right, are the corresponding PSDs generated by these CARMA models normalized and compared to the Fourier periodogram. A collection of all figures are made publicly available\footnote{DOI:\dataset[10.5281/zenodo.11490675]{ https://doi.org/10.5281/zenodo.11490675}}. On the left, we provide the \texttt{quaver} regressed light curve (black) for blazar 1ES2322-409 along with the replicated predictive light curves generated from the best-fit DRW (orange) and DHO (green) autoregressive models. 

    \begin{figure}[H]
        \centering
        \includegraphics[width=0.5\textwidth]{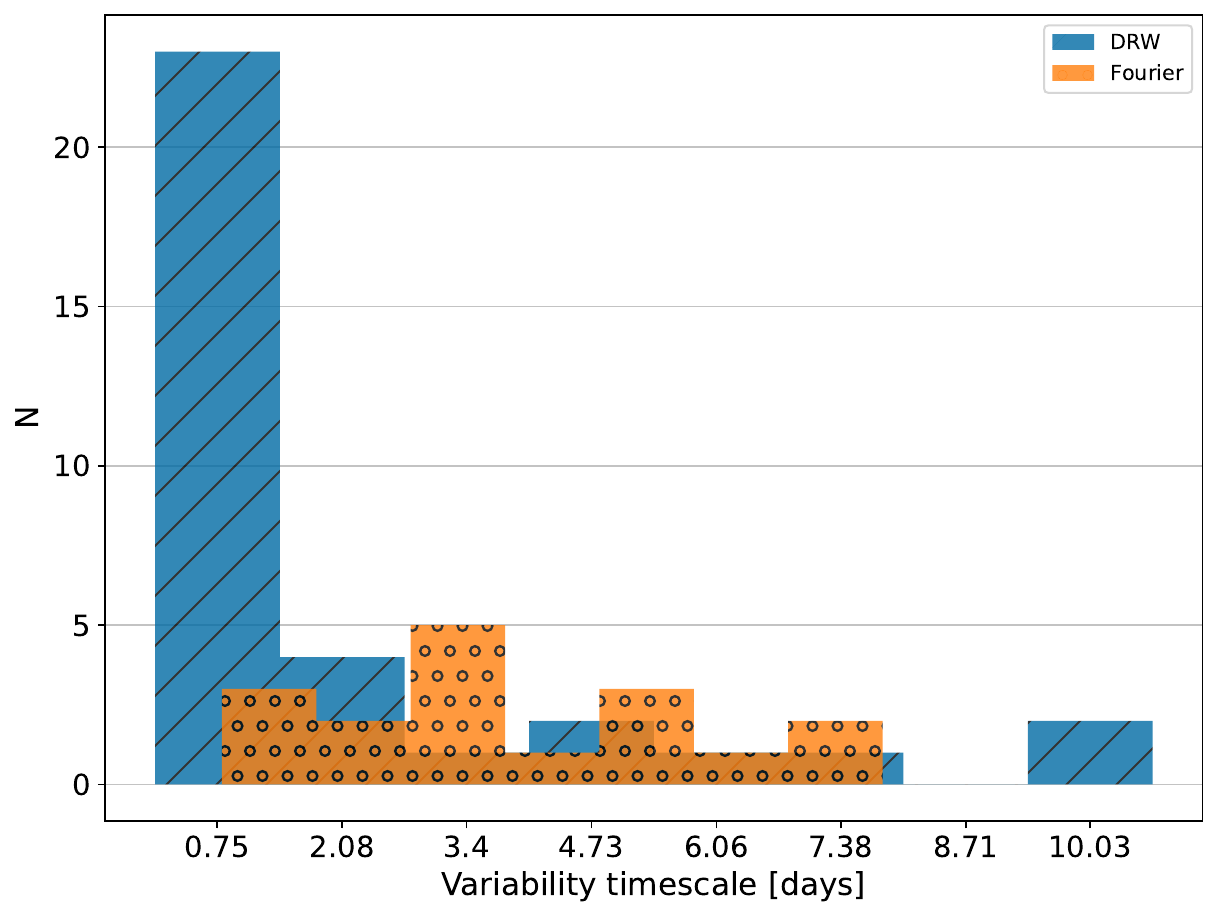}
        \caption{CARMA timescales}
        \label{fig:CARMAhist}
    \end{figure}

    \begin{deluxetable}{lccccccccc}
    \centering
    \tabletypesize{\scriptsize}
    \tablewidth{0pt}
    \tablenum{8}
    \tablecaption{CARMA modeling\label{tab:carma_results}}
    \tablehead{
    \colhead{Target} & \multicolumn{3}{c}{DRW}  & \multicolumn{5}{c}{DHO} & \colhead{best fit}\\
    \colhead{~} & \colhead{Amp.} & \colhead{$\tau$ [days]} & \colhead{$\chi^{2}$} & \colhead{$\alpha_{1}$} & \colhead{$\alpha_{2}$} & \colhead{$\beta_{0}$} & \colhead{$\beta_{1}$} & \colhead{$\chi^{2}$} & \colhead{~}}
    \startdata
    1ES2322-409 & 10.60 & 0.68 & 1.42 & 711.25 & 6.26$\times10^{-4}$ & 5.01$\times10^{3}$ & 63.64 & 0.78 & DHO\\
    1RXSJ054357.3-553206 & 7.60 & 5.36 & 61.03 & 1.22 & 0.28 & 5.90 & 4.55 & 53.36 & DHO\\
    3C120 & 10.11 & 0.59 & 0.54 & 2.93 & 1.13 & 16.75 & 18.71 & 0.53 & DHO\\
    NGC1218 & 7.29 & 5.31 & 1.27 & 0.60 & 0.20 & 3.36 & 4.06 & 0.86 & DHO\\
    PKS0035-252 & 29.89 & 1.11 & 0.26 & 22.92 & 50.13 & 1.47$\times10^{3}$ & 32.95 & 0.44 & DRW\\
    PKS0130-17 & 7.21 & 0.09 & 0.76 & 11.05 & 0.37 & 5.65$\times10^{-4}$ & 33.90 & 0.57 & DHO\\
    PKS0208-512 & 25.04 & 6.07 & 4.77 & 0.18 & 0.15 & 1.93 & 14.28 & 4.43 & DHO\\
    PKS0226-559 & 7.85 & 0.50 & 19.87 & 2.05 & 0.90 & 2.17$\times10^{-5}$ & 15.69 & 14.19 & DHO\\
    PKS0235-618 & 1.69 & 27.59 & 29.34 & 0.04 & 1.41$\times10^{-2}$ & 7.97$\times10^{-5}$ & 0.42 & 29.51 & DRW\\
    PKS0301-243 & 2.80 & 1.70 & 4.53 & 1.12$\times10^{3}$ & 526.32 & 3023.02 & 18.58 & 4.38 & DHO\\
    PKS0336-177 & 1.74 & 2.15 & 2.44 & 0.50 & 0.23 & 9.70$\times10^{-5}$ & 1.64 & 2.33 & DHO\\
    PKS0346-27 & 8.97 & 1.31 & 6.44 & 0.70 & 2.43 & 6.84$\times10^{-5}$ & 10.74 & 6.81 & DRW\\
    PKS0420-01 & 0.86 & 2.81 & 208.77 & 0.29 & 0.45 & 7.82$\times10^{-5}$ & 0.63 & 211.43 & DRW\\
    PKS0422+00 & 2.87 & 2.51 & 12.19 & 0.47 & 0.14 & 0.43 & 2.51 & 10.65 & DHO\\
    PKS0426-380 & 1.60 & 1.40 & 37.79 & 5.16 & 5.49 & 11.43 & 1.65 & 39.30 & DRW\\
    PKS0454-234 & 1.25 & 1.50 & 14.50 & 0.61 & 1.31 & 0.74 & 1.29 & 13.02 & DHO\\
    PKS0521-36 & 16.95 & 0.89 & 0.75 & 1.16 & 3.78$\times10^{-7}$ & 2.84$\times10^{-3}$ & 25.43 & 0.75 & DHO\\
    PKS0637-75 & 0.24 & 0.60 & 2.30 & 2.97 & 0.45 & 0.23 & 0.49 & 2.29 & DHO\\
    PKS0637-75 & 1.76 & 0.78 & 3.14 & 787.09 & 560.62 & 1.60$\times10^{3}$ & 15.95 & 2.73 & DHO\\
    PKS0637-75 & 0.51 & 0.70 & 2.21 & 2.85 & 5.77 & 2.34 & 0.71 & 2.27 & DRW\\
    PKS0736+01 & 25.80 & 0.56 & 0.21 & 30.92 & 133.23 & 2.34$\times10^{3}$ & 35.17 & 0.32 & DRW\\
    PKS0829+046 & 13.75 & 7.34 & 1.88 & 48.72 & 10.94 & 454.03 & 2.82 & 3.37 & DRW\\
    PKS1244-255 & 3.71 & 1.39 & 13.13 & 1.07 & 2.33 & 5.24 & 4.24 & 15.46 & DRW\\
    PKS2155-304 & 16.38 & 0.26 & 0.27 & 981.32 & 440.85 & 1.58$\times10^{4}$ & 179.49 & 0.20 & DHO\\
    PKS2155-83 & 5.79 & 0.64 & 79.59 & 292.61 & 1.02$\times10^{-3}$ & 2.05$\times10^{3}$ & 21.22 & 69.32 & DHO\\
    PKS2255-282 & 5.27 & 0.10 & 3.40 & 782.92 & 909.92 & 5.70$\times10^{3}$ & 69.34 & 2.42 & DHO\\
    PKS2326-502 & 1.06 & 1.40 & 374.19 & 0.71 & 0.16 & 1.35$\times10^{-4}$ & 1.24 & 374.92 & DRW\\
    PKS2345-16 & 4.13 & 0.17 & 18.82 & 776.10 & 654.81 & 4.19$\times10^{3}$ & 49.18 & 10.85 & DHO\\
    PMNJ0948+0022 & 8.49 & 0.38 & 3.00 & 20.62 & 28.49 & 268.76 & 20.06 & 2.99 & DHO\\
    PMNJ2141-6411 & 1.16 & 1.32 & 182.69 & 0.63 & 1.24 & 0.00 & 1.35 & 184.13 & DRW\\
    PMNJ2345-1555 & 4.53 & 0.28 & 14.55 & 3.61 & 0.29 & 0.00 & 12.16 & 13.40 & DHO\\
    WISEJ031612.72+090443.3 & 1.55 & 9.78 & 23.99 & 0.08 & 0.28 & 0.14 & 0.49 & 24.62 & DRW\\
    WISEJ085009.64-121335.3 & 4.14 & 0.53 & 15.41 & 1.90 & 0.44 & 1.18$\times10^{-2}$ & 8.06 & 13.12 & DHO\\
    WISEJ095302.70-084018.2 & 6.35 & 10.70 & 4.06 & 4.27 & 1.37 & 24.20 & 1.00 & 5.42 & DRW\\
    WISEJ105912.43-113422.6 & 5.31 & 0.13 & 2.75 & 137.12 & 82.83 & 807.49 & 35.52 & 1.97 & DHO\\
    \enddata
\end{deluxetable}

    \begin{figure}
        \centering
        \subfloat{
         \includegraphics[width=0.6\textwidth]{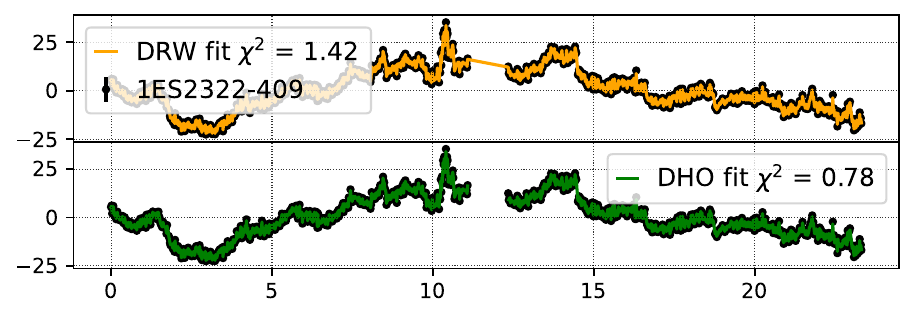}}
        \subfloat{
         \includegraphics[width=0.36\textwidth]{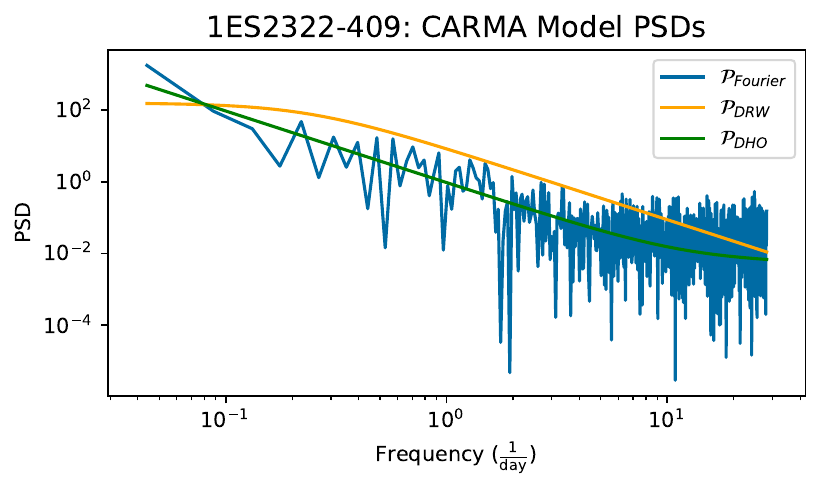}}
         \newline
         \subfloat{
         \includegraphics[width=0.6\textwidth]{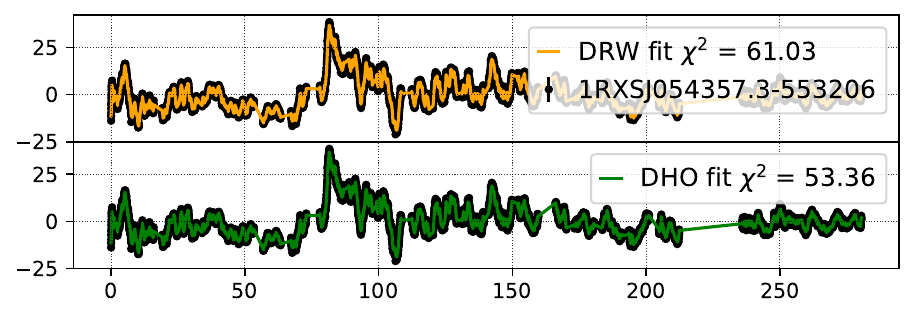}}
        \subfloat{
         \includegraphics[width=0.36\textwidth]{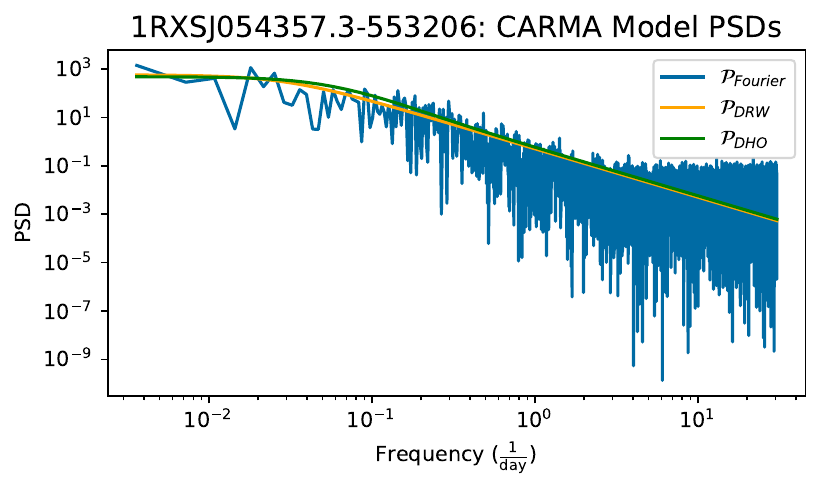}}
         \newline
         \subfloat{
         \includegraphics[width=0.6\textwidth]{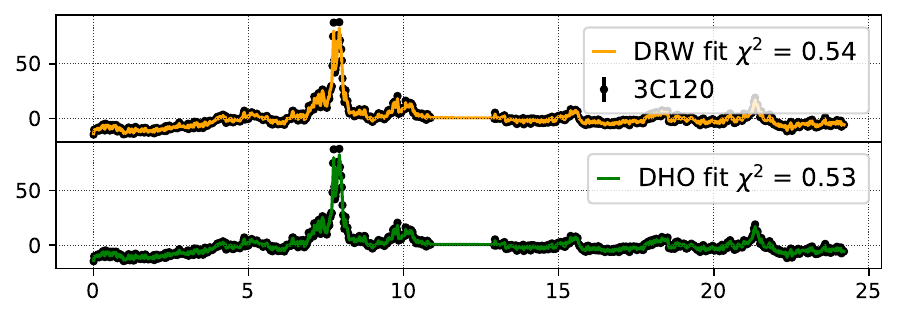}}
        \subfloat{
         \includegraphics[width=0.36\textwidth]{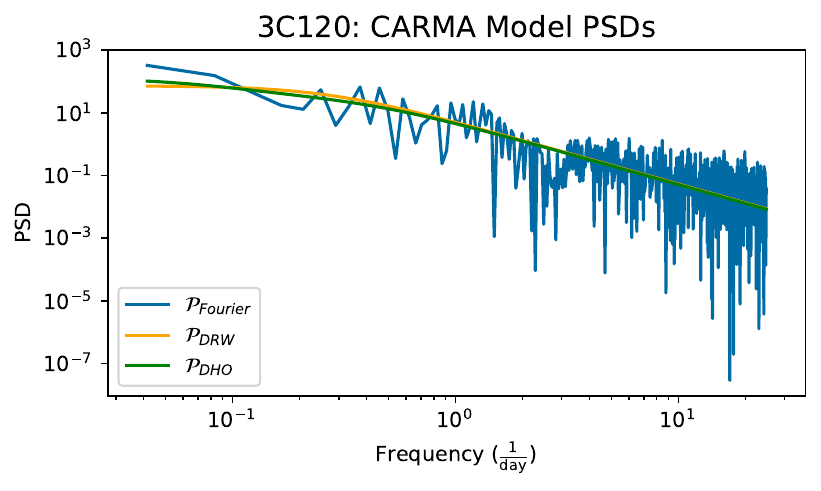}}
         \newline
         \subfloat{
         \includegraphics[width=0.6\textwidth]{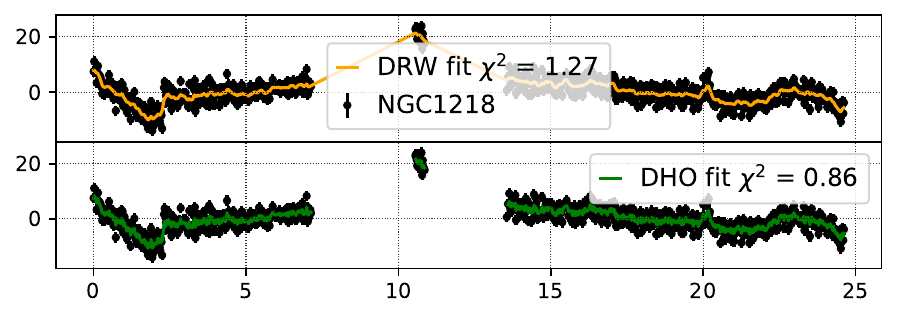}}
        \subfloat{
         \includegraphics[width=0.36\textwidth]{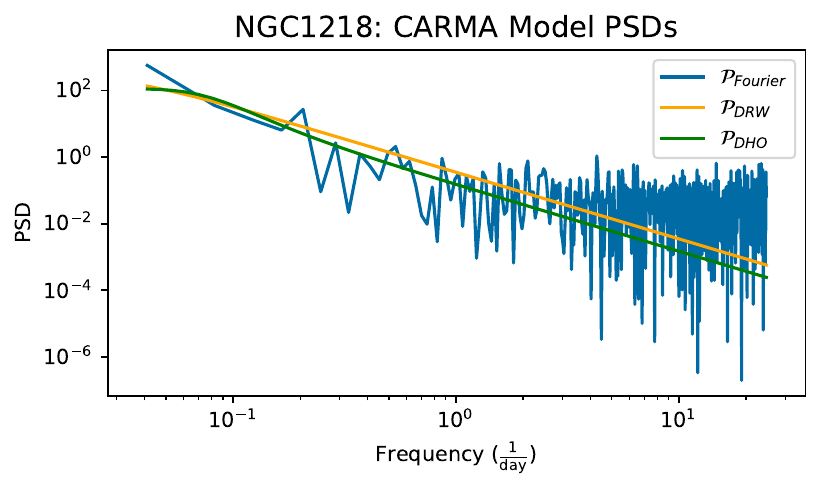}}
         \newline\subfloat{
         \includegraphics[width=0.6\textwidth]{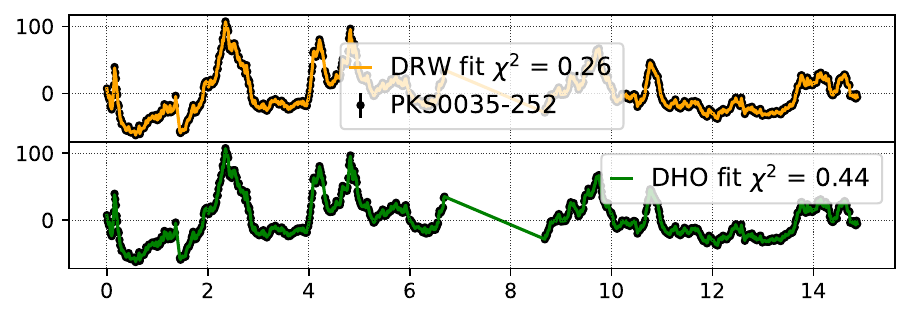}}
        \subfloat{
         \includegraphics[width=0.36\textwidth]{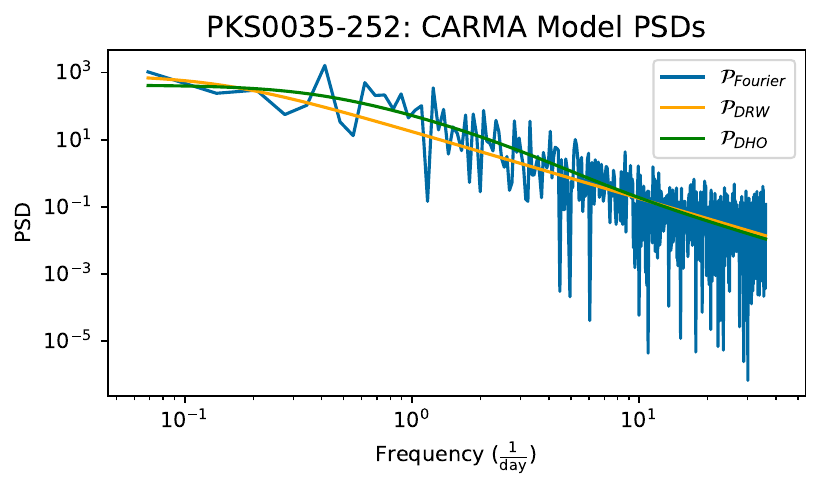}}
    \caption{}
    \label{fig:CARMAlcs+psd1}
    \end{figure}
    




    \begin{figure}
        \label{fig:CARMAlcs+psd2}
        \centering
        \subfloat{
         \includegraphics[width=0.6\textwidth]{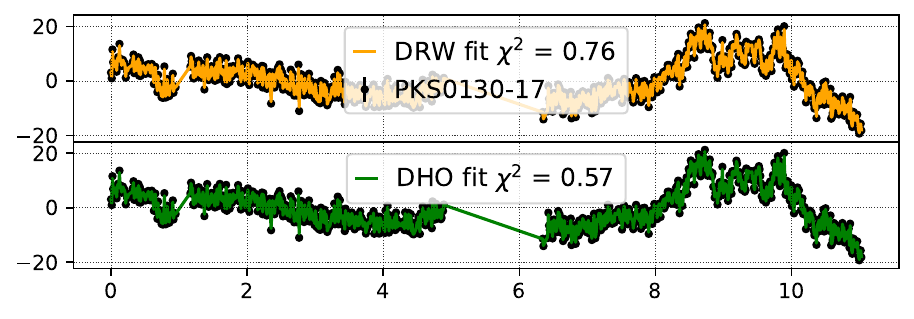}}
        \subfloat{
         \includegraphics[width=0.36\textwidth]{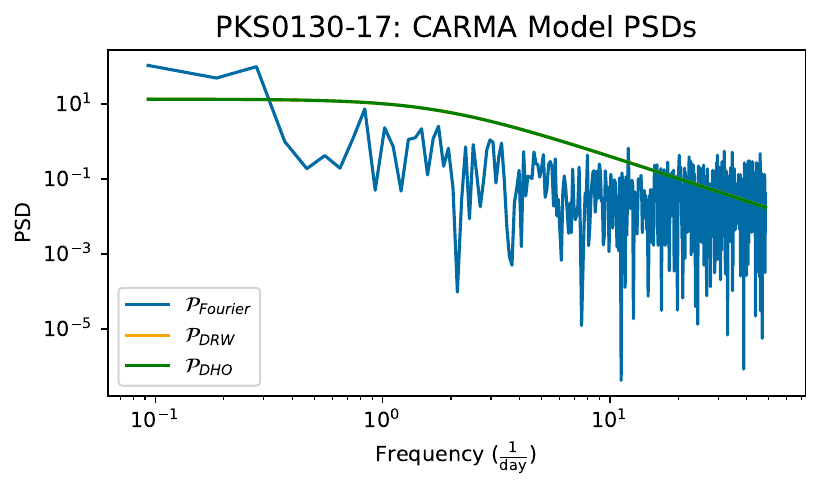}}
         \newline
         \subfloat{
         \includegraphics[width=0.6\textwidth]{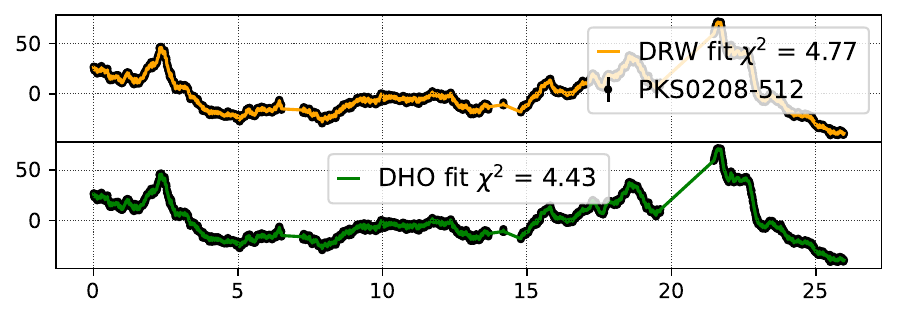}}
        \subfloat{
         \includegraphics[width=0.36\textwidth]{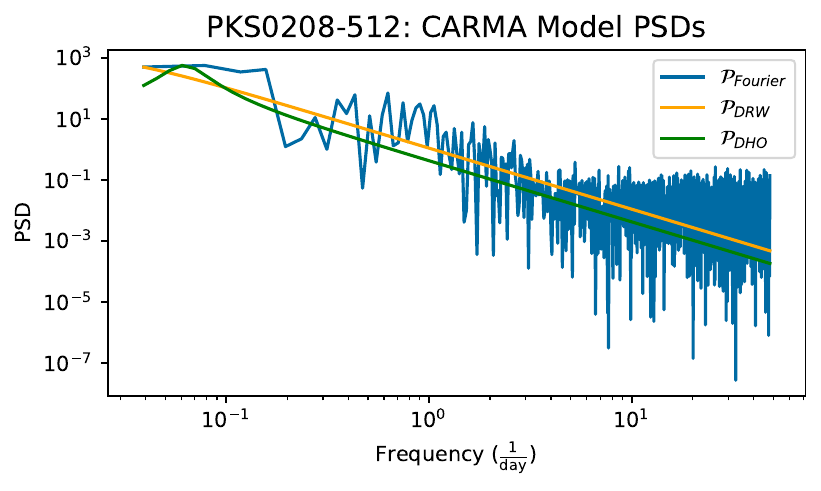}}
         \newline
        \subfloat{
         \includegraphics[width=0.6\textwidth]{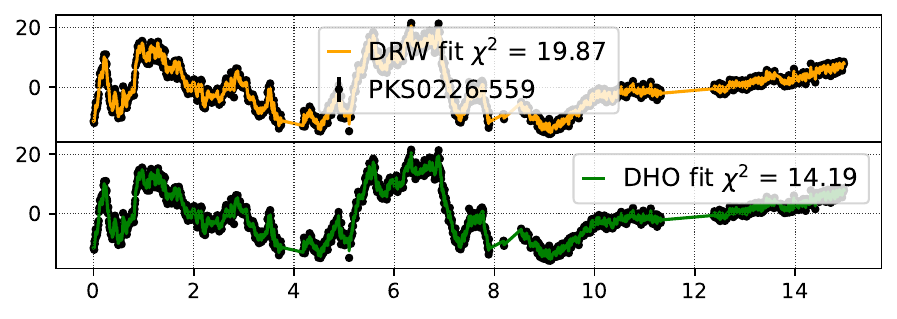}}
        \subfloat{
         \includegraphics[width=0.36\textwidth]{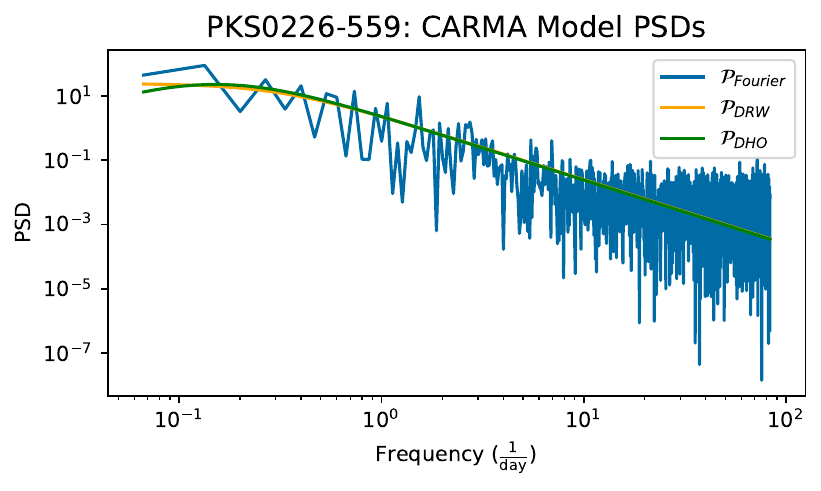}}
         \newline
        \subfloat{
         \includegraphics[width=0.6\textwidth]{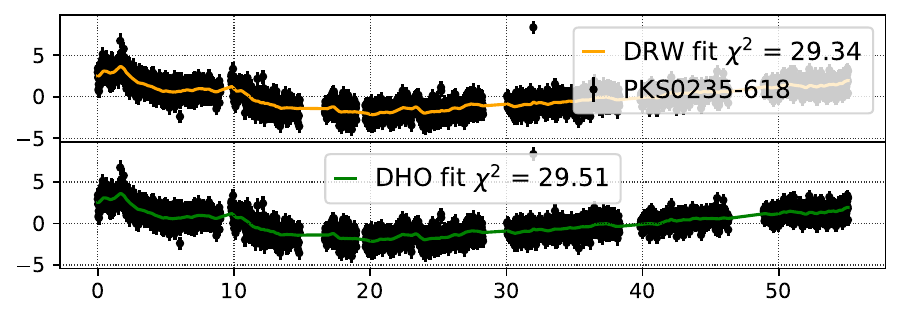}}
        \subfloat{
         \includegraphics[width=0.36\textwidth]{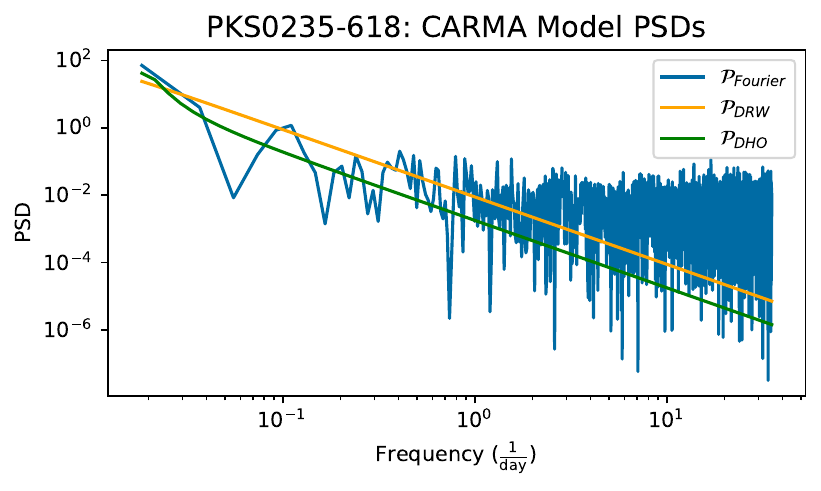}}
         \newline
        \subfloat{
         \includegraphics[width=0.6\textwidth]{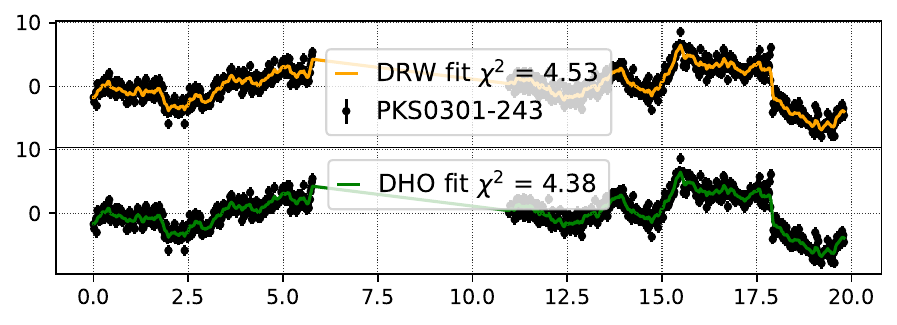}}
        \subfloat{
         \includegraphics[width=0.36\textwidth]{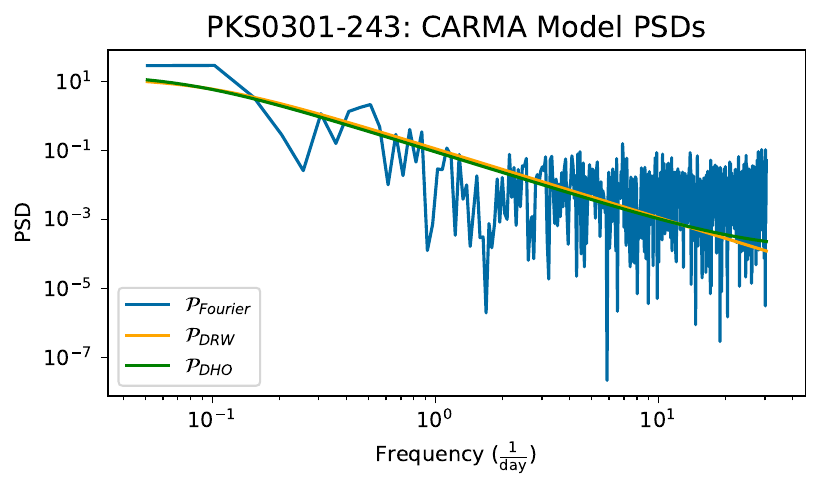}}
    \caption{}
    \end{figure}





    \begin{figure}
        \label{fig:CARMAlcs+psd3}
        \centering
        \subfloat{
         \includegraphics[width=0.6\textwidth]{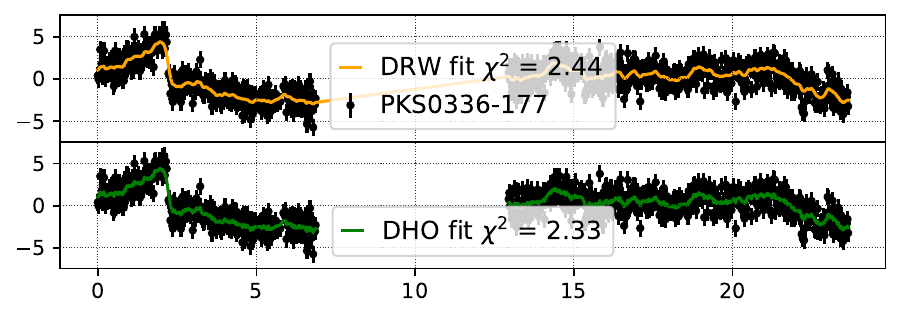}}
        \subfloat{
         \includegraphics[width=0.36\textwidth]{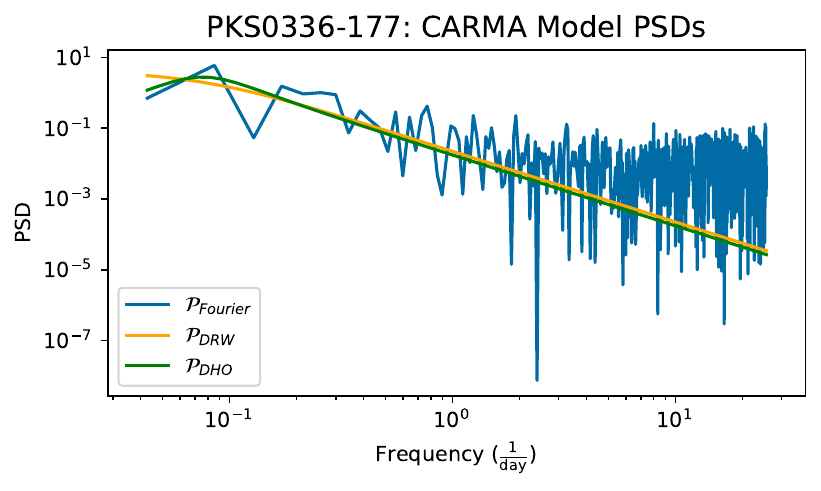}}
         \newline
         \subfloat{
         \includegraphics[width=0.6\textwidth]{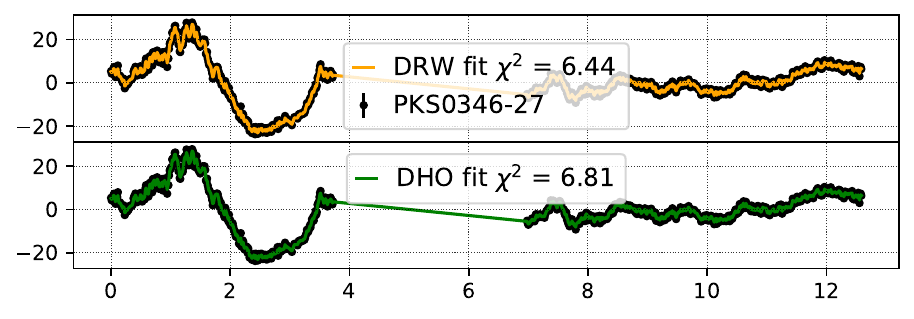}}
        \subfloat{
         \includegraphics[width=0.36\textwidth]{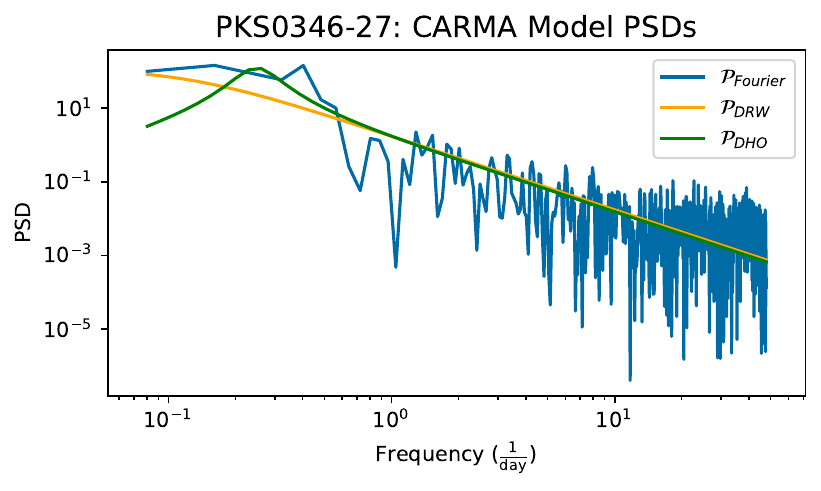}}
         \newline
        \subfloat{
         \includegraphics[width=0.6\textwidth]{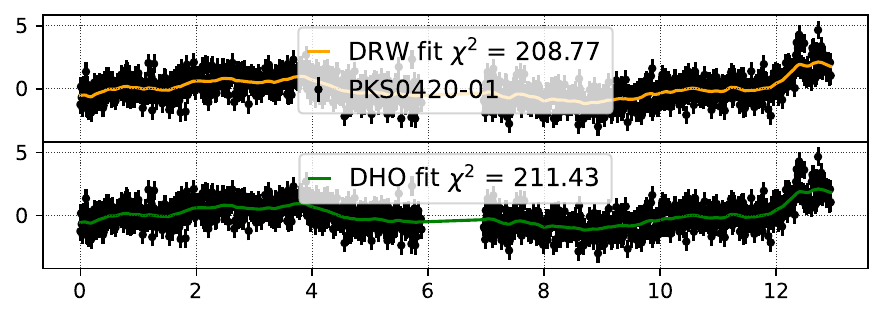}}
        \subfloat{
         \includegraphics[width=0.36\textwidth]{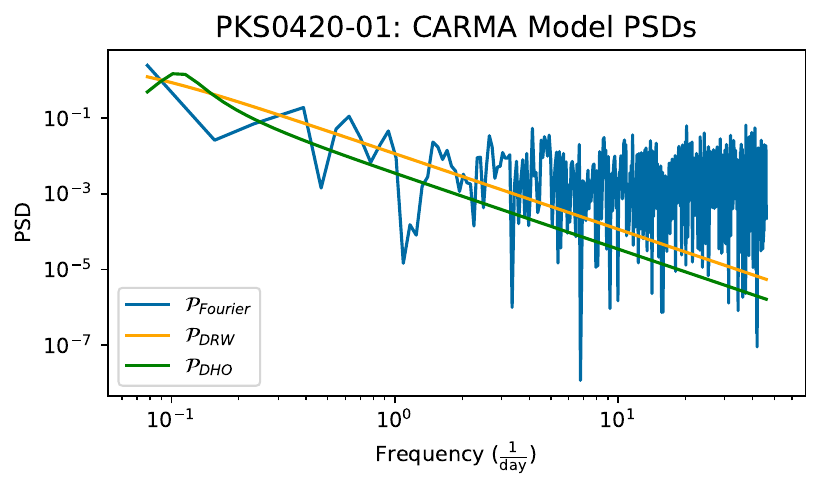}}
         \newline
        \subfloat{
         \includegraphics[width=0.6\textwidth]{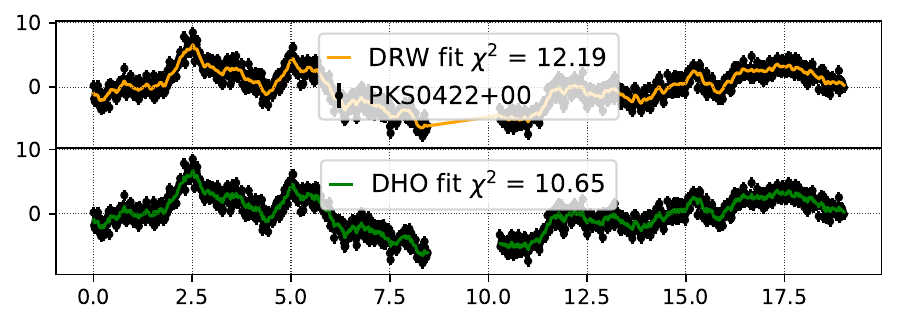}}
        \subfloat{
         \includegraphics[width=0.36\textwidth]{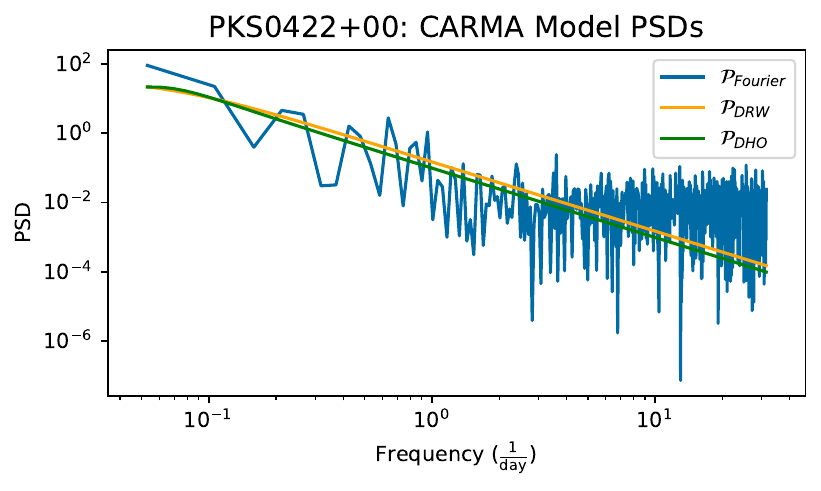}}
         \newline
        \subfloat{
         \includegraphics[width=0.6\textwidth]{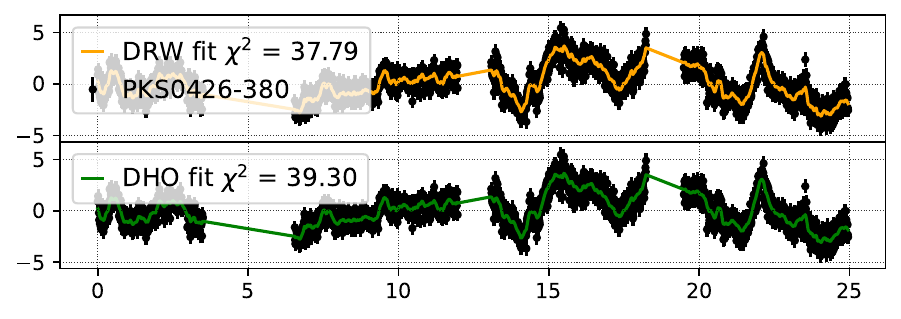}}
        \subfloat{
         \includegraphics[width=0.36\textwidth]{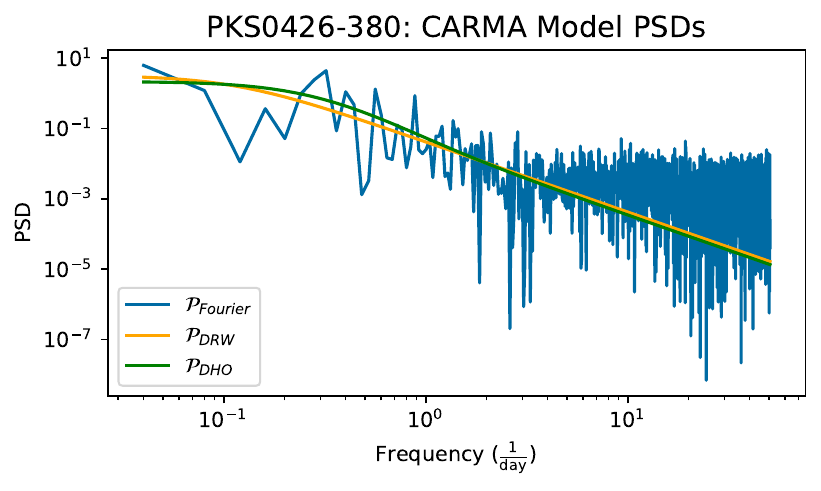}}
    \caption{}
    \end{figure}





    \begin{figure}
        \label{fig:CARMAlcs+psd4}
        \centering
        \subfloat{
         \includegraphics[width=0.6\textwidth]{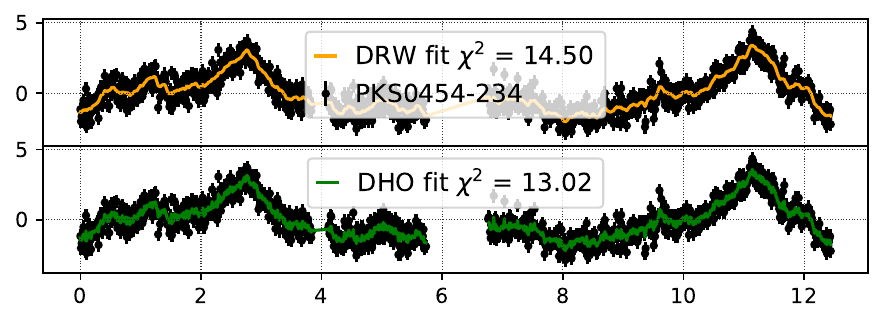}}
        \subfloat{
         \includegraphics[width=0.36\textwidth]{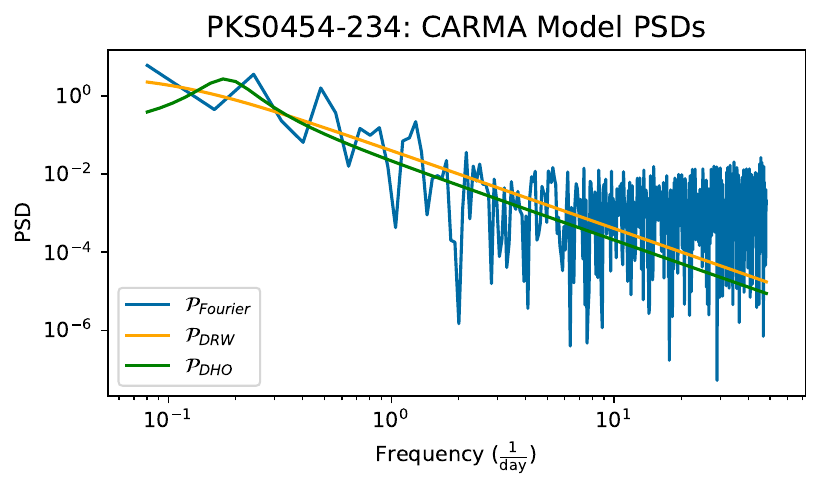}}
         \newline
         \subfloat{
         \includegraphics[width=0.6\textwidth]{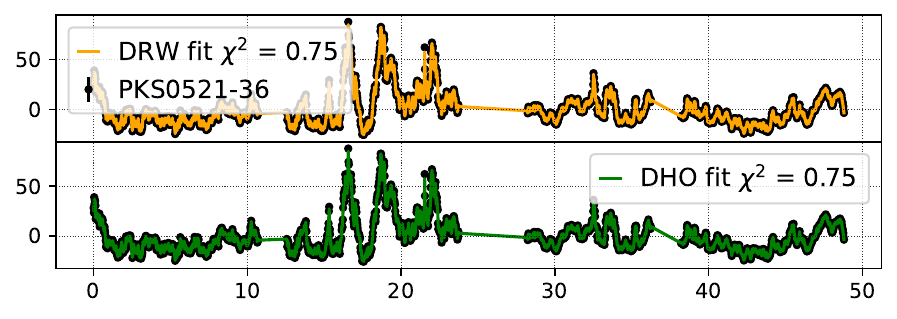}}
        \subfloat{
         \includegraphics[width=0.36\textwidth]{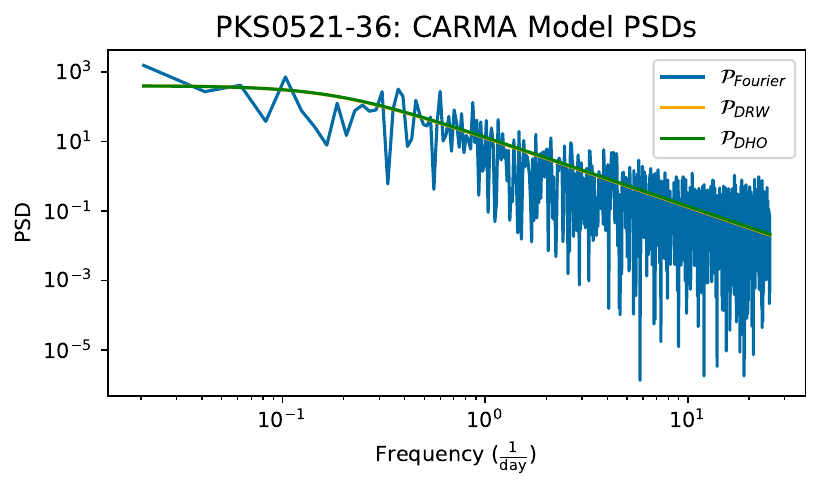}}
         \newline
        \subfloat{
         \includegraphics[width=0.6\textwidth]{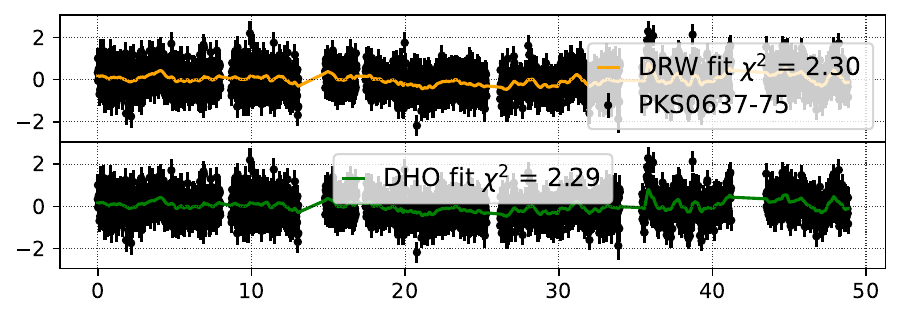}}
        \subfloat{
         \includegraphics[width=0.36\textwidth]{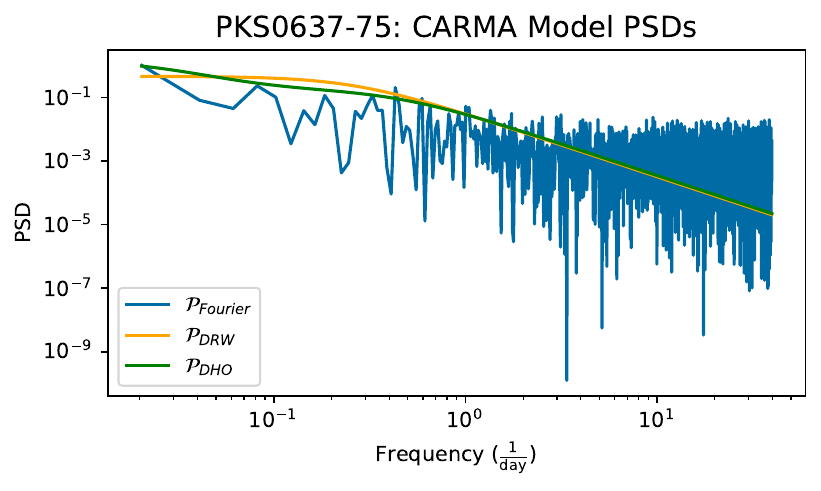}}
         \newline
        \subfloat{
         \includegraphics[width=0.6\textwidth]{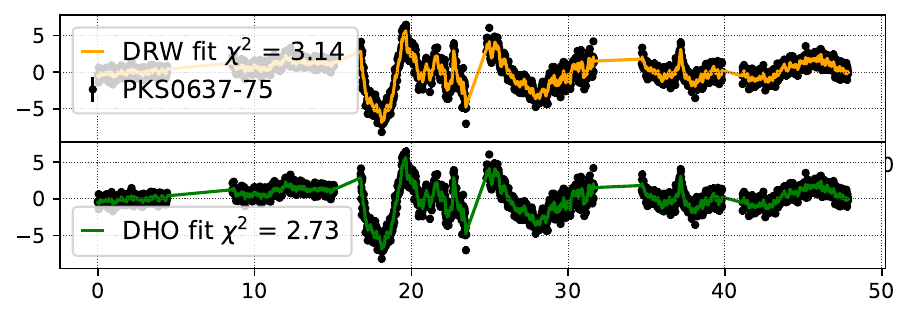}}
        \subfloat{
         \includegraphics[width=0.36\textwidth]{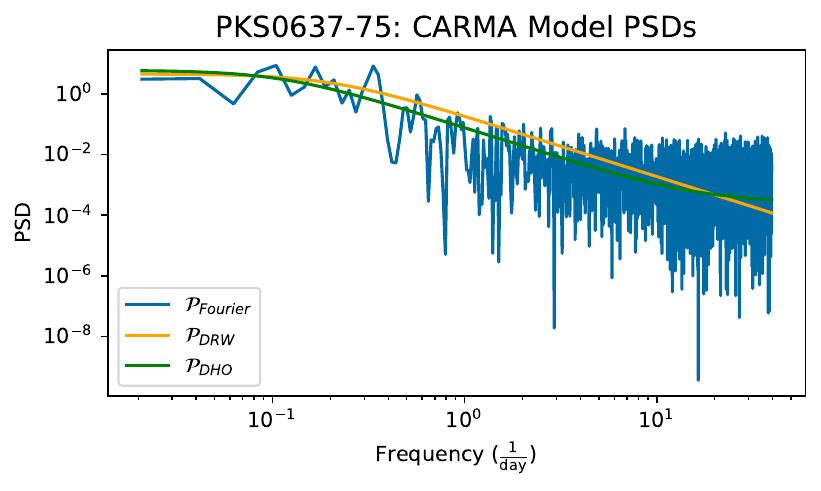}}
         \newline
        \subfloat{
         \includegraphics[width=0.6\textwidth]{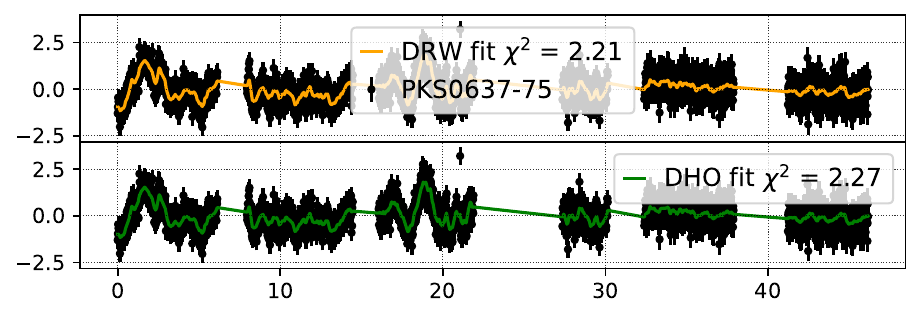}}
        \subfloat{
         \includegraphics[width=0.36\textwidth]{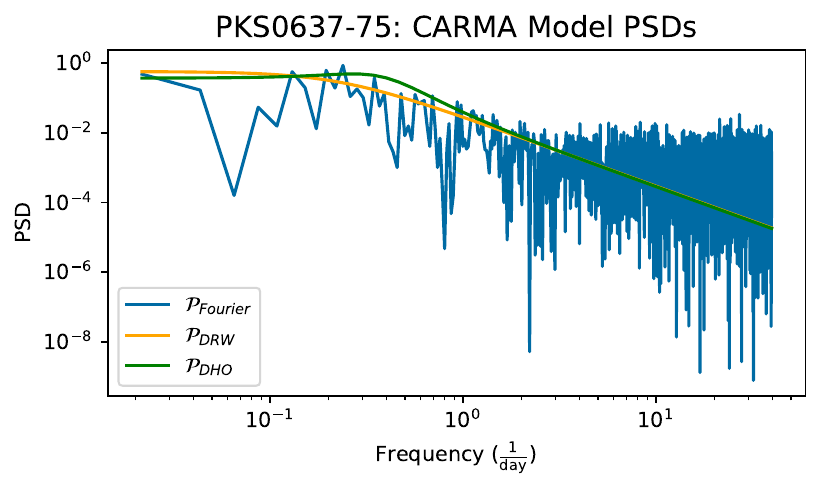}}
    \caption{}
    \end{figure}





    \begin{figure}
        \label{fig:CARMAlcs+psd5}
        \centering
        \subfloat{
         \includegraphics[width=0.6\textwidth]{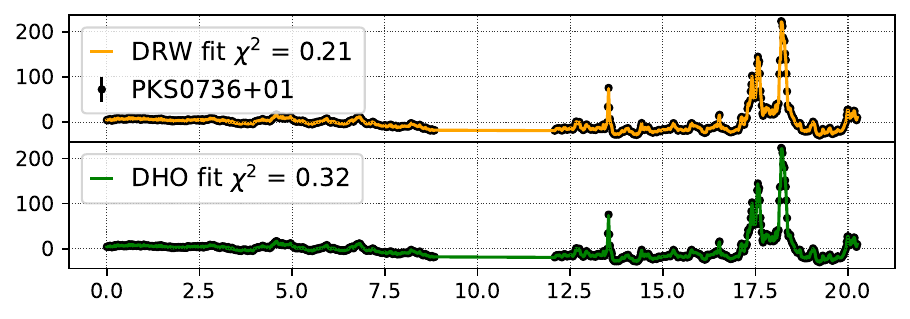}}
        \subfloat{
         \includegraphics[width=0.36\textwidth]{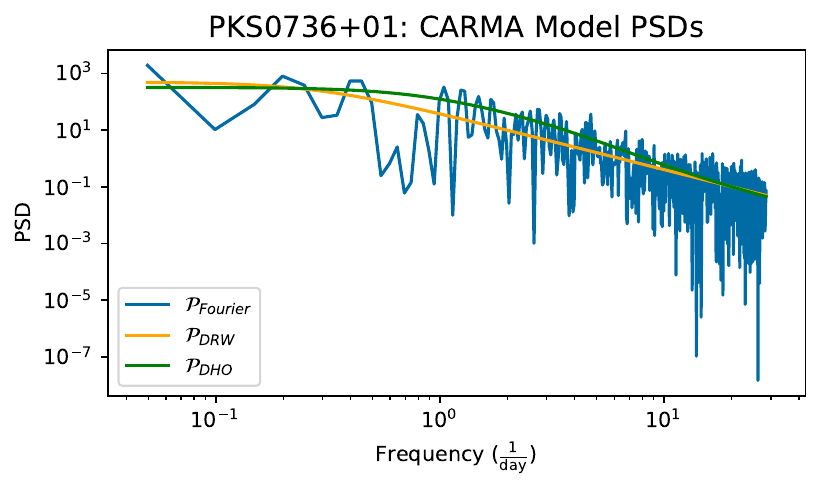}}
         \newline
         \subfloat{
         \includegraphics[width=0.6\textwidth]{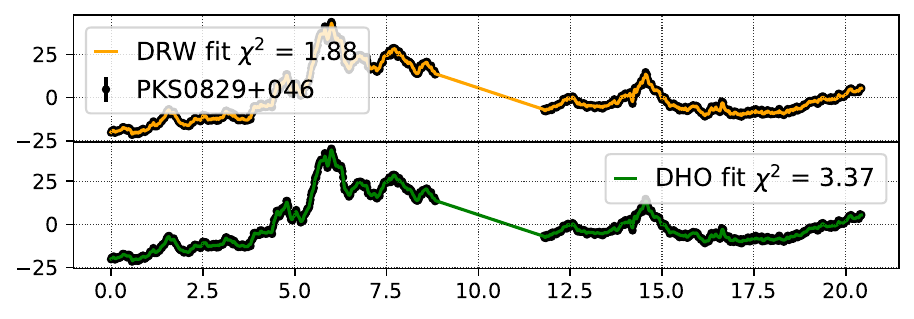}}
        \subfloat{
         \includegraphics[width=0.36\textwidth]{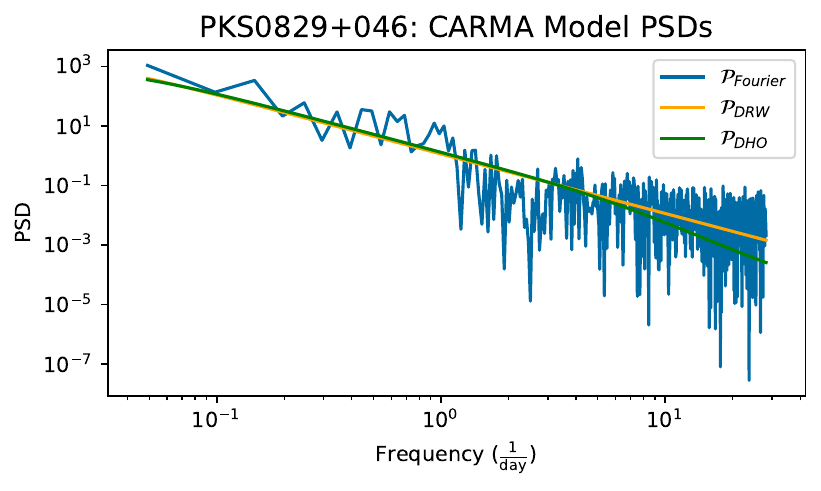}}
         \newline
        \subfloat{
         \includegraphics[width=0.6\textwidth]{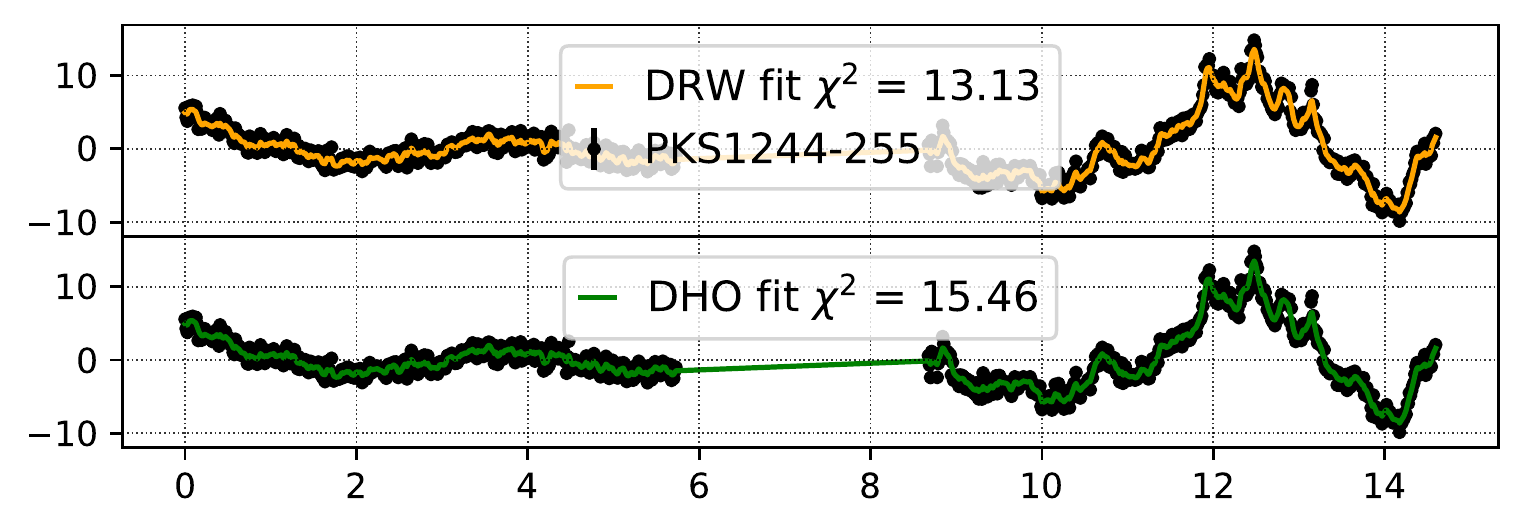}}
        \subfloat{
         \includegraphics[width=0.36\textwidth]{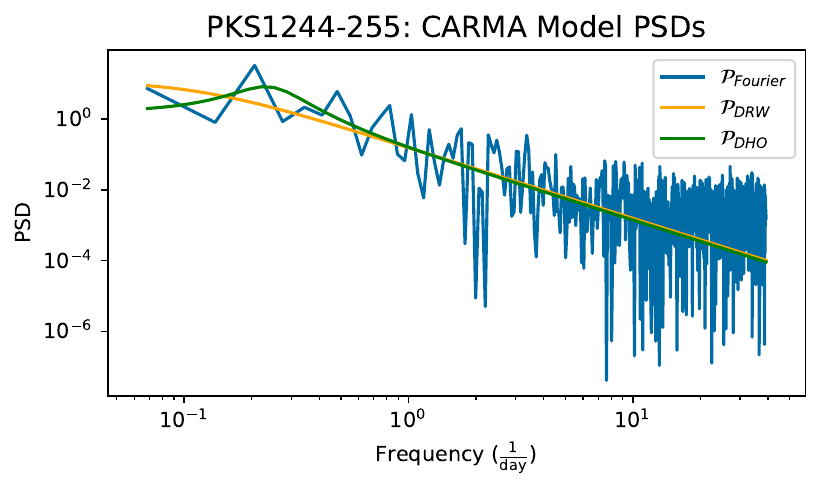}}
         \newline
        \subfloat{
         \includegraphics[width=0.6\textwidth]{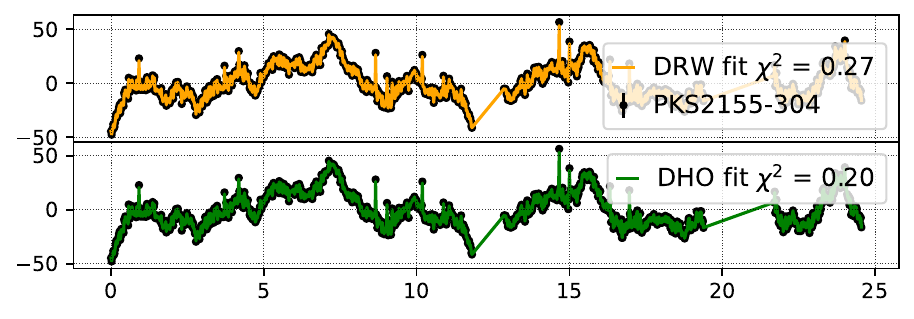}}
        \subfloat{
         \includegraphics[width=0.36\textwidth]{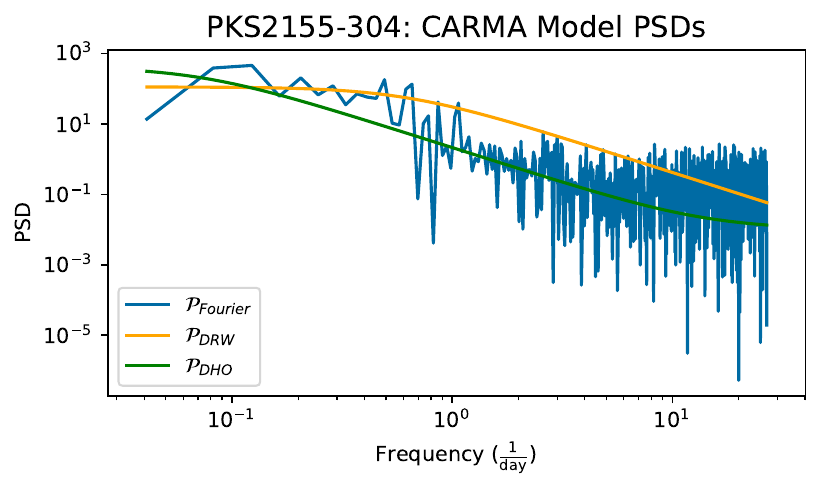}}
         \newline
        \subfloat{
         \includegraphics[width=0.6\textwidth]{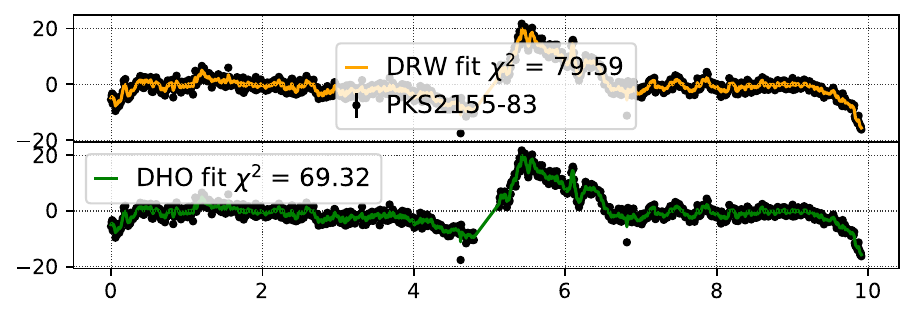}}
        \subfloat{
         \includegraphics[width=0.36\textwidth]{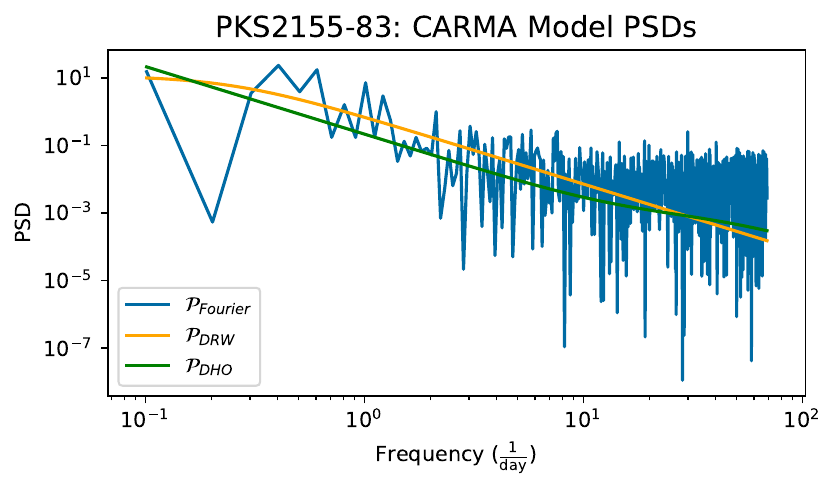}}
    \caption{}
    \end{figure}





    \begin{figure}
        \label{fig:CARMAlcs+psd6}
        \centering
        \subfloat{
         \includegraphics[width=0.6\textwidth]{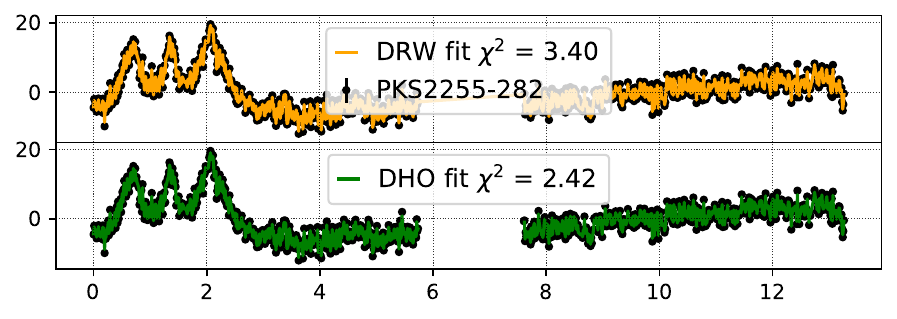}}
        \subfloat{
         \includegraphics[width=0.36\textwidth]{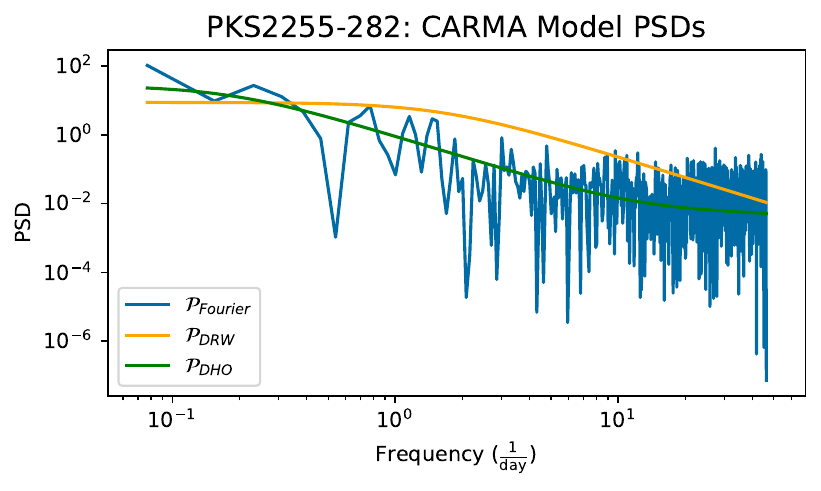}}
         \newline
         \subfloat{
         \includegraphics[width=0.6\textwidth]{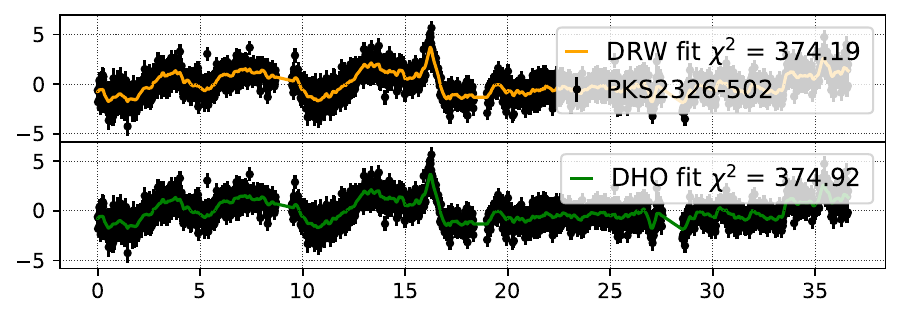}}
        \subfloat{
         \includegraphics[width=0.36\textwidth]{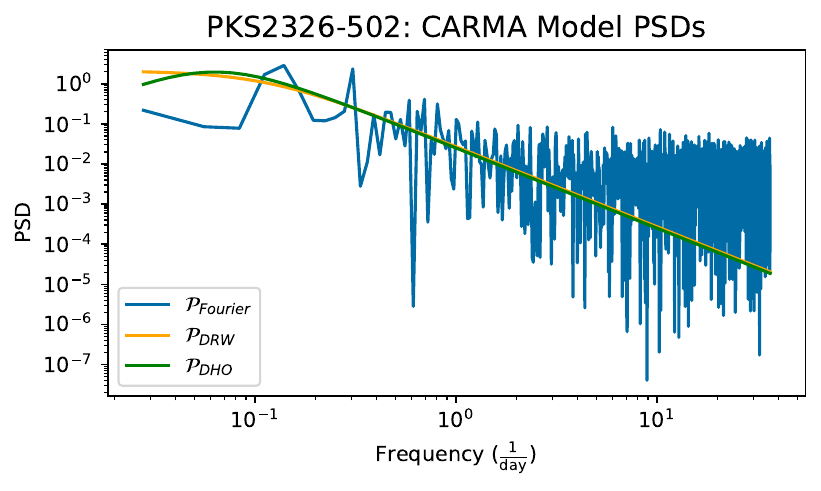}}
         \newline
        \subfloat{
         \includegraphics[width=0.6\textwidth]{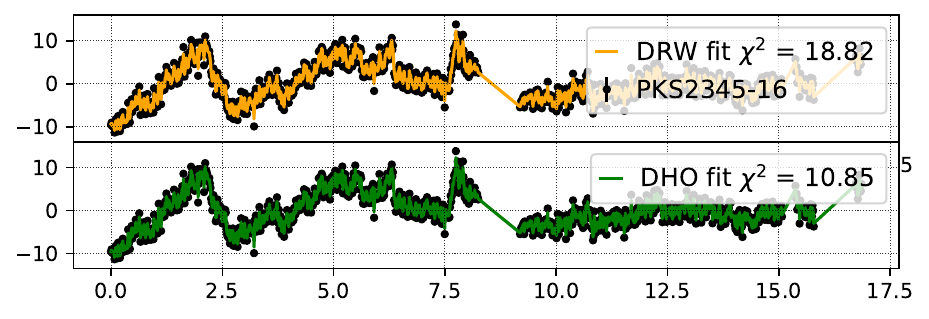}}
        \subfloat{
         \includegraphics[width=0.36\textwidth]{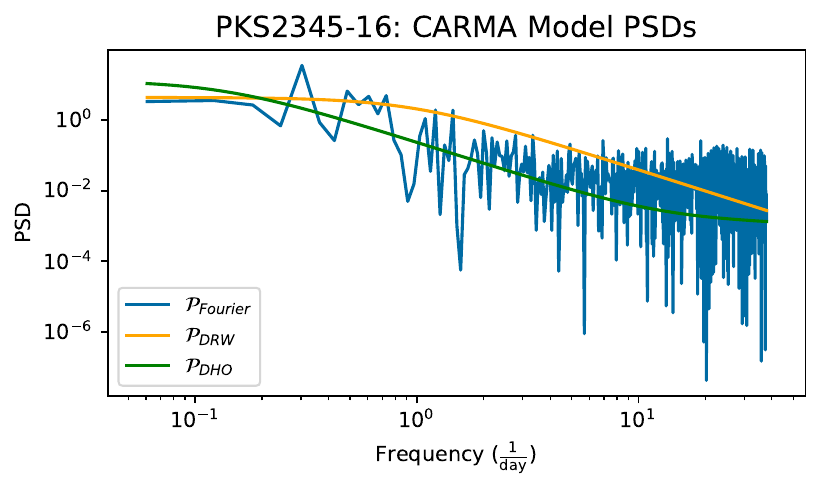}}
         \newline
        \subfloat{
         \includegraphics[width=0.6\textwidth]{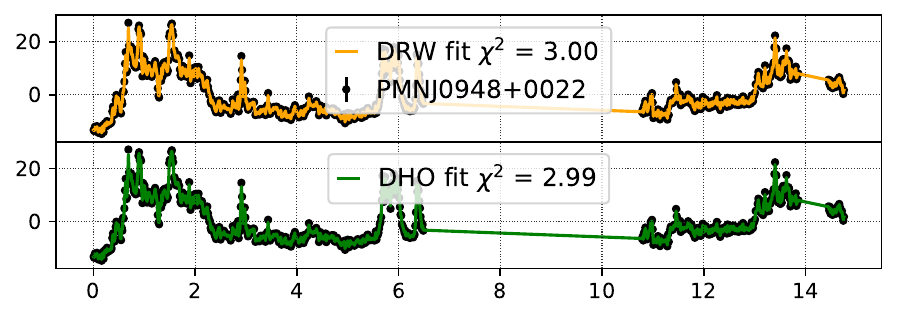}}
        \subfloat{
         \includegraphics[width=0.36\textwidth]{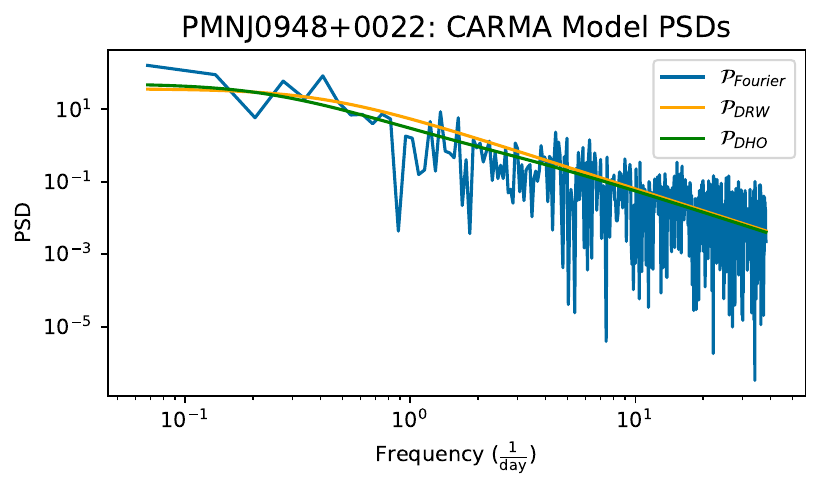}}
         \newline
        \subfloat{
         \includegraphics[width=0.6\textwidth]{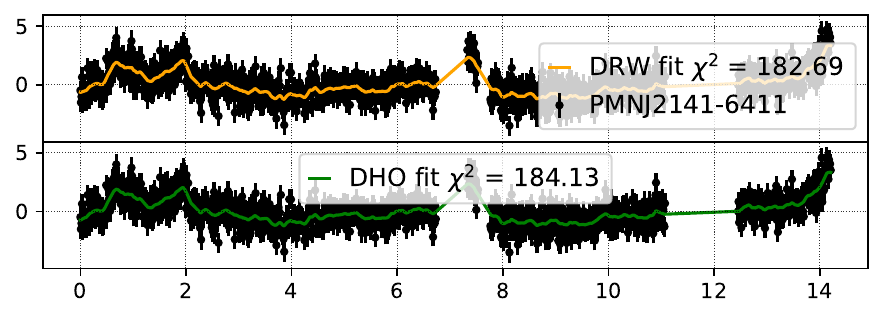}}
        \subfloat{
         \includegraphics[width=0.36\textwidth]{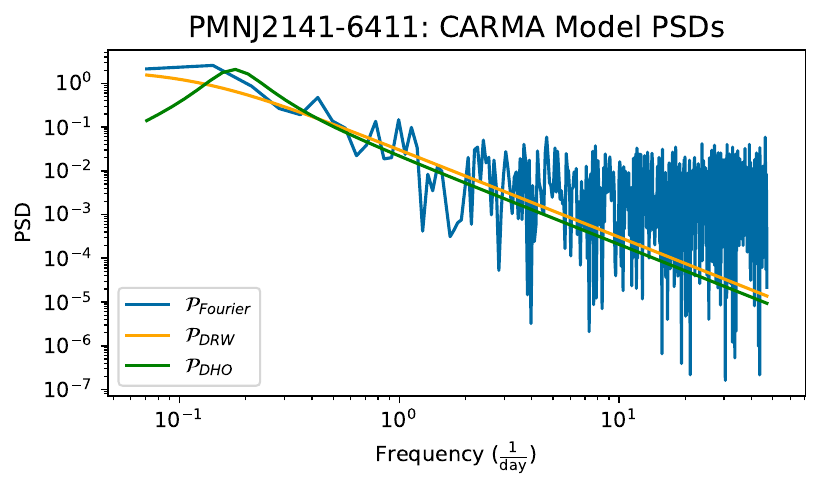}}
    \caption{}
    \end{figure}





    \begin{figure}
        \label{fig:CARMAlcs+psd7}
        \centering
        \subfloat{
         \includegraphics[width=0.6\textwidth]{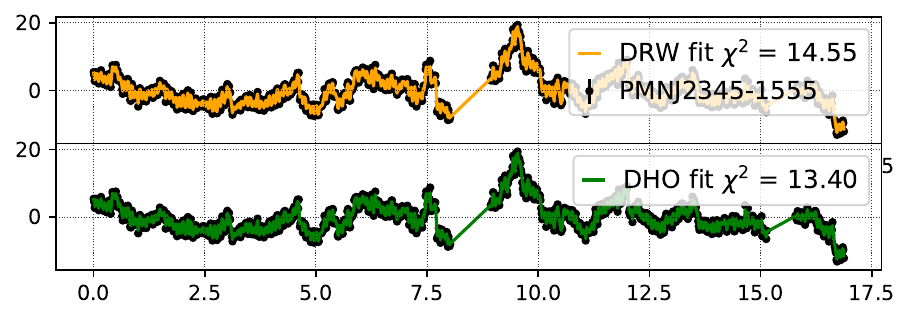}}
        \subfloat{
         \includegraphics[width=0.36\textwidth]{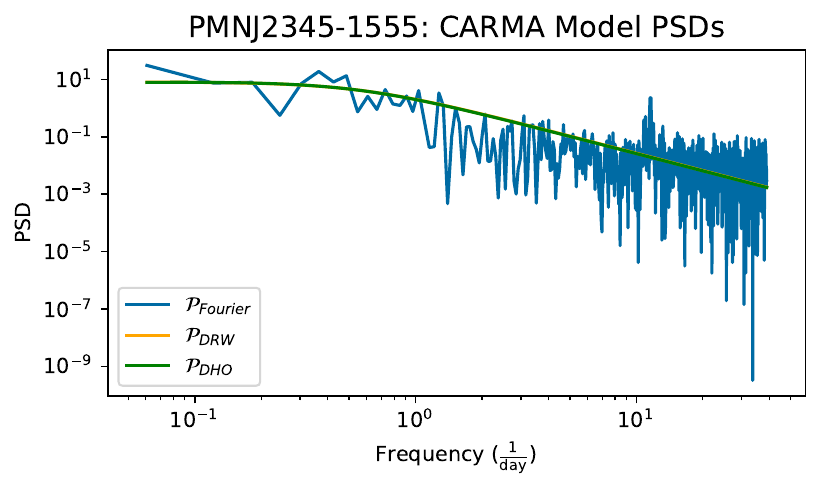}}
         \newline
         \subfloat{
         \includegraphics[width=0.6\textwidth]{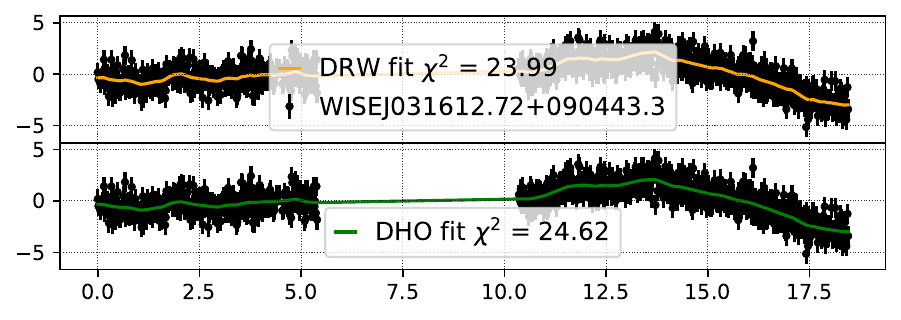}}
        \subfloat{
         \includegraphics[width=0.36\textwidth]{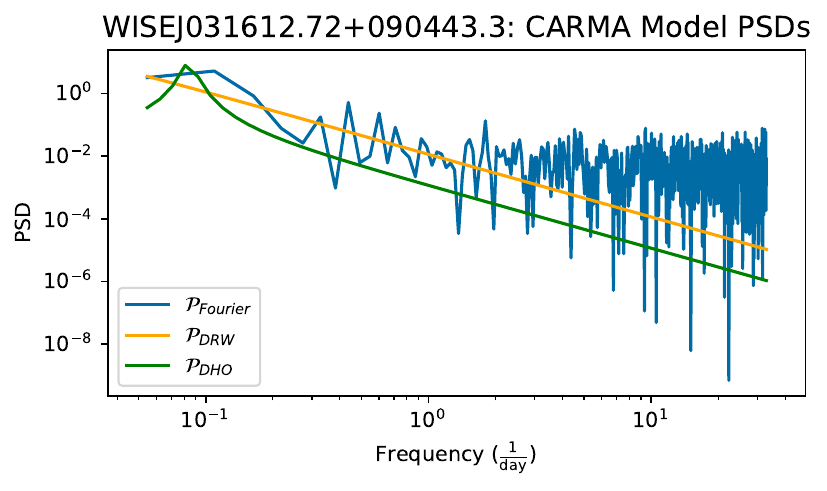}}
         \newline
        \subfloat{
         \includegraphics[width=0.6\textwidth]{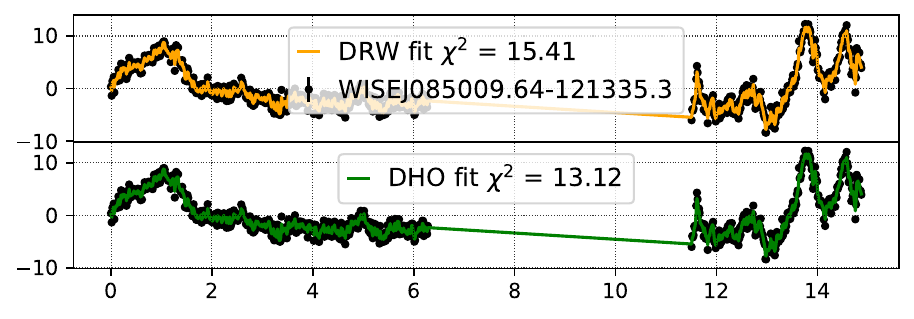}}
        \subfloat{
         \includegraphics[width=0.36\textwidth]{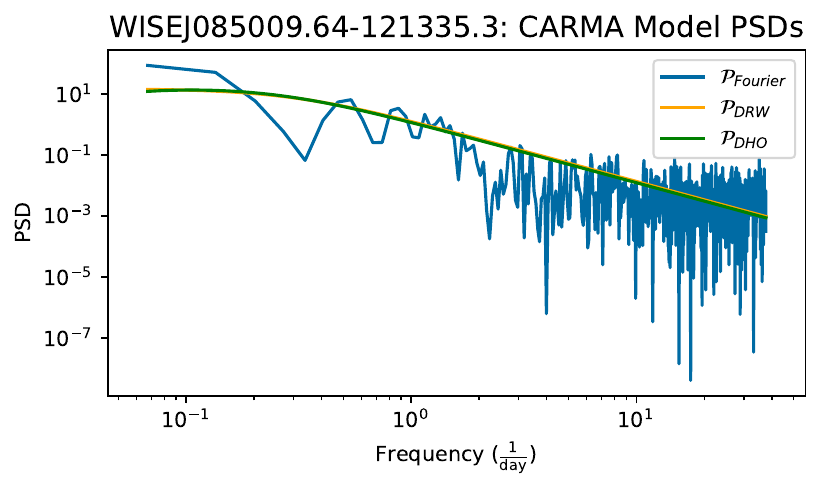}}
         \newline
        \subfloat{
         \includegraphics[width=0.6\textwidth]{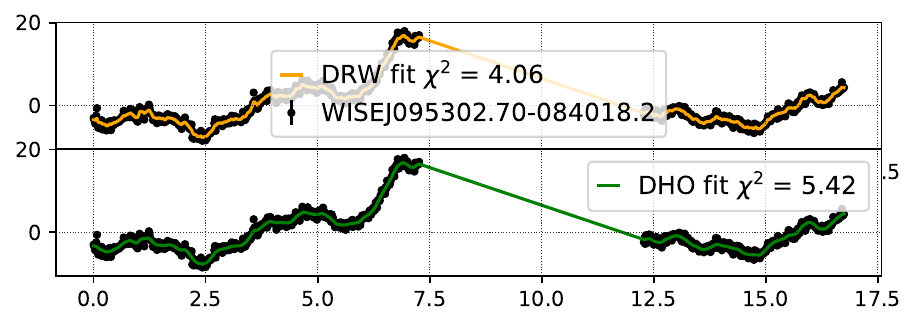}}
        \subfloat{
         \includegraphics[width=0.36\textwidth]{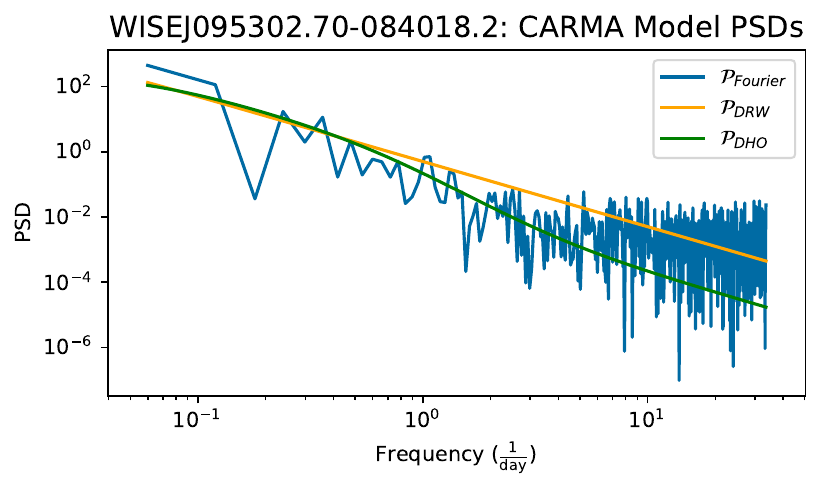}}
         \newline
        \subfloat{
         \includegraphics[width=0.6\textwidth]{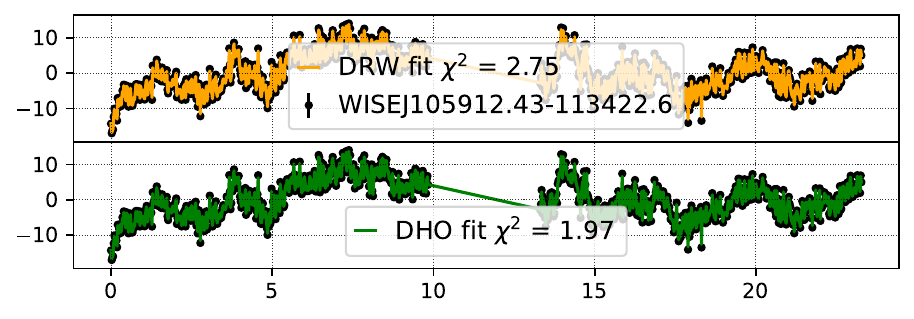}}
        \subfloat{
         \includegraphics[width=0.36\textwidth]{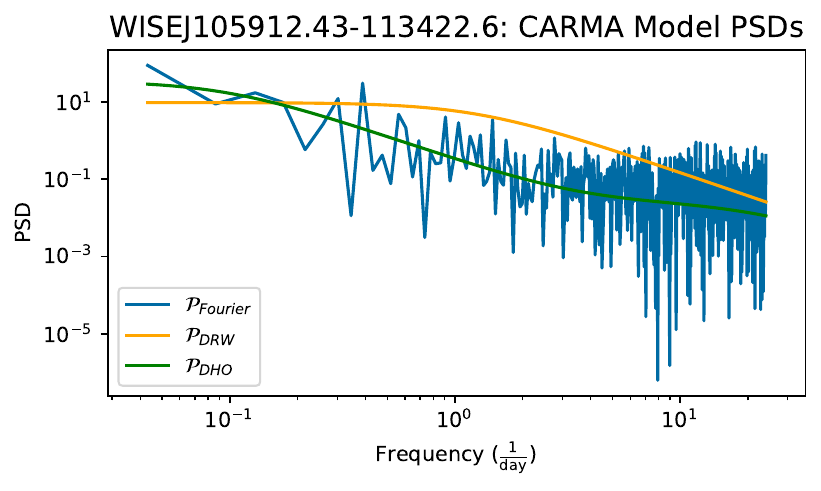}}
    \caption{}
    \end{figure}





}

\newpage

\begin{widetext}
        
\begin{longrotatetable}
    \centering
    \movetabledown=5mm
    \begin{deluxetable}{lcrrrrrrrrrrrrrr}
        \tabletypesize{\scriptsize}
        \tablewidth{0pt} 
        \tablenum{2}
        \tablecaption{Variability statistics for all \textit{unbinned} light curve. For each target, we provide the observation sectors, the mean and standard deviations for the flux and subsequent flux, the logarithms of the excess variance and RMS-scatter with associated errors, the shortest timescales of variability for increasing and decreasing flux respectively, estimated radius of emission region size associated with the shortest timescale, and the $\rm \chi^{2}$/dof for each light curve. \label{tab:variability1}}
        \tablehead{
        \colhead{Target} & \colhead{sector(s)} & \colhead{$\rm \overline{F}$ } & \colhead{$\rm \sigma_{F}$ } &
        \colhead{$\rm \overline{\Delta F} $} & \colhead{$\rm \sigma_{\Delta F}$} & \colhead{$\log{\sigma_{XS}^{2}}$} &
        \colhead{$\log{}$err($\sigma_{XS}^{2})$} & \colhead{$\sigma_{rms}$} &
        \colhead{err($\sigma_{rms})$} & \colhead{$\rm \tau_{+}$} & \colhead{$\rm \Delta t_{+}$} & \colhead{$\rm \tau_{-}$} &
        \colhead{$\rm \Delta t_{-}$} & \colhead{$R_{em}$} &\colhead{$\rm \chi^{2}$/dof}\\
        \colhead{} & \colhead{} & \colhead{[$\times10^{2}$ cts/s]} & \colhead{} & \colhead{[cts/s]} & \colhead{} & \colhead{[$10^{x}$]} &
        \colhead{[$10^{x}$]} & \colhead{[$e^{-}s^{-1}$]} & \colhead{[$e^{-}s^{-1}$]} & \colhead{[days]} & \colhead{[days]} & \colhead{[days]} &
        \colhead{[days]} & \colhead{ [$\times 10^{12}$ km]} & \colhead{}
        }
        \startdata
        1ES2322-409 & 2 & 109.866 & 10.824 & 0.119 & 2.628 & 2.067 & -0.329 & 10.798 & 0.022 & 3.255 & 0.586 & 2.554 & 0.391 & 0.066 & 201.763\\
        1RXSJ054357.3-553206 & 1-13 & 112.375 & 21.079 & 0.006 & 1.399 & 2.647 & -0.509 & 21.064 & 0.007 & 5.819 & 0.540 & 8.445 & 0.720 & 0.151 & 729.825\\
        3C120 & 5 & 854.346 & 7.518 & -0.026 & 2.218 & 1.245 & -0.488 & 4.192 & 0.039 & 2.303 & 0.168 & 2.341 & 0.168 & 0.060 & 66.209\\
        NGC1218 & 4 & 2936.545 & 1.961 & 0.009 & 2.424 & 0.191 & -0.769 & 1.245 & 0.068 & 96.639 & 0.446 & 106.603 & 0.446 & 2.503 & 1.670\\
        PKS0035-252 & 3 & 161.100 & 30.508 & 0.129 & 3.499 & 2.968 & 0.285 & 30.493 & 0.032 & 0.903 & 0.306 & 1.386 & 0.459 & 0.023 & 1026.563\\
        PKS0130-17 & 3 & 130.146 & 11.044 & -0.009 & 3.832 & 2.084 & -0.190 & 11.010 & 0.029 & 2.878 & 0.258 & 2.874 & 0.388 & 0.074 & 161.640\\
        PKS0208-512 & 2,3 & 107.053 & 23.721 & -0.009 & 2.076 & 2.750 & -0.100 & 23.709 & 0.017 & 2.994 & 0.344 & 2.652 & 0.229 & 0.069 & 950.121\\
        PKS0226-559 & 2,3 & 31.175 & 21.310 & -0.025 & 1.753 & 2.657 & -0.171 & 21.298 & 0.016 & 0.760 & 0.397 & 0.560 & 0.926 & 0.015 & 915.577\\
        PKS0235-618 & 1,2,3 & 62.404 & 2.757 & -0.006 & 1.092 & 0.844 & -1.121 & 2.642 & 0.014 & 5.599 & 1.250 & 8.940 & 2.188 & 0.145 & 11.640\\
        PKS0301-243 & 4 & 112.534 & 5.231 & 0.043 & 1.267 & 1.426 & -0.539 & 5.167 & 0.028 & 17.028 & 0.905 & 8.230 & 0.905 & 0.213 & 41.112\\
        PKS0336-177 & 4 & 354.678 & 2.807 & 0.029 & 1.466 & 0.836 & -0.730 & 2.617 & 0.036 & 21.729 & 0.430 & 19.659 & 0.430 & 0.509 & 7.684\\
        PKS0346-27 & 4 & 37.068 & 17.790 & 0.020 & 1.602 & 2.500 & -0.043 & 17.775 & 0.025 & 1.724 & 0.230 & 1.354 & 0.345 & 0.035 & 548.205\\
        PKS0420-01 & 5 & 3.726 & 1.551 & -0.008 & 1.196 & 0.268 & -1.180 & 1.361 & 0.024 & 0.655 & 1.559 & 1.942 & 18.244 & 0.017 & 4.369\\
        PKS0422+00 & 5 & 61.250 & 4.700 & 0.069 & 1.376 & 1.329 & -0.611 & 4.620 & 0.027 & 8.900 & 1.185 & 8.134 & 0.911 & 0.211 & 29.770\\
        PKS0426-380 & 4,5 & 20.898 & 3.054 & 0.003 & 0.982 & 0.948 & -1.030 & 2.978 & 0.016 & 3.155 & 1.089 & 2.364 & 0.980 & 0.061 & 20.030\\
        PKS0454-234 & 5 & 18.945 & 1.683 & 0.002 & 0.818 & 0.415 & -1.313 & 1.613 & 0.015 & 7.822 & 2.631 & 9.945 & 2.059 & 0.203 & 12.214\\
        PKS0521-36 & 5,6 & 412.689 & 19.307 & 0.007 & 2.952 & 2.570 & -0.051 & 19.279 & 0.023 & 2.781 & 0.434 & 5.296 & 0.651 & 0.072 & 337.767\\
        \multirow{3}{*}{PKS0637-75} & 1,2,3 & 321.446 & 0.530 & -0.002 & 0.689 & -1.423 & -2.124 & 0.194 & 0.019 & 140.218 & 1.941 & 273.000 & 2.634 & 3.632 & 1.153\\
        ~ & 4,5,6 & 253.437 & 2.499 & 0.002 & 0.908 & 0.776 & -1.314 & 2.444 & 0.010 & 22.251 & 0.277 & 15.466 & 0.555 & 0.401 & 23.119\\
        ~ & 7,8,9 & 231.882 & 0.975 & 0.007 & 0.656 & -0.264 & -1.793 & 0.738 & 0.011 & 160.017 & 2.218 & 102.883 & 1.664 & 2.665 & 3.321\\
        PKS0736+01 & 7 & 31.948 & 29.713 & 0.075 & 1.906 & 2.999 & 0.212 & 31.583 & 0.026 & 0.284 & 0.401 & 0.337 & 0.401 & 0.007 & 1733.016\\
        PKS0829+046 & 7 & 66.263 & 15.037 & -0.005 & 1.104 & 2.354 & -0.208 & 15.024 & 0.021 & 1.300 & 0.211 & 1.787 & 0.211 & 0.034 & 528.206\\
        PKS1244-255 & 10 & 22.634 & 6.913 & 0.014 & 1.058 & 1.674 & -0.464 & 6.872 & 0.025 & 2.057 & 0.421 & 2.687 & 0.702 & 0.053 & 85.123\\
        PKS2155-304 & 1 & 562.525 & 22.366 & -0.053 & 5.174 & 2.698 & 0.196 & 22.335 & 0.035 & 4.323 & 0.410 & 4.167 & 0.410 & 0.108 & 384.894\\
        PKS2155-83 & 13 & 19.275 & 5.634 & -0.022 & 1.737 & 1.488 & -0.518 & 5.548 & 0.027 & 0.693 & 0.400 & 0.207 & 0.880 & 0.005 & 32.390\\
        PKS2255-282 & 2 & 45.770 & 11.733 & 0.011 & 2.918 & 2.137 & -0.228 & 11.705 & 0.025 & 1.611 & 0.595 & 0.934 & 0.238 & 0.024 & 211.143\\
        PKS2326-502 & 1,2 & 4.422 & 1.267 & 0.002 & 1.095 & 0.044 & -1.435 & 1.052 & 0.017 & 0.982 & 2.264 & 2.074 & 4.378 & 0.025 & 2.861\\
        PKS2345-16 & 2 & 15.394 & 5.668 & 0.016 & 2.219 & 1.498 & -0.564 & 5.613 & 0.024 & 1.892 & 0.877 & 1.147 & 1.315 & 0.030 & 51.352\\
        PMNJ0948+0022 & 8 & 13.927 & 9.045 & 0.026 & 1.764 & 1.910 & -0.304 & 9.014 & 0.028 & 0.386 & 1.566 & 0.250 & 0.783 & 0.006 & 142.868\\
        PMNJ2141-6411 & 1 & 4.600 & 2.165 & 0.001 & 1.260 & 0.598 & -0.977 & 1.991 & 0.026 & 1.918 & 4.679 & 1.765 & 7.019 & 0.046 & 6.323\\
        PMNJ2345-1555 & 2 & 11.327 & 4.718 & -0.040 & 1.708 & 1.366 & -0.689 & 4.819 & 0.021 & 1.605 & 1.166 & 0.737 & 1.166 & 0.019 & 46.329\\
        WISEJ031612.72+090443.3 & 4 & 27.817 & 1.999 & -0.017 & 1.345 & 0.312 & -0.985 & 1.432 & 0.036 & 113.325 & 30.066 & 33.511 & 14.365 & 0.868 & 3.221\\
        WISEJ085009.64-121335.3 & 8 & 11.019 & 8.399 & 0.015 & 1.449 & 1.846 & -0.410 & 8.375 & 0.023 & 0.737 & 2.377 & 1.519 & 1.189 & 0.019 & 179.347\\
        WISEJ095302.70-084018.2 & 8 & 54.726 & 6.205 & 0.035 & 0.919 & 1.580 & -0.527 & 6.169 & 0.024 & 6.713 & 0.818 & 10.598 & 0.491 & 0.174 & 84.077\\
        WISEJ105912.43-113422.6 & 9 & 26.920 & 5.902 & -0.027 & 3.087 & 1.459 & -0.585 & 5.366 & 0.024 & 2.224 & 1.706 & 2.695 & 1.137 & 0.058 & 53.074
        \enddata
    \end{deluxetable}
\end{longrotatetable}
\begin{longrotatetable}
    \centering
    \begin{deluxetable}{lcrrrrrrrrrrrrr}
        \centering
        \tabletypesize{\scriptsize}
        \tablewidth{0pt} 
        \tablenum{3}
        \tablecaption{Variability statistics for all \textit{binned} light curve. Following the sectors of observations, mean flux and standard deviation, mean subsequent flux and standard deviation, the logarithm of the excess variance, and the RMS-scatter.\label{tab:variability2}}
        \tablehead{
        \colhead{Target} & \colhead{sector(s)} & \multicolumn{2}{c}{$\rm \overline{F}$} & \multicolumn{2}{c}{$\rm \sigma_{F}$} &
        \multicolumn{2}{c}{$\rm \overline{\Delta F}$} & \multicolumn{2}{c}{$\rm \sigma_{\Delta F}$} & \multicolumn{2}{c}{$\rm \log{\sigma^{2}_{XS}}$} & \multicolumn{2}{c}{$\rm \sigma_{rms}$}\\
        \colhead{} & \colhead{} & \multicolumn{2}{c}{[$\times10^{2}$ cts/s]} & \multicolumn{2}{c}{} & \multicolumn{2}{c}{[cts/s]} & \multicolumn{2}{c}{} & \multicolumn{2}{c}{[$10^{x}$]} & \multicolumn{2}{c}{[$e^{-}s^{-1}$]}\\
        \colhead{} & \colhead{} & \colhead{6 hr} & \colhead{12 hr} & \colhead{6 hr} & \colhead{12 hr} & \colhead{6 hr} & \colhead{12 hr} & \colhead{6 hr} & \colhead{12 hr} & \colhead{6 hr} & \colhead{12 hr} & \colhead{6 hr} & \colhead{12 hr}}
        \startdata
        1ES2322-409 & 2 & 1.093 & 1.103 & 12.915 & 10.465 & -0.121 & 0.118 & 2.788 & 4.592 & 2.055 & 2.038 & 10.656 & 10.449\\
        1RXSJ054357.3-553206 & 1-13 & 1.127 & 1.126 & 21.171 & 20.985 & -0.125 & -0.206 & 1.313 & 2.030 & 2.651 & 2.644 & 21.168 & 20.983\\
        3C120 & 5 & 8.541 & 8.538 & 3.609 & 5.111 & 0.469 & -0.310 & 2.317 & 3.196 & 1.073 & 1.921 & 3.441 & 9.135\\
        NGC1218 & 4 & 29.365 & 29.364 & 0.768 & 0.642 & -0.062 & 0.128 & 0.839 & 0.511 & -0.046 & -0.073 & 0.949 & 0.920\\
        PKS0035-252 & 3 & 1.514 & 1.566 & 25.524 & 25.925 & 3.106 & 2.242 & 10.347 & 24.775 & 2.934 & 2.896 & 29.296 & 28.051\\
        PKS0130-17 & 3 & 1.297 & 1.297 & 10.570 & 10.514 & 0.215 & 0.156 & 1.634 & 2.284 & 2.045 & 2.040 & 10.526 & 10.477\\
        PKS0208-512 & 2,3 & 1.068 & 0.982 & 23.625 & 20.133 & -0.183 & -0.657 & 2.517 & 5.336 & 2.746 & 2.765 & 23.618 & 24.127\\
        PKS0226-559 & 2,3 & 0.317 & 0.310 & 21.256 & 21.183 & -0.087 & -0.380 & 1.352 & 2.102 & 2.655 & 2.652 & 21.250 & 21.179\\
        PKS0235-618 & 1,2,3 & 0.624 & 0.624 & 2.649 & 2.631 & -0.027 & -0.082 & 0.330 & 0.270 & 0.843 & 0.838 & 2.639 & 2.625\\
        PKS0301-243 & 4 & 1.124 & 1.124 & 5.211 & 5.168 & 0.050 & 0.217 & 0.936 & 1.628 & 1.432 & 1.426 & 5.202 & 5.162\\
        PKS0336-177 & 4 & 3.546 & 3.545 & 2.640 & 2.645 & 0.075 & 0.163 & 0.643 & 0.716 & 0.836 & 0.841 & 2.619 & 2.632\\
        PKS0346-27 & 4 & 0.378 & 0.376 & 18.064 & 17.813 & -0.097 & 0.775 & 1.918 & 3.228 & 2.513 & 2.501 & 18.056 & 17.806\\
        PKS0420-01 & 5 & 0.038 & 0.037 & 1.378 & 1.354 & -0.056 & -0.040 & 0.421 & 0.321 & 0.267 & 0.256 & 1.360 & 1.343\\
        PKS0422+00 & 5 & 0.614 & 0.613 & 4.678 & 4.689 & -0.197 & -0.272 & 0.974 & 1.702 & 1.338 & 1.341 & 4.669 & 4.684\\
        PKS0426-380 & 4,5 & 0.210 & 0.218 & 2.937 & 1.954 & -0.039 & -0.145 & 0.523 & 0.752 & 0.933 & 0.928 & 2.929 & 2.912\\
        PKS0454-234 & 5 & 0.183 & 0.181 & 1.121 & 0.878 & 0.083 & 0.229 & 0.518 & 0.752 & 0.391 & 0.393 & 1.569 & 1.572\\
        PKS0521-36 & 5,6 & 4.123 & 4.124 & 18.736 & 18.026 & 0.634 & 0.069 & 6.259 & 7.299 & 2.543 & 2.509 & 18.677 & 17.968\\
        \multirow{3}{*}{PKS0637-75} & 1,2,3 & 3.214 & 3.214 & 0.285 & 0.235 & -0.007 & -0.015 & 0.236 & 0.195 & -1.268 & -1.336 & 0.232 & 0.215\\
        ~ & 4,5,6 & 2.534 & 2.536 & 2.398 & 3.593 & -0.024 & -0.029 & 0.373 & 0.657 & 0.756 & 0.752 & 2.387 & 2.378\\
        ~ & 7,8,9 & 2.319 & 2.319 & 0.748 & 0.899 & 0.020 & 0.060 & 0.266 & 0.334 & -0.269 & -0.282 & 0.734 & 0.723\\
        PKS0736+01 & 7 & 0.324 & 0.319 & 32.461 & 31.584 & 0.659 & 1.259 & 4.942 & 3.618 & 3.017 & 2.995 & 32.239 & 31.438\\
        PKS0829+046 & 7 & 0.666 & 0.664 & 15.068 & 15.036 & 0.458 & 0.990 & 2.273 & 3.167 & 2.356 & 2.354 & 15.060 & 15.026\\
        PKS1244-255 & 10 & 0.225 & 0.225 & 6.766 & 6.765 & 0.138 & 0.171 & 1.079 & 1.028 & 1.660 & 1.660 & 6.759 & 6.760\\
        PKS2155-304 & 1 & 5.626 & 5.610 & 26.481 & 29.134 & 0.389 & -0.564 & 9.245 & 11.671 & 2.664 & 2.666 & 21.484 & 21.529\\
        PKS2155-83 & 13 & 0.181 & 0.177 & 4.064 & 4.345 & -0.368 & -0.595 & 1.148 & 1.573 & 1.448 & 1.425 & 5.297 & 5.159\\
        PKS2255-282 & 2 & 0.461 & 0.460 & 11.668 & 11.487 & -0.483 & -0.499 & 1.571 & 1.404 & 2.133 & 2.119 & 11.648 & 11.474\\
        PKS2326-502 & 1,2 & 0.043 & 0.043 & 1.096 & 1.081 & 0.015 & 0.018 & 0.424 & 0.551 & 0.029 & 0.005 & 1.033 & 1.006\\
        PKS2345-16 & 2 & 0.154 & 0.153 & 5.503 & 5.517 & 0.266 & 0.402 & 1.801 & 2.448 & 1.477 & 1.481 & 5.477 & 5.502\\
        PMNJ0948+0022 & 8 & 0.139 & 0.140 & 8.623 & 8.518 & -0.656 & -1.815 & 2.103 & 3.571 & 1.867 & 1.858 & 8.581 & 8.488\\
        PMNJ2141-6411 & 1 & 0.046 & 0.046 & 2.029 & 2.025 & -0.105 & -0.187 & 0.514 & 0.407 & 0.607 & 0.609 & 2.012 & 2.017\\
        PMNJ2345-1555 & 2 & 0.110 & 0.106 & 5.191 & 5.480 & -0.012 & -0.281 & 2.113 & 3.674 & 1.321 & 1.289 & 4.577 & 4.413\\
        WISEJ031612.72+090443.3 & 4 & 0.279 & 0.279 & 1.474 & 1.477 & -0.006 & -0.021 & 0.422 & 0.418 & 0.322 & 0.331 & 1.449 & 1.464\\
        WISEJ085009.64-121335.3 & 8 & 0.109 & 0.105 & 8.254 & 7.854 & 0.581 & 0.218 & 1.328 & 1.910 & 1.832 & 1.789 & 8.245 & 7.846\\
        WISEJ095302.70-084018.2 & 8 & 0.533 & 0.529 & 6.053 & 6.399 & 0.082 & 0.342 & 1.214 & 1.678 & 1.593 & 1.637 & 6.259 & 6.583\\
        WISEJ105912.43-113422.6 & 9 & 0.267 & 0.276 & 5.861 & 7.056 & 0.535 & 1.041 & 2.679 & 3.557 & 1.373 & 1.337 & 4.856 & 4.664
        \enddata
    \end{deluxetable}
\end{longrotatetable}

\end{widetext}

\begin{widetext}
        \begin{longrotatetable}
    \centering
    \begin{deluxetable}{lccrrccrr}
        \tabletypesize{\scriptsize}
        \tablewidth{0pt}
        \tablenum{5}
        \tablecaption{Best Fit PSD Results\label{tab:psd_results}}
        \tablehead{
        \colhead{Target} & \colhead{sector(s)} & \colhead{P-law model} & \colhead{$\alpha$} & \colhead{err($\alpha$)} & \colhead{$\log{(\nu_{b,k})}~[10 ^{x}$ Hz]} & \colhead{$R_{em}$ [mpc]} & \colhead{$\chi^{2}$} & \colhead{$p_{\alpha}$}}
        \startdata
        1ES2322-409 & 2 & single & 2.188 & 0.371 & ~ & ~ & 8.37 & 0.892\\
        1RXSJ054357.3-553206 & 1-13 & broken & 2.791 & 0.327 & -5.68 & 4.65 & 13.65 & 0.792\\
        3C120 & 5 & single & 1.635 & 0.220 & ~ & ~ & 33.52 & 0.388\\
        NGC1218 & 4 & single & 2.170 & 0.415 & ~ & ~ & 8.84 & 0.846\\
        PKS0035-252 & 3 & broken & 2.606 & 0.449 & -5.14 & 1.34 & 31.37 & 0.508\\
        PKS0130-17 & 3 & single & 2.133 & 0.419 & ~ & ~ & 17.33 & 0.532\\
        PKS0208-512 & 2,3 & broken & 3.405 & 0.847 & -4.99 & 0.95 & 23.89 & 0.462\\
        PKS0226-559 & 2,3 & bending & 2.822 & 0.943 & -4.84 & 0.67 & 19.88 & 0.570\\
        PKS0235-618 & 1,2,3 & bending & 2.017 & 0.817 & -5.83 & 6.57 & 24.33 & 0.314\\
        PKS0301-243 & 4 & single & 1.938 & 0.613 & ~ & ~ & 4.37 & 0.940\\
        PKS0336-177 & 4 & single & 1.885 & 0.562 & ~ & ~ & 3.40 & 0.996\\
        PKS0346-27 & 4 & broken & 2.400 & 0.434 & -5.69 & 4.76 & 12.29 & 0.816\\
        PKS0420-01 & 5 & single & 2.057 & 0.883 & ~ & ~ & 6.04 & 0.708\\
        PKS0422+00 & 5 & single & 2.274 & 0.495 & ~ & ~ & 17.49 & 0.470\\
        PKS0426-380 & 4,5 & broken & 2.116 & 0.893 & -5.43 & 2.61 & 28.99 & 0.316\\
        PKS0454-234 & 5 & single & 2.797 & 0.423 & ~ & ~ & 5.39 & 0.914\\
        PKS0521-36 & 5,6 & broken & 2.041 & 0.306 & -5.35 & 2.18 & 16.24 & 0.890\\
        \multirow{3}{*}{PKS0637-75} & 1,2,3 & single & 1.381 & 0.357 & ~ & ~ & 17.93 & 0.588\\
        ~ & 4,5,6 & broken & 2.337 & 0.396 & -5.71 & 4.98 & 33.87 & 0.262\\
        ~ & 7,8,9 & broken & 2.111 & 0.710 & -5.56 & 3.53 & 13.46 & 0.778\\
        PKS0736+01 & 7 & single & 2.163 & 0.272 & ~ & ~ & 143.55 & 0.164\\
        PKS0829+046 & 7 & broken & 2.478 & 0.536 & -5.25 & 1.73 & 19.61 & 0.710\\
        PKS1244-255 & 10 & broken & 2.126 & 0.570 & -5.62 & 4.05 & 12.14 & 0.906\\
        PKS2155-304 & 1 & bending & 3.173 & 0.820 & -5.4 & 2.44 & 11.59 & 0.878\\
        PKS2155-83 & 13 & bending & 2.219 & 0.744 & -5.48 & 2.93 & 16.45 & 0.692\\
        PKS2255-282 & 2 & single & 1.973 & 0.469 & ~ & ~ & 11.74 & 0.736\\
        PKS2326-502 & 1,2 & bending & 2.061 & 0.550 & -5.78 & 5.85 & 18.83 & 0.574\\
        PKS2345-16 & 2 & broken & 2.296 & 0.573 & -5.46 & 2.80 & 18.61 & 0.678\\
        PMNJ0948+0022 & 8 & single & 1.722 & 0.319 & ~ & ~ & 6.82 & 0.984\\
        PMNJ2141-6411 & 1 & single & 2.037 & 0.736 & ~ & ~ & 3.66 & 0.930\\
        PMNJ2345-1555 & 2 & broken & 1.737 & 0.471 & -5.39 & 2.38 & 14.15 & 0.902\\
        WISEJ031612.72+090443.3 & 4 & single & 2.300 & 0.703 & ~ & ~ & 11.23 & 0.602\\
        WISEJ085009.64-121335.3 & 8 & single & 2.205 & 0.521 & ~ & ~ & 17.45 & 0.510\\
        WISEJ095302.70-084018.2 & 8 & single & 2.870 & 0.429 & ~ & ~ & 15.20 & 0.642\\
        WISEJ105912.43-113422.6 & 9 & single & 1.607 & 0.393 & ~ & ~ & 18.02 & 0.652\\
        \enddata
    \end{deluxetable}
\end{longrotatetable}

\end{widetext}

\end{document}